\documentclass[final,3p,times,twocolumn,nonatbib]{elsarticle}
\usepackage[utf8]{inputenc}
\usepackage[american]{babel}
\usepackage{csquotes}
\usepackage[backend=biber,style=apa]{biblatex}
\DeclareDelimAlias[cite]{finalnamedelim}[parencite]{finalnamedelim}
\usepackage[T1]{fontenc}
\usepackage[utf8]{inputenc}
\usepackage{textcomp}
\usepackage{newtxtext,newtxmath}
\usepackage{graphicx,booktabs,array,tabularx,longtable,rotating}
\usepackage{microtype}
\usepackage{xurl}
\usepackage[dvipsnames]{xcolor}
\usepackage[colorlinks=true,linkcolor=MidnightBlue,citecolor=MidnightBlue,urlcolor=MidnightBlue,breaklinks=true]{hyperref}
\usepackage{caption}
\newcolumntype{L}[1]{>{\raggedright\arraybackslash}p{#1}}
\newcolumntype{Y}{>{\raggedright\arraybackslash}X}
\usepackage{accsupp}
\usepackage{flushend}
\journal{Preprint}

\makeatletter
\def\ps@pprintTitle{\let\@oddhead\@empty\let\@evenhead\@empty\def\@oddfoot{\footnotesize\itshape Preprint\hfill 16 October 2026}\let\@evenfoot\@oddfoot}
\makeatother
\hypersetup{pdftitle={Usability, User Experience, and Acceptability of the Trail Making Test in Virtual Reality in Younger and Older Adults},pdfauthor={Panagiotis Kourtesis, Dante O'Connor and Sarah E. MacPherson}}
\begin{document}\raggedbottom\begin{frontmatter}
\title{Usability, User Experience, and Acceptability of the Trail Making Test in Virtual Reality in Younger and Older Adults}
\author[aff1,aff2,aff3,aff4]{\href{https://orcid.org/0000-0002-2914-1064}{Panagiotis Kourtesis}\corref{cor1}}
\author[aff4]{\href{https://orcid.org/0009-0006-8239-4008}{Dante O'Connor}}
\author[aff4]{\href{https://orcid.org/0000-0001-8676-6514}{Sarah E. MacPherson}}
\address[aff1]{Department of Informatics and Telecommunications, National and Kapodistrian University of Athens, Athens, Greece}
\address[aff2]{Department of Psychology, National and Kapodistrian University of Athens, Athens, Greece}
\address[aff3]{Department of Psychology, The American College of Greece, Athens, Greece}
\address[aff4]{Department of Psychology, University of Edinburgh, Edinburgh, United Kingdom}
\cortext[cor1]{Correspondence: Panagiotis Kourtesis, Department of Psychology, National and Kapodistrian University of Athens, 157 84 Athens, Greece; \href{mailto:pkourtesis@psych.uoa.gr}{pkourtesis@psych.uoa.gr}}
\begin{abstract}
Immersive virtual reality (VR) adaptations of neuropsychological tests change how a test operates. Before their scores are interpreted, the individuals they are intended to assess should find them usable, acceptable and tolerable, and for older adults this evidence is usually missing. This study evaluated the experience of the Trail Making Test in Virtual Reality (TMT-VR), a head-gaze adaptation of the Trail Making Test that requires no hand-held controller, in 47 younger (20--38 years) and 44 older adults (64--81 years). After the task, participants rated usability (System Usability Scale), pragmatic and hedonic user experience (short User Experience Questionnaire), acceptability (adapted Service User Technology Acceptability Questionnaire) and cybersickness (Cybersickness in Virtual Reality Questionnaire). The study had three aims: to compare the two age groups on these ratings; to examine whether the ratings were associated with age group, gaming and technology experience, and whether cybersickness was associated with motion sickness susceptibility; and to examine whether the ratings were related to TMT-VR completion and error-adjusted times. Both groups rated the TMT-VR as highly usable, acceptable and pleasant to use. Older adults gave slightly lower usability and pragmatic ratings, but neither the age-group differences nor the associations with age group or previous experience were statistically significant after correction. Cybersickness was low, with 38 of the 91 participants reporting no symptoms; it did not differ between the age groups and was not associated with motion sickness susceptibility. No correlation between the ratings and TMT-VR performance survived correction. The TMT-VR with head-gaze selection is therefore usable, acceptable and well tolerated by older as well as younger adults. Because the ratings were favourable across the whole range of TMT-VR scores, slower or less accurate performance is unlikely to reflect difficulty with the interface.
\end{abstract}
\end{frontmatter}\clearpage

\section{Introduction}

\subsection{Evaluating the experience of immersive cognitive assessment}

Immersive virtual reality (VR) allows neuropsychologists to present naturalistic tasks under experimental control \parencite{ref7,ref51}. Low-cost consumer head-mounted displays have brought immersive VR into clinical and research use \parencite{ref12,ref57}. The case for ecologically valid measures of executive function \parencite{ref11,ref66} has motivated function-led virtual environments \parencite{ref52} and technology-based adaptations of established tests \parencite{ref17}. An adaptation nevertheless changes the demands of the test, including the response. A pen movement becomes head, gaze or controller selection, and the display, feedback and layout change. The joint position paper of the American Academy of Clinical Neuropsychology and the National Academy of Neuropsychology therefore treats a computerised adaptation as a new instrument whose properties must be demonstrated before clinical use \parencite{ref5}. An immersive adaptation must additionally be operable, acceptable and tolerable for its intended users. Adverse symptoms have been absent or minimal where displays have met minimum technological criteria and the interactions have been ergonomic \parencite{ref31}. The development guidelines for immersive neuropsychological software translate these criteria into software characteristics, namely ergonomic and naturalistic interaction, in-application tutorials, prompts and feedback, and rendering that holds the frame rate at the refresh rate of the display. These characteristics limit cybersickness and secure usability for gamers and non-gamers alike, and the guidelines place the evaluation of usability and adverse symptoms before any cognitive use \parencite{ref34}. Our previous evaluations of immersive applications for cognitive assessment and training have followed this order \parencite{ref35,ref21}. The present paper follows the same procedures for older adults, the population for whom the Trail Making Test is most often administered. It asks whether older adults find the Trail Making Test in Virtual Reality (TMT-VR), an immersive adaptation of the test, usable, acceptable and tolerable. It also asks whether their evaluation differs from that of younger adults.

The constructs involved in an adaptation overlap but are not interchangeable. Usability is the perceived ease and manageability of operating a system. It is usually rated after use on a standardised questionnaire; several such questionnaires exist, and their scores are not directly comparable \parencite{ref43}. User experience is broader. It separates pragmatic qualities (i.e., how well the system supports the task) from hedonic qualities (i.e., how stimulating and novel it feels), a distinction built into the User Experience Questionnaire and its short form \parencite{ref41,ref61}. Acceptability is whether the technology and its intended use are satisfactory to the user. It is approached through models of technology acceptance \parencite{ref68} and, for health technologies, through questionnaires completed by the service user \parencite{ref22}. Cybersickness comprises the nausea, vestibular and oculomotor symptoms provoked by immersion. It is most often measured with the Simulator Sickness Questionnaire (SSQ; \cite{ref28}). The SSQ was developed for flight simulators, and its psychometric properties are inadequate for cybersickness in consumer VR \parencite{ref64} as its VR-specific derivative omits nausea \parencite{ref29}. Questionnaires developed for VR that cover all three symptom types have been devised more recently (e.g., Cybersickness in Virtual Reality Questionnaire, CSQ-VR; \cite{ref36}). The instruments chosen for the present study are described in the Materials and methods. Favourable ratings do not show that a task measures the intended construct, and better performance does not imply a better experience. Middle-aged adults performed the TMT-VR more slowly than young adults yet rated it as more engaging \parencite{ref33}. Experience and performance are therefore distinct questions about the same assessment.

\subsection{The Trail Making Test in Virtual Reality and its users}

The Trail Making Test (TMT; \cite{ref54,ref55,ref8}) requires numbered circles to be connected in ascending order in Part A, and numbers and letters to be connected in alternating, ascending order in Part B. While the TMT is typically and widely used to assess executive dysfunction \parencite{ref56,ref44,ref16}, administering both Parts A and B allows the clinician or researcher to assess visual search, sequencing, psychomotor speed, working memory and task switching in one test \parencite{ref3,ref58,ref46}. TMT performance consistently slows with age \parencite{ref27,ref67}, and older adults are the population for whom the test is most often administered and normed. An immersive TMT adaptation should therefore be evaluated with older adults, not only the young adults who usually take part in VR studies.

The TMT-VR was developed to retain the established structure of the TMT while adding what immersive, automated test administration offers: standardised presentation and instruction, precise recording of completion times and errors, and a search space that extends around the user \parencite{ref19,ref33}. It preserves the sequencing structure of the TMT but replaces pen movements with other ways of selecting targets in an immersive environment. In young adults, eye-tracking, head-gaze and controller-based selection have all yielded high usability, user experience and acceptability, and general computer and smartphone use, rather than gaming, correlated positively with these ratings \parencite{ref19}. Across young and middle adulthood, head-gaze selection was the fastest mode on Part B with the fewest selection errors, and acceptability was uniformly high \parencite{ref33}. The only usability difference in that study was that middle-aged adults with little gaming experience rated usability lower than young adults did. In young adults with attention-deficit/hyperactivity disorder (ADHD) and neurotypical controls, the TMT-VR was rated as usable and acceptable \parencite{ref21}. These studies identified head-gaze selection as the mode open to the widest range of participants: it needs no eye-tracking calibration, which can fail with progressive lenses or in poor lighting, and no hand-held controller \parencite{ref33}. In generic selection tasks outside the TMT-VR, head pointing was slower than gaze or controller pointing when targets appeared in unexpected positions \parencite{ref65}. However, none of these studies have examined how older adults experience the task. Older adults are less familiar with digital interfaces and fine motor control declines with age \parencite{ref24}, so favourable evaluations in younger users cannot be assumed to extend to older age. To our knowledge, no study has evaluated the usability, acceptability, user experience or cybersickness of the TMT-VR in older adults.

\subsection{Age, technology experience and the acceptance of immersive VR}

How people experience new technology in later life varies greatly. In the CREATE study involving 1,204 adults, the association between age and technology adoption was mediated by cognitive abilities, computer self-efficacy and computer anxiety \parencite{ref14}. Attitudes and abilities are among the strongest predictors of technology use in later life \parencite{ref13}. Older adults' willingness to adopt a technology depends mainly on its perceived value and on the confidence and effort involved in learning it \parencite{ref48}. Digital device use is also related to cognitive function in older adults: daily users of computers and touchscreen devices outperformed non-daily users on global cognition, processing speed, short-term memory and executive tasks, including the TMT \parencite{ref70}. Familiarity, ability and age are therefore intertwined. As immersive VR reaches older adults \parencite{ref63}, the first experience of a head-mounted display becomes part of the assessment. Reviews of immersive VR for older users agree that the task should be self-paced, with simplified interactions and structured practice \parencite{ref26,ref60}. This is the in-application assistance that the development guidelines prescribe for users without gaming experience \parencite{ref34}, and it is a property of the administration protocol as much as of the software.

Older adults have been found to accept head-mounted VR well after their first exposure. Among 76 VR-naive older adults, attitudes changed from neutral to positive after immersion, with minimal cybersickness, whereas watching time-lapse videos produced no such change \parencite{ref25}. Older adults with cognitive or physical impairments reported immediate favourable experiences during brief immersive nature viewing \parencite{ref2}. Older adults with mild cognitive impairment or subjective cognitive decline accepted an immersive shopping task with negligible side effects, although those with cognitive impairment made more selection errors and felt less in control \parencite{ref4}. The two passive viewing studies \parencite{ref25,ref2} show that older adults are willing to wear a headset. Only active tasks, such as the shopping task of \textcite{ref4}, expose the interaction difficulties that matter for an assessment. A sequential selection task such as the TMT-VR, which requires continuous search, aiming at targets and target confirmation, has to be evaluated under those demands.

Previous experience with digital technology has been found to be a more specific predictor of how a VR task is experienced than age. Gaming experience, as recorded by the Gaming Skills Questionnaire (GSQ; \cite{ref72,ref42}), combines exposure to interactive environments with self-rated proficiency. Routine computer, smartphone and VR use reflect a broader familiarity with digital interfaces. Gaming experience accompanies differences in cognitive and spatial performance \parencite{ref49,ref40}. Presence and prior exposure also influence performance on VR assessments \parencite{ref47}, and older adults can engage productively with game-like tasks \parencite{ref1}. Habitual gaming is nevertheless uncommon in later life, so gaming and technology experience have to be examined separately. In the TMT-VR, familiarity with technology rather than gaming accompanied more favourable evaluations \parencite{ref33}. Immersive working-memory and psychomotor assessments were largely independent of computing and gaming experience and were rated as more usable than desktop versions \parencite{ref37}, and greater gaming and smartphone experience accompanied milder cybersickness \parencite{ref39}. When age group, gaming experience and technology experience are examined in the same model, it becomes clearer which of the three is associated with the assessment ratings once the other two are held constant.

\subsection{Cybersickness and individual differences}

An acceptable neuropsychological assessment in immersive VR also has to limit adverse symptoms. Cybersickness is generally attributed to a conflict between visual motion and the absence of corresponding vestibular and proprioceptive signals \parencite{ref15}. It depends on the exposure as well as on the person. Across head-mounted VR studies, symptoms have varied with the content, visual stimulation, locomotion and exposure duration. Gaming content produced the most symptoms, with older samples reporting fewer symptoms than younger adults \parencite{ref59}. Adequate hardware \parencite{ref31} and software design, above all ergonomic interaction, navigation without joystick-driven optic flow and a stable frame rate \parencite{ref34}, are preconditions of tolerability. In older adults, a systematic review of cybersickness reported mostly minor symptoms, but also small samples, weak designs and inconsistent symptom reporting \parencite{ref18}. Active joystick navigation increased symptoms relative to passive navigation in older but not in younger adults \parencite{ref62}. A selection task without locomotion avoids these triggers, so low symptoms can be expected, but they should still be measured.

Symptom intensity also differs between individuals exposed to the same content. Motion sickness susceptibility, measured with the short form of the Motion Sickness Susceptibility Questionnaire (MSSQ-Short; \cite{ref20}), was the strongest predictor of cybersickness in adults aged 20--45 years exposed to a virtual roller-coaster ride \parencite{ref39}. In that study, gaming and smartphone experience accompanied milder symptoms. Gaming experience predicted lower symptom intensity in two further exposure studies \parencite{ref30,ref50}, and an apparent sex difference disappeared once gaming experience was taken into account \parencite{ref30}. Age, technology familiarity and susceptibility may therefore contribute to the symptoms reported after an assessment in immersive VR. A task with a low likelihood of cybersickness allows these individual contributions to be examined under realistic assessment conditions rather than under an exposure designed to induce symptoms.

\subsection{The present study}

The present study had three aims in evaluating the experience of the TMT-VR with head-gaze selection in 47 younger adults aged 20--38 years and 44 older adults aged 64--81 years. The first was to characterise and compare usability, pragmatic and hedonic user experience, acceptability and cybersickness across the two age groups. The second was to examine whether these ratings were associated with age group and reported gaming and technology experience, and whether cybersickness was associated with motion sickness susceptibility, adjusting for education and sex. The third aim was to examine whether the ratings were related to performance on the TMT-VR, indexed by its completion and error-adjusted times.

Favourable usability, user experience and acceptability ratings were expected in both age groups, with somewhat lower usability ratings in older adults and more favourable ratings among participants with greater technology experience \parencite{ref33}. Low cybersickness was expected under a head-gaze task without locomotion, and participants with higher motion sickness susceptibility were expected to report somewhat more symptoms. The ratings were not expected to be strongly related to performance \parencite{ref33}, so the third aim was exploratory.

\section{Materials and methods}

\subsection{Participants}

The sample comprised 91 adults with 5--22 years of education (M = 16.4, SD = 2.9): 47 younger adults aged 20--38 years (M = 25.1, SD = 4.1; 25 women) and 44 older adults aged 64--81 years (M = 71.8, SD = 4.9; 24 women). There were no participants aged 39--63 years, so age was analysed as the contrast between the two groups. Seventy-one participants were tested at the University of Edinburgh. They were recruited through the volunteer panel of the School of Philosophy, Psychology and Language Sciences, postgraduate mailing lists, word of mouth and posters, and received \pounds{}20 for their participation. Twenty participants (10 younger and 10 older adults), all native English speakers, were tested at The American College of Greece (ACG) under the identical protocol. They were recruited through the ACG volunteer panel and the Open Care Centre for the Elderly (KAPI) of Aghia Paraskevi, and received no payment. No one had a history of neurological or psychiatric conditions, learning difficulties, conditions affecting balance, or serious susceptibility to motion sickness, defined by an MSSQ-Short total score above 30 (observed maximum = 29.5). All participants provided written informed consent under the approvals given in the Ethics Statement, and the study was performed in compliance with the Declaration of Helsinki.

No participant withdrew or was excluded after screening, and all 91 records were complete for the analysed variables. No participant was excluded on the basis of questionnaire scores, task performance or model influence.

\subsection{Apparatus and TMT-VR}

The TMT-VR (build TMT-VR\_2025, version 0.1, 19 May 2025; Unity 2021.3.0f1, Windows 64-bit, OpenXR with the SteamVR runtime) was presented on an HTC VIVE Pro Eye head-mounted display (HTC Corporation, Taoyuan, Taiwan; 1440 \ensuremath{\times} 1600 pixels per eye, 90 Hz, 110\textdegree{} field of view), tracked by two SteamVR base stations and driven by a laptop with an Intel Core i7-8750H processor, 16 GB of RAM and an NVIDIA GeForce RTX 2070 Max-Q graphics processor. Targets were selected by head gaze. A ray cast from the headset positioned a central reticle, and a target was registered when the reticle had rested on it for 0.5 s, the dwell threshold set on every target in the build. Selection therefore used head orientation rather than eye tracking and required no hand-held controller. Participants stood throughout the assessment, with the experimenter beside them.

Each part of the TMT-VR presented numbered or numbered and lettered blocks at randomly generated positions in a plain virtual room (Figure 1). The 25 positions were drawn from a pool of 168 candidate positions. Part A required the numbers 1 to 25 to be selected in ascending order, and Part B required alternation between numbers and letters (1-A-2-B \ldots\unkern{}). A short tone marked the arrival of the reticle on a block. A confirmation tone indicated a correct selection, the block turned yellow and a yellow trail was drawn from the previous block. An incorrect selection was indicated by a buzzer and turned the block briefly red, after which the participant went back to the last correct target. The feedback sounds were spatialised at the position of the block. The application recorded the completion time and the number of wrong-target selections for each part. The design and interaction principles of the task are described in the published implementations \parencite{ref19,ref33}, and the software follows the development guidelines for immersive neuropsychological applications \parencite{ref34}. That is, it offers an ergonomic interaction (here, head-gaze selection, which needs neither a controller nor locomotion), a static low-detail environment, in-application video tutorials and practice trials, and immediate visual and auditory feedback.

\begin{figure}[!tp]\centering
\includegraphics[width=\columnwidth]{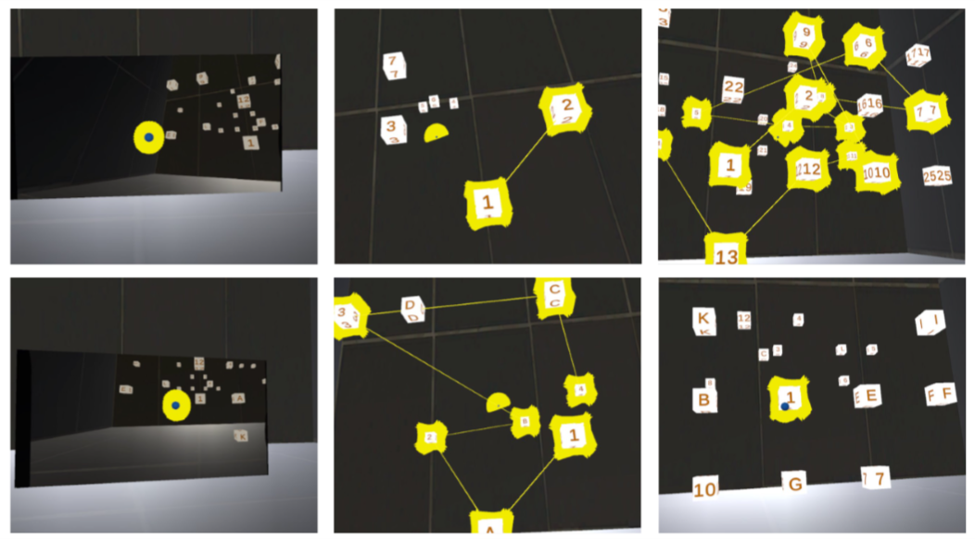}
\caption{Screenshots from the TMT-VR at different stages of Part A and Part B. Top row: Part A video tutorial (left), practice trial (middle) and main test (right). Bottom row: Part B video tutorial (left), practice trial (middle) and main test (right). Targets are selected by resting the central head-gaze reticle on a block; a correct selection turns the block yellow and draws the trail from the previous target. English-language demonstrations: \href{https://www.youtube.com/watch?v=npki7i4OnwY}{Task A video} and \href{https://www.youtube.com/watch?v=immvIkOyVuA}{Task B video}.}\label{fig:1}
\end{figure}

\subsection{Procedure}

Participants first provided demographic information and completed the technology-experience questionnaires and the MSSQ-Short. Before the TMT-VR, the experimenter explained head-controlled selection, the required sequences and the importance of both speed and accuracy, and warned participants that blocks could obscure one another, so that they might need to turn to locate a target. Narrated in-environment videos provided the instructions, and a practice trial preceded each part (Figure 1). The TMT-VR took approximately 10--15 min to administer. No participant required assistance beyond this briefing, and no session was interrupted or stopped. Immediately after the TMT-VR, participants removed the headset and completed the CSQ-VR, which asked about cybersickness symptoms during or immediately after immersion, followed by the SUS, the UEQ-S and the acceptability questionnaire. Each participant contributed one set of ratings and one post-session cybersickness score. The same protocol was followed at both sites.

\subsection{Measures}

\subsubsection{Usability}

The System Usability Scale (SUS; \cite{ref9,ref10}), the most widely used usability questionnaire \parencite{ref43}, comprised ten statements rated from 1 (strongly disagree) to 5 (strongly agree). For odd-numbered items, the contribution was the response minus one, and for even-numbered items, it was five minus the response. The sum was multiplied by 2.5 to give a score ranging from 0 to 100 (maximum 100), with higher values indicating more favourable perceived usability. No numerical benchmark was used as a threshold of acceptable usability.

\subsubsection{User experience}

The User Experience Questionnaire-Short (UEQ-S; \cite{ref61}), the eight-item short form of the User Experience Questionnaire \parencite{ref41}, comprised eight semantic-differential pairs rated on seven-point scales: four pragmatic (obstructive--supportive, complicated--easy, inefficient--efficient, confusing--clear) and four hedonic (boring--exciting, not interesting--interesting, conventional--inventive, usual--leading edge). Responses ranged from 1 to 7, with the positive anchor at 7, and were recoded by subtracting four, so that each domain mean ranged from \ensuremath{-}3 to +3 (maximum +3), with higher values more favourable. The pragmatic and hedonic domain means were analysed separately.

\subsubsection{Acceptability}

Acceptability was assessed with the ten-item adaptation of the Service User Technology Acceptability Questionnaire (SUTAQ; \cite{ref22}), a questionnaire developed for the users of health technologies, in the form used in previous TMT-VR studies \parencite{ref19,ref33}. Statements about satisfaction with the assessment and its fit with everyday life were rated on six-point agreement scales. The mean of the ten items, ranging from 1 to 6 (maximum 6), with higher values more favourable, was analysed as a composite index of acceptability. No cut-off, factor structure or equivalence with the full SUTAQ total was assumed, because the ten-item adaptation has not been validated as a stand-alone short form.

\subsubsection{Gaming and technology experience}

The Gaming Skills Questionnaire (GSQ), validated in adolescents \parencite{ref72} and across adulthood \parencite{ref42}, recorded the frequency of play and self-rated ability for six game genres (sports, first-person shooter, role-playing, action--adventure, strategy and puzzle games), each on a six-point scale, giving a total ranging from 12 to 72. A technology composite summed the corresponding frequency and ability ratings for computers, smartphones and VR, giving a total from 6 to 36. Higher scores indicated higher frequency and ability. Both composites represented reported exposure and perceived proficiency rather than observed skill, and both were used as continuous predictors.

\subsubsection{Motion sickness susceptibility}

The Motion Sickness Susceptibility Questionnaire-Short (MSSQ-Short; \cite{ref20}) asked how often nine forms of transport or motion (i.e., cars, buses or coaches, trains, aircraft, small boats, ships, swings, roundabouts and funfair rides) caused sickness during childhood and during the last ten years. Each item was rated from 0 (never felt sick) to 3 (frequently felt sick). Forms of transport never experienced were not counted. For each section, the sum of the ratings was multiplied by nine and divided by the number of transport types experienced, and the two section scores were summed to give a total from 0 to 54. Higher scores indicated greater motion sickness susceptibility. Every participant had experienced at least one transport type in childhood and in recent adulthood.

\subsubsection{Cybersickness}

The Cybersickness in Virtual Reality Questionnaire (CSQ-VR; \cite{ref36}) comprised six items, two each for nausea, vestibular and oculomotor symptoms, rated on seven-point scales. The total score ranged from 6 to 42 (maximum 42), with 6 denoting no symptom endorsement, and each subscale from 2 to 14. The CSQ-VR was used rather than the Simulator Sickness Questionnaire because it was developed for head-mounted VR and covers all three symptom types. In its validation study, it showed better internal consistency than the Simulator Sickness Questionnaire and the Virtual Reality Sickness Questionnaire, and greater sensitivity to the performance decline that cybersickness produces \parencite{ref36}. The total was the principal symptom outcome and the three subscales were secondary outcomes. Only the post-session scores were analysed, so the results refer to the symptom level after VR immersion rather than to symptom change or induction.

\subsubsection{TMT-VR performance}

Six TMT-VR scores indexed performance: the Part A, Part B and total completion times, and the corresponding error-adjusted times \parencite{ref32}. An error-adjusted time multiplied the completion time by one plus the proportion of the 24 transitions on which a wrong-target selection was made (the total used the 48 transitions of both parts), so that an error extended the time by the fraction of the sequence it represented. The adjusted times were functions of the completion times and error counts, so the six scores carried overlapping information and their associations were interpreted as dependent rather than as separate replications.

\subsection{Statistical analyses}

All analyses were conducted in R 4.4.1 \parencite{ref53} with the package `sandwich' \parencite{ref71}, except the sample comparisons of Table 1, which were computed in Python 3.13 with SciPy 1.18 \parencite{ref69} following the same test conventions. The main analysis script verified the checksum of the dataset before running, and every script is archived with the data \parencite{ref38}. A dated analysis plan preceded the reported models; the study was not preregistered. Questionnaire totals were checked against their component items. Tests were two-sided with \ensuremath{\upalpha} = .05, confidence intervals (CIs) were pointwise 95\% intervals, and bootstrap CIs used 5,000 paired-participant resamples. To correct for multiple comparisons, each set of related tests was treated as a family (e.g., the five age-group comparisons of the principal outcomes, or the 12 focal regression coefficients), and p values were adjusted within that family. The Holm procedure \parencite{ref23} was used for the group comparisons and the regression coefficients, and the Benjamini--Hochberg procedure \parencite{ref6} for the exploratory correlations between the ratings and performance. The family is identified with each result. Comparisons that did not meet the corrected criterion were not interpreted as evidence of equivalence. The Supplementary Material reports every model, correlation, sensitivity analysis and diagnostic in full.

Sex, education, gaming and technology experience and motion sickness susceptibility were compared between the age groups with Fisher's exact test (sex), Welch's t-tests (education, technology and susceptibility) and the Mann--Whitney U-test (gaming, which was concentrated at its minimum), with Holm adjustment across these five tests (Table 1). Usability, pragmatic UX, hedonic UX and acceptability were compared between the groups with Welch's t-tests. The CSQ-VR total, which was discrete and concentrated at its lower bound, was compared with the Mann--Whitney U-test with continuity correction. Holm adjustment covered these five comparisons as one family, and the three CSQ-VR subscales formed a secondary Mann--Whitney family with its own adjustment. The size of each group difference is described by the older-minus-younger difference in the rating with its Welch CI and by Hedges' g with a bootstrap CI, with rank-based tests and bootstrap intervals as sensitivity checks.

Four ordinary least-squares models regressed usability, pragmatic UX, hedonic UX and acceptability on age group (older versus younger), the gaming and technology composites (centred, per sample SD), education (centred) and sex (centred male indicator), with HC3 heteroscedasticity-consistent standard errors and t-based intervals \parencite{ref45}. Holm adjustment covered the 12 focal age-group, gaming and technology coefficients. A fifth model regressed the CSQ-VR total on the standardised MSSQ-Short total and the same covariates, with Holm adjustment across its six coefficients. As sensitivity analyses, the models were re-estimated with within-group age added, on rank-transformed variables, and, for the symptom model, with the recent-adulthood MSSQ section in place of the total and across susceptibility tertiles. Residuals, leverage, Cook's distance (i.e., the influence of each participant on the fitted coefficients), variance inflation factors and the range of fitted values were inspected without deleting any cases.

Each of the six TMT-VR scores was correlated (Pearson) with the five ratings (CSQ-VR total, usability, pragmatic UX, hedonic UX and acceptability), giving 30 tests in the whole sample and 30 within each age group, each family Benjamini--Hochberg-adjusted, with Spearman sensitivity families. Pooled partial correlations residualised the scores and ratings on age group, within-group age, education and sex, a further 30-test family whose bootstrap CIs refitted the adjustment within each resample.

\section{Results}

\subsection{Sample characteristics}

Table 1 presents the sample characteristics overall and by age group, with the age-group comparisons. The age groups did not differ significantly in sex composition, education or motion sickness susceptibility. Younger participants reported significantly more gaming and technology experience (both Holm p < .001; Table 1). Gaming scores were positively skewed in the older group, with 52 participants (57.1\%) across the sample obtaining the minimum score. Reported VR use was minimal in both groups: 89 of the 91 participants gave the lowest frequency rating for VR, and 41 of the 44 older participants reported no VR ability. Motion sickness susceptibility, which entered the analyses as a predictor, ranged from 0 to 29.5 on the 0--54 scale and had similar means in the two groups (Table 1).

\begin{table*}[!tp]
\caption{Sample characteristics overall and by age group.}\label{tab:1}
\centering\small\setlength{\tabcolsep}{4pt}\renewcommand{\arraystretch}{1.15}
\begin{tabular}{L{0.215332\dimexpr\textwidth-56pt\relax}L{0.123816\dimexpr\textwidth-56pt\relax}L{0.123816\dimexpr\textwidth-56pt\relax}L{0.123816\dimexpr\textwidth-56pt\relax}L{0.214040\dimexpr\textwidth-56pt\relax}L{0.096899\dimexpr\textwidth-56pt\relax}L{0.102283\dimexpr\textwidth-56pt\relax}}
\toprule
\textbf{\boldmath Characteristic} & \textbf{\boldmath All (N = 91)} & \textbf{\boldmath Younger (n = 47)} & \textbf{\boldmath Older (n = 44)} & \textbf{\boldmath Older \ensuremath{-} younger [95\% CI]} & \textbf{\boldmath Raw p} & \textbf{\boldmath Holm p} \\
\midrule
Age (years) & 47.69 (23.87) & 25.13 (4.07) & 71.80 (4.85) & --- & --- & --- \\
Age range (years) & 20--81 & 20--38 & 64--81 & --- & --- & --- \\
Education (years) & 16.40 (2.86) & 16.64 (3.00) & 16.14 (2.71) & \ensuremath{-}0.50 [\ensuremath{-}1.69, 0.69] & .404 & 1.000 \\
Women, n (\%) & 49 (53.8) & 25 (53.2) & 24 (54.5) & --- & --- & --- \\
Men, n (\%) & 42 (46.2) & 22 (46.8) & 20 (45.5) & OR 0.95 [0.38, 2.34] & 1.000 & 1.000 \\
Gaming (12--72) & 16.24 (7.08) & 20.00 (8.22) & 12.23 (0.86) & \ensuremath{-}7.77 [\ensuremath{-}10.20, \ensuremath{-}5.35] & < .001 & < .001 \\
Technology (6--36) & 20.74 (3.72) & 23.32 (2.04) & 17.98 (3.06) & \ensuremath{-}5.34 [\ensuremath{-}6.44, \ensuremath{-}4.25] & < .001 & < .001 \\
MSSQ-Short total (0--54) & 11.01 (8.60) & 11.12 (8.72) & 10.90 (8.57) & \ensuremath{-}0.22 [\ensuremath{-}3.82, 3.38] & .903 & 1.000 \\
\bottomrule\end{tabular}
\par\vspace{4pt}\begin{minipage}{\textwidth}\footnotesize Note. Values are mean (SD) unless otherwise indicated. Gaming = Gaming Skills Questionnaire total (12--72); Technology = technology composite (6--36); MSSQ-Short = Motion Sickness Susceptibility Questionnaire short form total (0--54). Differences are older minus younger with pointwise Welch 95\% CIs; for sex, the older-versus-younger odds ratio (OR) of being male with its 95\% CI. Tests: Welch's t for education (t(88.88) = \ensuremath{-}0.84), technology (t(74.29) = \ensuremath{-}9.72) and MSSQ-Short (t(88.78) = \ensuremath{-}0.12); the Mann--Whitney U-test with continuity correction for gaming (U = 302.0), whose p value tests a rank contrast rather than the displayed mean difference; Fisher's exact test for sex. Holm adjustment covers the five sample comparisons as one family. The age groups are defined by age, which is therefore not tested.\end{minipage}\end{table*}

\subsection{Usability, user experience and acceptability}

Participants rated the TMT-VR favourably on every instrument (Table 2; Figure 2). Usability scores had a median of 92.5 on the 0--100 SUS, ranged from 60 to 100, and 12 participants (13.2\%) gave the maximum rating of 100. Nineteen participants gave the maximum pragmatic score of +3 and 14 participants gave the maximum hedonic score of +3, whereas four participants gave a negative pragmatic score and seven a negative hedonic score. No acceptability score fell below 3.7 on the 1--6 scale.

Older participants gave lower usability and pragmatic ratings than younger participants, whereas the hedonic and acceptability means were closer (Table 2). The usability and pragmatic differences were of moderate size (Hedges' g = \ensuremath{-}0.49 and \ensuremath{-}0.47) and were significant before correction (raw p = .023 and .028), but no difference remained statistically significant after Holm adjustment across the five outcome comparisons (both Holm p = .116), and the rank-based sensitivity tests led to the same decision. The mean ratings of both age groups lay in the favourable range of every scale (Table 2).

\begin{table*}[!tp]
\caption{Ratings and cybersickness by age group.}\label{tab:2}
\centering\small\setlength{\tabcolsep}{4pt}\renewcommand{\arraystretch}{1.15}
\begin{tabular}{L{0.215332\dimexpr\textwidth-56pt\relax}L{0.123816\dimexpr\textwidth-56pt\relax}L{0.123816\dimexpr\textwidth-56pt\relax}L{0.123816\dimexpr\textwidth-56pt\relax}L{0.214040\dimexpr\textwidth-56pt\relax}L{0.096899\dimexpr\textwidth-56pt\relax}L{0.102283\dimexpr\textwidth-56pt\relax}}
\toprule
\textbf{\boldmath Outcome} & \textbf{\boldmath All M (SD)} & \textbf{\boldmath Younger M (SD)} & \textbf{\boldmath Older M (SD)} & \textbf{\boldmath Difference [95\% CI]} & \textbf{\boldmath Raw p} & \textbf{\boldmath Holm p} \\
\midrule
SUS (0--100) & 89.42 (7.99) & 91.28 (6.38) & 87.44 (9.08) & \ensuremath{-}3.83 [\ensuremath{-}7.13, \ensuremath{-}0.54] & .023 & .116 \\
Pragmatic UX (\ensuremath{-}3 to +3) & 1.87 (0.98) & 2.09 (0.81) & 1.64 (1.10) & \ensuremath{-}0.45 [\ensuremath{-}0.86, \ensuremath{-}0.05] & .028 & .116 \\
Hedonic UX (\ensuremath{-}3 to +3) & 1.59 (1.10) & 1.64 (1.09) & 1.52 (1.11) & \ensuremath{-}0.12 [\ensuremath{-}0.58, 0.34] & .602 & 1.000 \\
Acceptability (1--6) & 4.83 (0.45) & 4.85 (0.47) & 4.80 (0.44) & \ensuremath{-}0.05 [\ensuremath{-}0.24, 0.14] & .628 & 1.000 \\
CSQ-VR total (6--42) & 8.70 (3.66) & 8.89 (3.46) & 8.50 (3.90) & \ensuremath{-}0.39 [\ensuremath{-}1.93, 1.15] & .350 & 1.000 \\
CSQ-VR nausea (2--14) & 2.74 (1.14) & 2.89 (1.20) & 2.57 (1.07) & \ensuremath{-}0.33 [\ensuremath{-}0.80, 0.15] & .130 & .390 \\
CSQ-VR vestibular (2--14) & 3.02 (1.53) & 3.00 (1.41) & 3.05 (1.66) & 0.05 [\ensuremath{-}0.60, 0.69] & .920 & .920 \\
CSQ-VR oculomotor (2--14) & 2.95 (1.73) & 3.00 (1.35) & 2.89 (2.07) & \ensuremath{-}0.11 [\ensuremath{-}0.85, 0.62] & .286 & .572 \\
\bottomrule\end{tabular}
\par\vspace{4pt}\begin{minipage}{\textwidth}\footnotesize Note. Differences are older minus younger with pointwise Welch 95\% CIs. Tests are Welch's t for the four ratings and Mann--Whitney with continuity correction for the CSQ-VR total and subscales, whose p values test a rank contrast rather than the displayed mean difference. Holm adjustment covers the five principal outcomes (SUS, pragmatic UX, hedonic UX, acceptability and the CSQ-VR total) as one family and the three subscales as a secondary family. SUS = System Usability Scale; UX = user experience on the short User Experience Questionnaire; acceptability = mean of the ten adapted SUTAQ items; CSQ-VR = Cybersickness in Virtual Reality Questionnaire, completed after the session. Hedges' \textit{g}, bootstrap intervals and rank-based sensitivity tests are in Supplementary Section S3.\end{minipage}\end{table*}

\begin{figure*}[!tp]\centering
\includegraphics[width=.85\textwidth,height=.72\textheight,keepaspectratio]{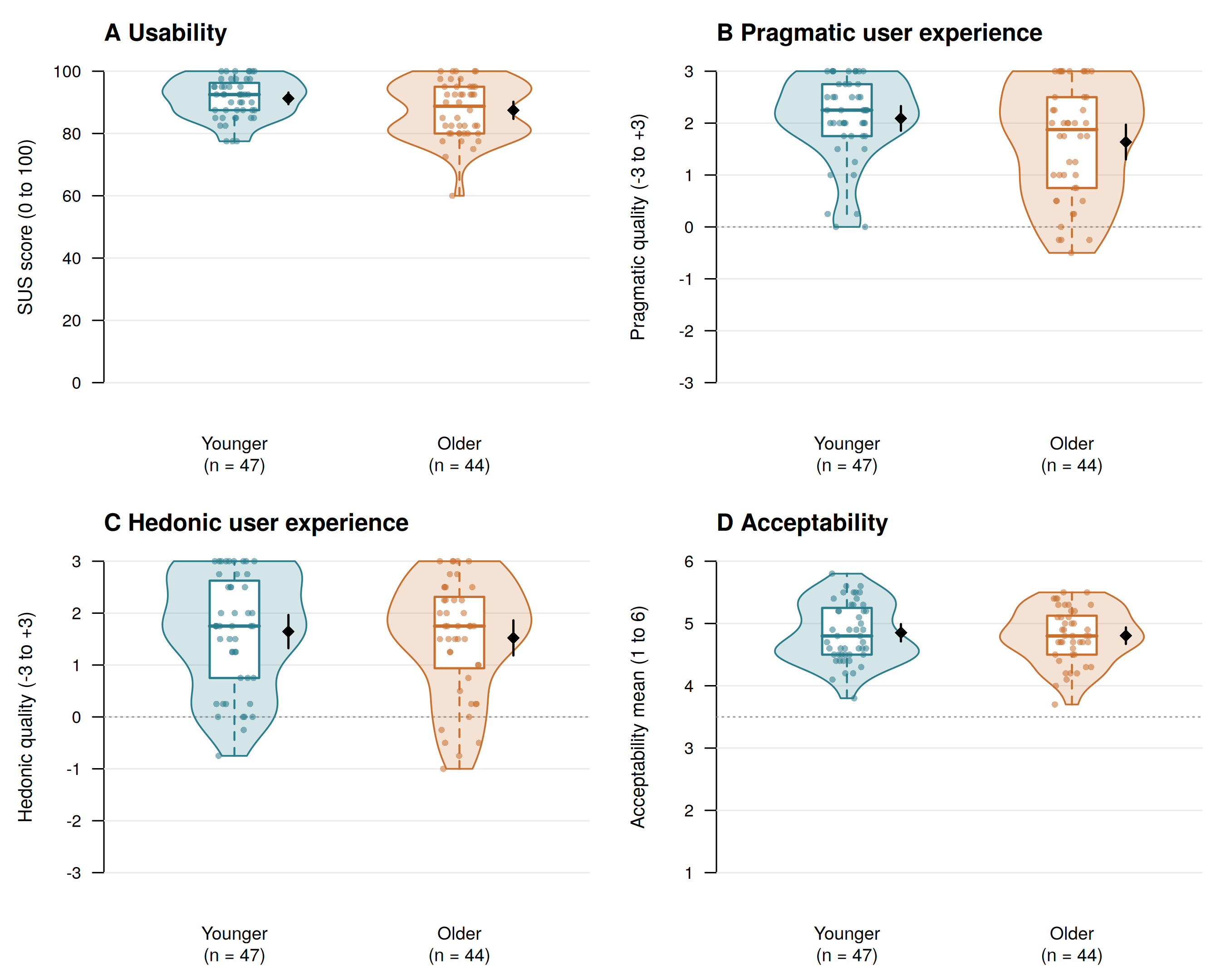}
\caption{Ratings by age group. (A) Usability (System Usability Scale, 0--100). (B) Pragmatic and (C) hedonic user experience (short User Experience Questionnaire, \ensuremath{-}3 to +3). (D) Acceptability (mean of the ten adapted SUTAQ items, 1--6). Violin outlines show kernel-density estimates truncated to the observed range; internal boxes show medians and interquartile ranges, with whiskers to the most extreme observations within 1.5 interquartile ranges; black diamonds and bars show group means and 95\% confidence intervals. The full theoretical range of each scale is shown; the reference lines mark zero for user experience and the midpoint for acceptability and are not thresholds. The violins smooth discrete scores and are a distributional aid.}\label{fig:2}
\end{figure*}

\subsection{Associations with age group and reported experience}

In the adjusted models (Table 3), higher technology scores corresponded to higher usability (2.32 SUS points per SD; raw p = .043) and pragmatic UX (0.32 points per SD; raw p = .050), and the technology coefficients remained positive under rank transformation. None of the 12 focal coefficients met the corrected criterion (smallest Holm p = .519), so the pointwise intervals do not establish an association once the 12 coefficients are corrected as a family. Once technology experience, which differed markedly between the groups, was held constant, the adjusted older-minus-younger difference in usability was \ensuremath{-}0.91 points (95\% CI [\ensuremath{-}6.30, 4.49]) and the pragmatic difference was close to zero (0.00 points, 95\% CI [\ensuremath{-}0.64, 0.64]). The unadjusted group differences were therefore absorbed by the difference in technology experience. The hedonic coefficient of age group was small and imprecise (\ensuremath{-}0.42 points, 95\% CI [\ensuremath{-}1.09, 0.25]), and the acceptability coefficient was close to zero. The gaming coefficients were small and imprecise for every rating, as expected given the restricted gaming scores of the older group. Adding within-group age changed no result (Supplementary Section S4).

\begin{table*}[!tp]
\caption{Adjusted associations of age group and reported experience with the ratings.}\label{tab:3}
\centering\small\setlength{\tabcolsep}{4pt}\renewcommand{\arraystretch}{1.15}
\begin{tabular}{L{0.161499\dimexpr\textwidth-48pt\relax}L{0.204565\dimexpr\textwidth-48pt\relax}L{0.129199\dimexpr\textwidth-48pt\relax}L{0.224806\dimexpr\textwidth-48pt\relax}L{0.139966\dimexpr\textwidth-48pt\relax}L{0.139966\dimexpr\textwidth-48pt\relax}}
\toprule
\textbf{\boldmath Outcome} & \textbf{\boldmath Predictor} & \textbf{\boldmath b} & \textbf{\boldmath 95\% CI} & \textbf{\boldmath Raw p} & \textbf{\boldmath Holm p} \\
\midrule
SUS & Older vs younger & \ensuremath{-}0.91 & [\ensuremath{-}6.30, 4.49] & .739 & 1.000 \\
SUS & Gaming per SD & \ensuremath{-}0.59 & [\ensuremath{-}2.23, 1.05] & .476 & 1.000 \\
SUS & Technology per SD & 2.32 & [0.07, 4.57] & .043 & .519 \\
Pragmatic UX & Older vs younger & 0.00 & [\ensuremath{-}0.64, 0.64] & .995 & 1.000 \\
Pragmatic UX & Gaming per SD & 0.01 & [\ensuremath{-}0.18, 0.21] & .893 & 1.000 \\
Pragmatic UX & Technology per SD & 0.32 & [0.00, 0.63] & .050 & .549 \\
Hedonic UX & Older vs younger & \ensuremath{-}0.42 & [\ensuremath{-}1.09, 0.25] & .216 & 1.000 \\
Hedonic UX & Gaming per SD & \ensuremath{-}0.22 & [\ensuremath{-}0.50, 0.06] & .122 & 1.000 \\
Hedonic UX & Technology per SD & \ensuremath{-}0.03 & [\ensuremath{-}0.40, 0.34] & .860 & 1.000 \\
Acceptability & Older vs younger & \ensuremath{-}0.02 & [\ensuremath{-}0.35, 0.31] & .907 & 1.000 \\
Acceptability & Gaming per SD & 0.02 & [\ensuremath{-}0.13, 0.17] & .808 & 1.000 \\
Acceptability & Technology per SD & 0.00 & [\ensuremath{-}0.13, 0.13] & 1.000 & 1.000 \\
\bottomrule\end{tabular}
\par\vspace{4pt}\begin{minipage}{\textwidth}\footnotesize Note. Ordinary least-squares models with age group (older versus younger), gaming and technology (per sample SD; SDs 7.08 and 3.72 points), education and sex as predictors; \textit{N} = 91, residual df = 85. b = unstandardised coefficient in outcome units with its pointwise HC3 95\% CI. Holm adjustment covers the 12 focal age-group, gaming and technology coefficients. The education and sex coefficients, the sensitivity specifications and the diagnostics are in Supplementary Section S4.\end{minipage}\end{table*}

\subsection{Cybersickness}

CSQ-VR totals were concentrated at the lower end of the 6--42 scale (median 7, interquartile range 6--10, observed range 6--20), with 38 participants (41.8\%) at the minimum score; the same held for each subscale (Table 2). Neither the total (U = 920.5, Holm p = 1.000; Figure 3A) nor any subscale (Holm p \ensuremath{\geq} .390, adjusted as a family of three) differed significantly between the age groups. The largest standardised difference concerned nausea, which was lower in older adults (g = \ensuremath{-}0.28, 95\% CI [\ensuremath{-}0.72, 0.11]). A minority of participants in each group nevertheless reported appreciably more than the minimum level of cybersickness symptoms.

In the adjusted symptom model, motion sickness susceptibility had a positive but imprecise association with the CSQ-VR total (0.54 points per susceptibility SD, 95\% CI [\ensuremath{-}0.32, 1.39]; Holm p = .648; Figure 3B), and the susceptibility tertiles differed by less than one point. Older age group (\ensuremath{-}2.04 points, 95\% CI [\ensuremath{-}4.01, \ensuremath{-}0.08]), higher technology scores (\ensuremath{-}1.25 points per SD, 95\% CI [\ensuremath{-}2.24, \ensuremath{-}0.26]) and male sex (\ensuremath{-}1.81 points, 95\% CI [\ensuremath{-}3.32, \ensuremath{-}0.31]) were associated with lower symptoms in the pointwise analysis, but none met the six-test corrected criterion (Holm p = .167, .084 and .093, respectively), and the rank-based sensitivity analysis was less supportive. Gaming and education were unrelated to symptoms. The linear model produced one fitted value marginally below the scale minimum and skewed residuals, so the rank-based and grouped sensitivity models are retained in the Supplementary Material (Section S6).

\begin{figure*}[!tp]\centering
\includegraphics[width=\textwidth,height=.72\textheight,keepaspectratio]{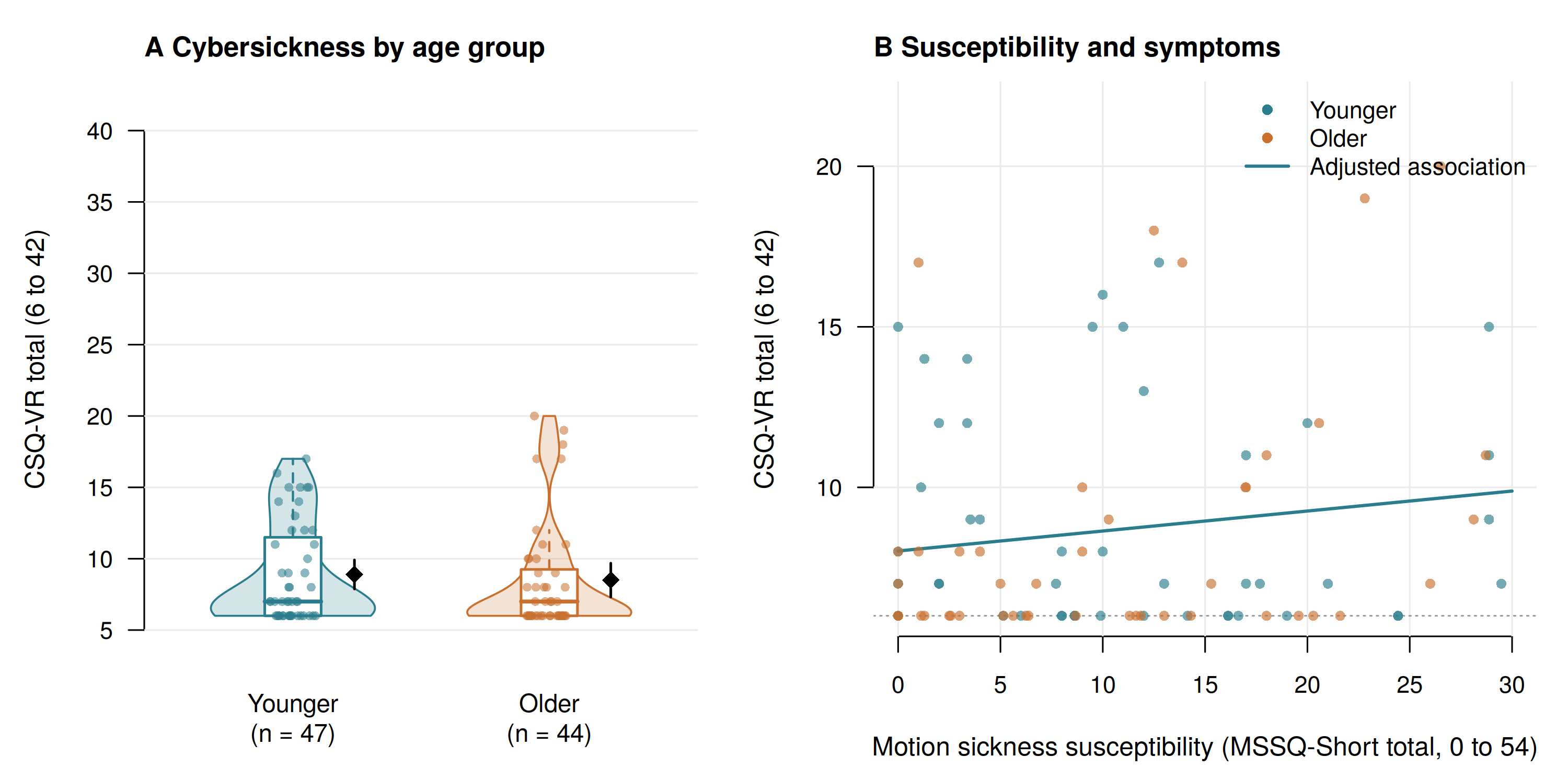}
\caption{Cybersickness. (A) CSQ-VR totals by age group on the theoretical 6--42 scale: violin outlines show kernel-density estimates truncated to the observed range, boxes show medians and interquartile ranges with whiskers to 1.5 interquartile ranges, points are individual participants and black diamonds and bars show group means with 95\% confidence intervals. (B) CSQ-VR total against motion sickness susceptibility (MSSQ-Short total), with the adjusted linear association of the symptom model drawn at the sample means of the other predictors. The plotted association describes the symptom level reported after the session; its pointwise interval is given in the text.}\label{fig:3}
\end{figure*}

\subsection{Ratings and TMT-VR performance}

Table 4 presents the correlations between the five rating scales and the six TMT-VR scores for the whole sample. None of the 30 correlations survived Benjamini--Hochberg adjustment in the whole sample (smallest adjusted p = .659), the younger group (.446) or the older group (.951). The same held for the Spearman sensitivity families and for the 30 pooled partial correlations adjusted for age group, within-group age, education and sex (smallest adjusted p = .783). The coefficients were of modest size (|r| \ensuremath{\leq} .23 in the whole sample and \ensuremath{\leq} .34 within the age groups). In the whole sample, the younger group and the older group, two, three and zero of the 30 raw p values, respectively, fell below .05, against the 1.5 expected by chance alone. Slower Part B times accompanied slightly lower usability in the whole sample (r = \ensuremath{-}.23 and \ensuremath{-}.21 for the completion and adjusted times), and slower Part A times accompanied slightly higher hedonic ratings in the younger group (r = .30 and .34; Supplementary Sections S10--S11). These results identify no robust performance correlate of usability, user experience, acceptability or cybersickness.

\begin{table*}[!tp]
\caption{Correlations between the ratings and the TMT-VR completion and error-adjusted times in the whole sample.}\label{tab:4}
\centering\small\setlength{\tabcolsep}{4pt}\renewcommand{\arraystretch}{1.15}
\begin{tabular}{L{0.192506\dimexpr\textwidth-48pt\relax}L{0.161499\dimexpr\textwidth-48pt\relax}L{0.161499\dimexpr\textwidth-48pt\relax}L{0.161499\dimexpr\textwidth-48pt\relax}L{0.161499\dimexpr\textwidth-48pt\relax}L{0.161499\dimexpr\textwidth-48pt\relax}}
\toprule
\textbf{\boldmath TMT-VR score} & \textbf{\boldmath SUS r [95\% CI]} & \textbf{\boldmath Pragmatic UX r [95\% CI]} & \textbf{\boldmath Hedonic UX r [95\% CI]} & \textbf{\boldmath Acceptability r [95\% CI]} & \textbf{\boldmath CSQ-VR total r [95\% CI]} \\
\midrule
Part A time & \ensuremath{-}.01 [\ensuremath{-}.23, .17] & \ensuremath{-}.06 [\ensuremath{-}.29, .14] & .10 [\ensuremath{-}.13, .29] & \ensuremath{-}.10 [\ensuremath{-}.28, .12] & .00 [\ensuremath{-}.24, .25] \\
Part B time & \ensuremath{-}.23 [\ensuremath{-}.39, \ensuremath{-}.03] & \ensuremath{-}.09 [\ensuremath{-}.34, .10] & \ensuremath{-}.02 [\ensuremath{-}.22, .19] & .09 [\ensuremath{-}.09, .26] & \ensuremath{-}.04 [\ensuremath{-}.21, .15] \\
Total time & \ensuremath{-}.18 [\ensuremath{-}.36, .02] & \ensuremath{-}.10 [\ensuremath{-}.32, .10] & .03 [\ensuremath{-}.17, .24] & .03 [\ensuremath{-}.17, .22] & \ensuremath{-}.03 [\ensuremath{-}.22, .17] \\
Part A adjusted time & .01 [\ensuremath{-}.23, .18] & \ensuremath{-}.03 [\ensuremath{-}.28, .16] & .13 [\ensuremath{-}.10, .31] & \ensuremath{-}.11 [\ensuremath{-}.28, .11] & \ensuremath{-}.03 [\ensuremath{-}.24, .21] \\
Part B adjusted time & \ensuremath{-}.21 [\ensuremath{-}.37, \ensuremath{-}.03] & \ensuremath{-}.05 [\ensuremath{-}.34, .14] & \ensuremath{-}.01 [\ensuremath{-}.21, .17] & .07 [\ensuremath{-}.10, .23] & \ensuremath{-}.06 [\ensuremath{-}.20, .15] \\
Total adjusted time & \ensuremath{-}.16 [\ensuremath{-}.35, .04] & \ensuremath{-}.06 [\ensuremath{-}.33, .13] & .04 [\ensuremath{-}.17, .23] & .01 [\ensuremath{-}.18, .20] & \ensuremath{-}.05 [\ensuremath{-}.23, .16] \\
\bottomrule\end{tabular}
\par\vspace{4pt}\begin{minipage}{\textwidth}\footnotesize Note. \textit{N} = 91. Values are Pearson r with pointwise paired-participant bootstrap 95\% CIs from 5,000 draws. Benjamini--Hochberg adjustment covers the 30 correlations as one family; none meets the corrected criterion (smallest adjusted \textit{p} = .659). Adjusted time = completion time \ensuremath{\times} (1 + wrong-target selections/24) for each part and \ensuremath{\times} (1 + wrong-target selections/48) for the total. SUS = System Usability Scale; UX = user experience on the short User Experience Questionnaire; acceptability = mean of the ten adapted SUTAQ items; CSQ-VR = Cybersickness in Virtual Reality Questionnaire. The within-group, Spearman and partial-correlation families are in Supplementary Sections S10--S11.\end{minipage}\end{table*}

\section{Discussion}

This study asked how younger and older adults rate the TMT-VR under head-gaze selection, whether those ratings and cybersickness are associated with age group, previous experience and motion sickness susceptibility, and whether the ratings are related to TMT-VR performance. Three findings emerged. First, both age groups rated the assessment as highly usable, acceptable and pleasant to use. Older adults gave slightly lower usability and pragmatic ratings, and technology-experienced participants gave slightly higher ones, but no group difference or experience association survived correction, and the adjusted group differences were close to zero once technology experience was held constant. Second, cybersickness was low in both groups and on every subscale, with no supported age-group difference or susceptibility association. Third, no TMT-VR score was associated with any rating after correction. The TMT-VR therefore proved usable, acceptable and well tolerated by older as well as younger adults, the prerequisite for examining what its scores measure in this population. Because the ratings were favourable across the whole range of performance, the completion and error-adjusted times are unlikely to reflect difficulty with the interface.

\subsection{A usable and acceptable assessment for younger and older adults}

The favourable ratings extend the evaluation of the TMT-VR from young and middle-aged adults \parencite{ref19,ref33} and young adults with ADHD \parencite{ref21} to older adults, using the head-gaze mode that those studies identified as the mode open to the widest range of users. The older group's acceptability and usability ratings were of the same order as those reported in young adults. The concentration of usability scores near the top of the scale, with more than a tenth of participants at the maximum, indicates that most participants encountered no difficulty with the TMT-VR. This concentration limits the variance available to the analyses, a point addressed in the limitations. For the question of whether older adults can operate the assessment, however, this is the central result. An unfamiliar immersive interface, operated by head movement alone while standing, was experienced as easy and acceptable by adults in their seventies, almost none of whom had used VR before. Usability, availability and cost are the pragmatic conditions of adoption for clinical VR \parencite{ref57}, and the first is met here for the population with the strongest reason to be sceptical of a new interface \parencite{ref13}.

These findings are consistent with the favourable initial responses of older adults to head-mounted VR reported for passive viewing \parencite{ref25,ref2}. The participants in the present study, however, faced a more demanding situation. Passive viewing shows that older adults are willing to wear a headset, whereas the TMT-VR, like other sequential selection tasks, requires individuals to search continuously, aim at targets and confirm them under time pressure. It is in active tasks that interaction difficulties have appeared, from selection errors and reduced perceived control in cognitively impaired users to a fall by a participant who tried to lean on a virtual shelf \parencite{ref4}. The TMT-VR imposes no locomotion or reaching. The protocol supported the interaction with a video explanation, practice and an experimenter beside the participant throughout, in line with the design recommendations for older users \parencite{ref26,ref60} and the in-application assistance that the development guidelines identify as decisive for users without gaming experience \parencite{ref34}. The ratings therefore include the assessment and the support; they do not isolate the interface, and they do not show that participants could complete the assessment unaided.

\subsection{Age group and previous experience}

The direction of the age and experience results matched the pattern across young and middle adulthood, in which older users rated the interface as slightly less usable and technology familiarity accompanied more favourable evaluations \parencite{ref33}. Here, however, neither the group differences nor the technology coefficients were statistically significant after correction for multiple comparisons. Once technology experience, which differed markedly between the groups, was held constant, the usability difference between the age groups decreased to a fraction of a point. The parsimonious reading of this finding is that any age-related difference in the experience of the TMT-VR is small and partly attributable to technology familiarity. This is compatible with evidence that the association between age and technology use runs through self-efficacy, anxiety and cognitive abilities rather than age itself \parencite{ref14,ref13}. It also aligns with the better cognitive performance of older adults who use digital devices daily \parencite{ref70} and with the lesser dependence of immersive than of desktop assessments on computing and gaming experience \parencite{ref37}. When an immersive assessment follows the development guidelines, as the TMT-VR does \parencite{ref34}, older adults rate it as favourably as younger adults do, with no statistically significant difference between the groups.

Gaming experience was uninformative, as almost every older participant reported the minimum gaming score. Therefore, the gaming coefficient could only describe variation among younger adults. The technology composite, which varied in both groups, is the more interpretable index of previous experience, as is the case in young and middle-aged adults \parencite{ref33}. Both composites combine exposure with self-rated proficiency, so their associations describe reported familiarity rather than interaction skill. In practice, age does not predict poor experience with the TMT-VR, whereas unfamiliarity with digital interfaces may. Given an appropriately designed interface, such as the controller-free head-gaze selection of the TMT-VR \parencite{ref34}, the experience of the assessment is largely independent of age. Familiarity and support needs should therefore be assessed for the individual rather than inferred from age.

\subsection{Cybersickness}

Symptoms were low in both age groups and on each subscale, consistent with the generally minor cybersickness reported for older adults \parencite{ref18} and with what the content of the task would predict. Participants with serious motion sickness susceptibility (an MSSQ-Short total above 30) had not been recruited, which will also have limited the symptoms observed. Symptoms across head-mounted VR studies depend on visual stimulation, locomotion and exposure duration \parencite{ref59}, and it was active navigation that increased symptoms in older adults \parencite{ref62}. The display met the technological criteria under which adverse symptoms have been absent or minimal \parencite{ref31}. The software has the characteristics that the development guidelines prescribe against cybersickness, namely ergonomic interaction, no artificial locomotion and immediate feedback \parencite{ref34}, and the TMT-VR is a brief selection task without optic flow. Participants stood during the task, so the low symptoms were not obtained under a protective seated condition. The psychomotor nature of the task may also contribute. Engagement in eye--hand coordination tasks after a virtual roller-coaster ride that had induced cybersickness accelerated the decline of nausea and vestibular symptoms \parencite{ref50}. The TMT-VR keeps the user continuously engaged in goal-directed head movements towards stationary targets, a form of active, self-generated motion unlike the passive visual motion that provokes symptoms. Whether this engagement protects against symptoms cannot be established here, because no alternative task or interaction mode was compared. The study shows that the assessment as administered was well tolerated by older adults, not why.

Motion sickness susceptibility, the strongest predictor of cybersickness in younger adults exposed to a virtual roller-coaster ride designed to induce symptoms \parencite{ref39}, had only a small and unsupported association here. The most plausible reason is that the exposure did not provoke enough symptoms for susceptibility to express itself: with two fifths of participants at the scale minimum, there was little symptom variance to predict. The lower pointwise symptoms among older, technology-experienced and male participants echo the fewer symptoms of older samples across studies \parencite{ref59}, the associations of gaming and smartphone experience with milder cybersickness \parencite{ref30,ref39,ref50} and the disappearance of the sex difference once gaming experience is taken into account \parencite{ref30}. None of these differences were statistically significant after correction, however, and none should guide screening. Of note, symptoms were measured once, after the headset had been removed, and nausea and vestibular symptoms decline after removal \parencite{ref39}, so the scores index residual rather than peak symptoms. This argues for measuring symptoms during immersion, which the CSQ-VR permits in its in-VR form \parencite{ref36}.

\subsection{Ratings and performance}

The ratings were not related to TMT-VR performance in any population or after adjustment, and the number of nominal associations was close to chance, consistent with the modest and precision-specific associations reported across young and middle adulthood \parencite{ref33}. The practical implication is direct. Neither the ratings nor the symptom scores can substitute for measured performance, and performance cannot be read as a proxy for how the assessment was experienced. The two kinds of measure describe different properties of the same session, and both belong in the evaluation of an immersive assessment. At the same time, the absence of any detectable association is informative. Participants reported a favourable experience across the whole range of performance, so slower or less accurate performance on the TMT-VR is unlikely to reflect difficulty with the interface, and the performance scores can be interpreted on their own terms.

\subsection{Implications for immersive cognitive assessment}

The principal contribution of this work was to establish, for older adults, the usability, acceptability and tolerability of the TMT-VR under the interaction mode that requires the least of the user. This is the evidence that our development guidelines place before the cognitive use of an immersive application \parencite{ref34}, that reviews of VR for older users identify as the first condition of adoption \parencite{ref26,ref60}, and that the position paper on computerised assessment requires before clinical use \parencite{ref5}. One consequence is that head-gaze selection can be retained as the default mode for older adults, because it was experienced as easy without calibration or a controller and produced minimal symptoms. A second is that the in-environment video, the practice trial and the presence of the examiner form part of what was rated and should be retained rather than treated as optional. A third is that, as the ratings and performance were unrelated, the human-factors and the cognitive properties of the assessment can be examined separately. What the study does not supply is the cognitive evidence itself. The correspondence of the TMT-VR scores with the paper-and-pencil test, their sensitivity to age, and their reliability and diagnostic accuracy in older adults should be the focus of future work.

\subsection{Limitations and future directions}

The evaluation of four experience constructs and cybersickness in the same participants, the measurement of gaming, technology and susceptibility, and the explicit correction for multiple comparisons within every family of tests are strengths of this study. Its limitations identify three directions for future research. First, the sample consisted of two separated age groups of relatively highly educated adults from two sites, with no participants in middle adulthood. The results therefore describe the contrast between the age groups and do not show how the experience of the assessment changes across adulthood or in less educated groups. Sampling across the adult range, with a record of how many people were approached and how many declined, is needed before acceptability can be described for a population. Second, the ratings were compressed towards the favourable end of every scale, which limits the variance available for detecting differences and associations. In addition, the ten-item adaptation of the SUTAQ has not been validated as a stand-alone short form of the full questionnaire. Studies that vary the interface, for example by comparing interaction modes within older adults, would generate the variance that a single supported protocol does not. Third, the evaluation rests on one supervised session, with symptoms only measured after the headset had been removed. Repeated sessions, symptom measurement during immersion and documentation of assistance and stopping decisions are required before independent or repeated use can be evaluated.

\section{Conclusion}

To our knowledge, this study provides the first evaluation of the usability, user experience, acceptability and cybersickness of the Trail Making Test in Virtual Reality in older adults, alongside younger adults, using head-gaze selection to avoid controller use and calibration. Both age groups rated the TMT-VR as highly usable, acceptable and pleasant to use, and cybersickness was low on every subscale, with 38 of the 91 participants reporting no symptoms at all. Older adults and participants with less technology experience gave slightly lower usability and pragmatic ratings, but no age-group difference or experience association was statistically significant after correction, and the small differences that did appear were accounted for by technology familiarity rather than by age. No rating was associated with TMT-VR performance, so slower or less accurate performance is unlikely to reflect difficulty with the interface. The study shows that an immersive adaptation of one of the most widely used neuropsychological tests can be operated comfortably by older adults, almost all of whom had no previous VR experience. It identifies head-gaze selection, in-VR video instruction, practice and the presence of an examiner as an administration protocol that older adults accept. Finally, the study supplies usability and tolerability evidence that the development guidelines and the clinical position papers require before the cognitive properties of an immersive test are examined, which can now be addressed for the TMT-VR in this population.

\section*{Data Availability Statement}

The de-identified dataset with its variable codebook, the analysis and figure scripts, the logs of the R analysis, and the archived outputs of this study are openly available in the Zenodo repository of the GERIATRIC programme (\cite{ref38}; \url{https://doi.org/10.5281/zenodo.23270971}). The data and outputs are shared under the CC BY 4.0 licence and the code under the MIT licence, as participant consent and the ethics approvals permit. The participant identifier in the deposited dataset is an arbitrary serial number that cannot be resolved outside the study. Reasonable requests for further information should be addressed to the corresponding author.

\section*{Ethics Statement}

The study involving human participants was conducted in accordance with the Declaration of Helsinki and with the local legislation and institutional requirements. The Edinburgh data collection was approved by the PPLS Research Ethics Committee of the University of Edinburgh (protocol no. 244-2324/13). The data collection at The American College of Greece was carried out within the GERIATRIC programme (Horizon Europe MSCA grant agreement no. 101149390) under the approvals of the Research Ethics and Deontology Committee (\BeginAccSupp{method=hex,unicode,ActualText=0395002E0397002E0394002E0395002E}\ensuremath{\mathrm{E}}.\ensuremath{\mathrm{H}}.\ensuremath{\Delta}.\ensuremath{\mathrm{E}}.\EndAccSupp{}) of the National and Kapodistrian University of Athens (protocol no. 20579/5.3.2024; approval decision no. 152/4.4.2024) and of the Institutional Review Board of The American College of Greece (expedited review, protocol \#202602564, approved 16 February 2026). All participants provided written informed consent.

\section*{Author Contributions}

PK: Conceptualization, Data curation, Formal analysis, Funding acquisition, Investigation, Methodology, Project administration, Resources, Software, Supervision, Validation, Visualization, Writing -- original draft, Writing -- review \& editing. DO: Validation, Investigation, Data curation, Writing -- review \& editing. SM: Validation, Writing -- review \& editing, Resources, Supervision, Project administration. All authors contributed to the article and approved the submitted version.

\section*{Funding}

This project has received funding from the European Union's Horizon Europe research and innovation programme under the Marie Sk\l{}odowska-Curie grant agreement No 101149390 (GERIATRIC). Funded by the European Union. Views and opinions expressed are however those of the author(s) only and do not necessarily reflect those of the European Union or the European Research Executive Agency (REA). Neither the European Union nor the granting authority can be held responsible for them.

\section*{Acknowledgments}

The authors thank all participants for their time and engagement.

\section*{Conflict of Interest}

The authors declare that the research was conducted in the absence of any commercial or financial relationships that could be construed as a potential conflict of interest.

\section*{Generative AI Statement}

During the preparation of this work, the authors used OpenAI Codex (GPT-6; September 2026) and Anthropic Claude (Claude Fable 5.1 and Claude Opus 5.5, via Claude Code/Cowork; September--October 2026) to assist with coding of the analysis scripts, language refinement, and manuscript editing. The authors reviewed, verified, and edited all outputs and take full responsibility for the content of this work. Generative AI tools are not authors and received no CRediT roles.

\section*{Author identifiers}

Panagiotis Kourtesis, \url{https://orcid.org/0000-0002-2914-1064}; Dante O'Connor, \url{https://orcid.org/0009-0006-8239-4008}; Sarah E. MacPherson, \url{https://orcid.org/0000-0001-8676-6514}.

\section*{Author contact details}

Dante O'Connor, dan.oconnor43@outlook.com; Sarah E. MacPherson, sarah.macpherson@ed.ac.uk.

\section*{Supplementary Material}

The Supplementary Material (Sections S1--S11) reports the questionnaire checks, descriptive statistics, group comparisons with sensitivity tests, the full rating and symptom models with their sensitivity specifications and diagnostics, and the within-group, Spearman and partial ratings--performance families.

\printbibliography[title={References}]
\end{document}


\raggedbottom
{\LARGE\bfseries Supplementary Material\par}

{\large\bfseries Usability, User Experience, and Acceptability of the Trail Making Test in Virtual Reality in Younger and Older Adults\par}

{\large Panagiotis Kourtesis, Dante O'Connor, Sarah E. MacPherson\par}

This Supplementary Material reports in full the analyses summarised in the article. Section S1 describes the data, the scoring of every questionnaire and the checks of the stored totals. Section S2 gives the descriptive statistics and score distributions, and Section S3 the age-group comparisons with their sensitivity tests. Sections S4--S8 report the rating and symptom models with their sensitivity specifications and diagnostics. Sections S9--S11 report the correlations between the ratings and TMT-VR performance: the whole-sample family, the within-group and Spearman families, and the partial correlations. Every value is taken from the archived outputs of the study (Zenodo, \url{https://doi.org/10.5281/zenodo.23270971}), except the model diagnostics of Supplementary Tables S12, S13, S15 and S23, which the code in the Appendix computes from the archived data.

\clearpage\section*{S1 Data, scoring and questionnaire checks}
The data set of record comprises 91 participants, 47 younger adults aged 20--38 years and 44 older adults aged 64--81 years, with complete records for every analysed variable. Nothing was imputed, winsorised or excluded. The analysis script verified the SHA-256 checksum of the data workbook before running. It then recomputed every stored questionnaire total from its items and would have stopped at any discrepancy (Supplementary Table S1). The CSQ-VR subscales were checked in the same way for this Supplementary Material. Every stored score equalled its item formula.

\Needspace{309pt}
\textbf{Supplementary Table S1.} Scoring of the questionnaires and the TMT-VR scores, and the check applied to each stored score.

\setlength{\tabcolsep}{2.5pt}\renewcommand{\arraystretch}{1.12}
\small
\begin{longtable}{L{0.151164\dimexpr\textwidth-25pt\relax}L{0.226746\dimexpr\textwidth-25pt\relax}L{0.261614\dimexpr\textwidth-25pt\relax}L{0.093016\dimexpr\textwidth-25pt\relax}L{0.267460\dimexpr\textwidth-25pt\relax}}
\toprule
\textbf{\boldmath Measure} & \textbf{\boldmath Items and responses} & \textbf{\boldmath Score} & \textbf{\boldmath Range} & \textbf{\boldmath Check} \\
\midrule\endfirsthead
\multicolumn{5}{l}{\textbf{Supplementary Table S1} (continued)}\\[3pt]\toprule
\textbf{\boldmath Measure} & \textbf{\boldmath Items and responses} & \textbf{\boldmath Score} & \textbf{\boldmath Range} & \textbf{\boldmath Check} \\
\midrule\endhead
\midrule\multicolumn{5}{r}{\footnotesize Continued on next page}\\\endfoot
\bottomrule\endlastfoot
SUS & 10 items, 1--5 & Odd items: response \ensuremath{-} 1; even items: 5 \ensuremath{-} response; sum \ensuremath{\times} 2.5 & 0--100 & Stored total = item formula (91/91) \\
UEQ-S & 8 semantic-differential pairs, 1--7 & Each item \ensuremath{-} 4; pragmatic = mean of items 1--4; hedonic = mean of items 5--8 & \ensuremath{-}3 to +3 & Both domain means = item formula (91/91) \\
Acceptability (adapted SUTAQ) & 10 items, 1--6 & Mean of the ten items & 1--6 & Stored mean = item formula (91/91) \\
Gaming (GSQ) & Frequency and ability for six game genres, 1--6 (12 items) & Sum & 12--72 & Stored total = item sum (91/91) \\
Technology composite & Frequency and ability for computers, smartphones and VR, 1--6 (6 items) & Sum & 6--36 & Stored total = item sum (91/91) \\
MSSQ-Short & 9 forms of transport or motion, childhood and last ten years, 0--3; not experienced = not counted & Each section: sum \ensuremath{\times} 9 / number of types experienced; total = childhood + recent adulthood & 0--54 & Both sections and total = item formula (91/91); all totals \ensuremath{\leq} 30 (eligibility; maximum 29.48) \\
CSQ-VR (after the session) & 6 items, 1--7: two each for nausea, vestibular and oculomotor symptoms & Total = sum of the six items; subscale = sum of its two items & 6--42; subscales 2--14 & Total = item sum (91/91); each subscale = its two items (91/91) \\
TMT-VR scores & Completion time (s) and wrong-target selections, Parts A and B & Adjusted time = time \ensuremath{\times} (1 + errors/24) per part; total adjusted = total time \ensuremath{\times} (1 + total errors/48) & --- & Total time = Part A + Part B time (91/91) \\
\end{longtable}\normalsize
Note. SUS~=~System Usability Scale; UEQ-S~=~short User Experience Questionnaire; SUTAQ~=~Service User Technology Acceptability Questionnaire; GSQ~=~Gaming Skills Questionnaire; MSSQ-Short~=~Motion Sickness Susceptibility Questionnaire short form; CSQ-VR~=~Cybersickness in Virtual Reality Questionnaire. Checks of the stored totals used a tolerance of 10\textsuperscript{\ensuremath{-}}\textsuperscript{9} (10\textsuperscript{\ensuremath{-}}\textsuperscript{3} s for the TMT-VR total time).

\textbf{Analysis settings. }All analyses were conducted in R 4.4.1 with the package `sandwich' 3.1-3, except the sample comparisons of Table 1 of the article, which were computed in Python 3.13 with SciPy 1.18. Tests were two-sided with \ensuremath{\upalpha}~=~.05, and every interval is a pointwise 95\% interval. Regression inference used HC3 heteroscedasticity-consistent standard errors with t-based intervals. Bootstrap intervals are percentile intervals from 5,000 resamples (seed 20260921). Participants were resampled within each age group for the group comparisons, and within the analysed sample for the correlations and regression coefficients. A dated analysis plan preceded the reported models; the study was not preregistered.

\textbf{Age. }No participant was aged 39--63 years, so the reported models enter age as the contrast between the age groups. Within-group age (each participant's age minus the mean age of their group, per 10 years) enters only as a sensitivity term. The models were also estimated with age as a continuous variable across the whole sample (per 10 years), as a sensitivity specification (Supplementary Tables S11 and S17). The rank-transformed, recent-adulthood MSSQ-Short, susceptibility-tertile and subscale models also enter age as a continuous variable. No decision after correction differs between the specifications.

\textbf{Predictors. }Age group was coded 1 for older and 0 for younger participants and centred. Gaming, technology and motion sickness susceptibility entered as standardised scores (per sample SD of 7.08, 3.72 and 8.60 points, respectively). Education was centred at the sample mean (16.40 years), and sex entered as a centred indicator of male sex. Because every predictor was centred, the intercept of each model is the predicted score at the sample means of the predictors.

\textbf{Model diagnostics. }The fit and influence diagnostics, variance inflation factors and predictor correlations (Supplementary Tables S12, S13, S15 and S23) were computed in R 4.4.1 with the code in the Appendix. The code reads the processed data set of the repository after verifying its checksum, and it gives the diagnostics of the age-group models and of the continuous-age models.

\clearpage\section*{S2 Descriptive statistics}
Supplementary Table S2 gives the descriptive statistics of the ratings, the cybersickness scores and the predictors for the whole sample and each age group. Supplementary Tables S3--S5 give the sex and VR-use counts, the number of participants at the minimum and maximum of each scale, and the distribution of the CSQ-VR totals. The ratings were concentrated towards the favourable end of every scale, and 38 of the 91 participants reported no cybersickness symptom at all.

\Needspace{160pt}
\textbf{Supplementary Table S2.} Descriptive statistics of the ratings, the cybersickness scores and the predictors.

\setlength{\tabcolsep}{2.5pt}\renewcommand{\arraystretch}{1.12}
\fontsize{7.6}{9}\selectfont
\begin{longtable}{L{0.218414\dimexpr\textwidth-45pt\relax}L{0.086189\dimexpr\textwidth-45pt\relax}L{0.045934\dimexpr\textwidth-45pt\relax}L{0.074731\dimexpr\textwidth-45pt\relax}L{0.068952\dimexpr\textwidth-45pt\relax}L{0.074731\dimexpr\textwidth-45pt\relax}L{0.143683\dimexpr\textwidth-45pt\relax}L{0.143683\dimexpr\textwidth-45pt\relax}L{0.143683\dimexpr\textwidth-45pt\relax}}
\toprule
\textbf{\boldmath Variable} & \textbf{\boldmath Group} & \textbf{\boldmath n} & \textbf{\boldmath M} & \textbf{\boldmath SD} & \textbf{\boldmath Mdn} & \textbf{\boldmath Q1 to Q3} & \textbf{\boldmath Min to max} & \textbf{\boldmath 95\% CI of M} \\
\midrule\endfirsthead
\multicolumn{9}{l}{\textbf{Supplementary Table S2} (continued)}\\[3pt]\toprule
\textbf{\boldmath Variable} & \textbf{\boldmath Group} & \textbf{\boldmath n} & \textbf{\boldmath M} & \textbf{\boldmath SD} & \textbf{\boldmath Mdn} & \textbf{\boldmath Q1 to Q3} & \textbf{\boldmath Min to max} & \textbf{\boldmath 95\% CI of M} \\
\midrule\endhead
\midrule\multicolumn{9}{r}{\footnotesize Continued on next page}\\\endfoot
\bottomrule\endlastfoot
SUS (0--100) & All & 91 & 89.42 & 7.99 & 92.50 & 82.50 to 95.00 & 60.00 to 100.00 & [87.76, 91.09] \\
 & Younger & 47 & 91.28 & 6.38 & 92.50 & 87.50 to 96.25 & 77.50 to 100.00 & [89.40, 93.15] \\
 & Older & 44 & 87.44 & 9.08 & 88.75 & 80.00 to 95.00 & 60.00 to 100.00 & [84.68, 90.20] \\
Pragmatic UX (\ensuremath{-}3 to +3) & All & 91 & 1.87 & 0.98 & 2.00 & 1.25 to 2.62 & \ensuremath{-}0.50 to 3.00 & [1.67, 2.08] \\
 & Younger & 47 & 2.09 & 0.81 & 2.25 & 1.75 to 2.75 & 0.00 to 3.00 & [1.85, 2.33] \\
 & Older & 44 & 1.64 & 1.10 & 1.88 & 0.75 to 2.50 & \ensuremath{-}0.50 to 3.00 & [1.30, 1.97] \\
Hedonic UX (\ensuremath{-}3 to +3) & All & 91 & 1.59 & 1.10 & 1.75 & 0.75 to 2.50 & \ensuremath{-}1.00 to 3.00 & [1.36, 1.81] \\
 & Younger & 47 & 1.64 & 1.09 & 1.75 & 0.75 to 2.62 & \ensuremath{-}0.75 to 3.00 & [1.32, 1.96] \\
 & Older & 44 & 1.52 & 1.11 & 1.75 & 0.94 to 2.31 & \ensuremath{-}1.00 to 3.00 & [1.18, 1.86] \\
Acceptability (1--6) & All & 91 & 4.83 & 0.45 & 4.80 & 4.50 to 5.20 & 3.70 to 5.80 & [4.73, 4.92] \\
 & Younger & 47 & 4.85 & 0.47 & 4.80 & 4.50 to 5.25 & 3.80 to 5.80 & [4.71, 4.99] \\
 & Older & 44 & 4.80 & 0.44 & 4.80 & 4.50 to 5.12 & 3.70 to 5.50 & [4.67, 4.94] \\
CSQ-VR total (6--42) & All & 91 & 8.70 & 3.66 & 7.00 & 6.00 to 10.00 & 6.00 to 20.00 & [7.94, 9.47] \\
 & Younger & 47 & 8.89 & 3.46 & 7.00 & 6.00 to 11.50 & 6.00 to 17.00 & [7.88, 9.91] \\
 & Older & 44 & 8.50 & 3.90 & 7.00 & 6.00 to 9.25 & 6.00 to 20.00 & [7.32, 9.68] \\
CSQ-VR nausea (2--14) & All & 91 & 2.74 & 1.14 & 2.00 & 2.00 to 3.00 & 2.00 to 6.00 & [2.50, 2.97] \\
 & Younger & 47 & 2.89 & 1.20 & 2.00 & 2.00 to 4.00 & 2.00 to 6.00 & [2.54, 3.25] \\
 & Older & 44 & 2.57 & 1.07 & 2.00 & 2.00 to 3.00 & 2.00 to 6.00 & [2.24, 2.89] \\
CSQ-VR vestibular (2--14) & All & 91 & 3.02 & 1.53 & 2.00 & 2.00 to 4.00 & 2.00 to 9.00 & [2.70, 3.34] \\
 & Younger & 47 & 3.00 & 1.41 & 2.00 & 2.00 to 4.00 & 2.00 to 7.00 & [2.58, 3.42] \\
 & Older & 44 & 3.05 & 1.66 & 2.00 & 2.00 to 4.00 & 2.00 to 9.00 & [2.54, 3.55] \\
CSQ-VR oculomotor (2--14) & All & 91 & 2.95 & 1.73 & 2.00 & 2.00 to 3.00 & 2.00 to 14.00 & [2.59, 3.30] \\
 & Younger & 47 & 3.00 & 1.35 & 2.00 & 2.00 to 4.00 & 2.00 to 6.00 & [2.60, 3.40] \\
 & Older & 44 & 2.89 & 2.07 & 2.00 & 2.00 to 3.00 & 2.00 to 14.00 & [2.26, 3.52] \\
Age (years) & All & 91 & 47.69 & 23.87 & 38.00 & 24.00 to 71.50 & 20.00 to 81.00 & [42.72, 52.66] \\
 & Younger & 47 & 25.13 & 4.07 & 24.00 & 22.50 to 26.50 & 20.00 to 38.00 & [23.93, 26.32] \\
 & Older & 44 & 71.80 & 4.85 & 72.00 & 67.75 to 75.00 & 64.00 to 81.00 & [70.32, 73.27] \\
Education (years) & All & 91 & 16.40 & 2.86 & 17.00 & 15.00 to 18.00 & 5.00 to 22.00 & [15.80, 16.99] \\
 & Younger & 47 & 16.64 & 3.00 & 17.00 & 16.50 to 18.00 & 5.00 to 22.00 & [15.76, 17.52] \\
 & Older & 44 & 16.14 & 2.71 & 16.00 & 14.75 to 18.00 & 10.00 to 21.00 & [15.31, 16.96] \\
Gaming (12--72) & All & 91 & 16.24 & 7.08 & 12.00 & 12.00 to 18.00 & 12.00 to 45.00 & [14.77, 17.72] \\
 & Younger & 47 & 20.00 & 8.22 & 18.00 & 12.50 to 25.00 & 12.00 to 45.00 & [17.59, 22.41] \\
 & Older & 44 & 12.23 & 0.86 & 12.00 & 12.00 to 12.00 & 12.00 to 17.00 & [11.97, 12.49] \\
Technology (6--36) & All & 91 & 20.74 & 3.72 & 21.00 & 19.00 to 23.00 & 11.00 to 29.00 & [19.96, 21.51] \\
 & Younger & 47 & 23.32 & 2.04 & 23.00 & 22.00 to 25.00 & 20.00 to 29.00 & [22.72, 23.92] \\
 & Older & 44 & 17.98 & 3.06 & 19.00 & 15.00 to 20.00 & 11.00 to 22.00 & [17.05, 18.91] \\
MSSQ-Short total (0--54) & All & 91 & 11.01 & 8.60 & 9.90 & 3.38 to 17.00 & 0.00 to 29.48 & [9.22, 12.80] \\
 & Younger & 47 & 11.12 & 8.72 & 9.90 & 3.46 to 16.81 & 0.00 to 29.48 & [8.56, 13.68] \\
 & Older & 44 & 10.90 & 8.57 & 9.64 & 3.00 to 17.25 & 0.00 to 28.71 & [8.29, 13.50] \\
MSSQ-Short childhood (0--27) & All & 91 & 6.53 & 5.12 & 5.62 & 2.12 to 9.50 & 0.00 to 19.80 & [5.47, 7.60] \\
 & Younger & 47 & 5.96 & 4.54 & 5.62 & 2.12 to 9.00 & 0.00 to 16.50 & [4.62, 7.29] \\
 & Older & 44 & 7.15 & 5.66 & 5.81 & 2.43 to 10.46 & 0.00 to 19.80 & [5.43, 8.87] \\
MSSQ-Short recent adulthood (0--27) & All & 91 & 4.48 & 4.73 & 3.00 & 0.00 to 6.71 & 0.00 to 18.00 & [3.49, 5.46] \\
 & Younger & 47 & 5.16 & 4.94 & 4.00 & 1.12 to 7.50 & 0.00 to 18.00 & [3.71, 6.61] \\
 & Older & 44 & 3.75 & 4.44 & 1.90 & 0.00 to 6.11 & 0.00 to 13.00 & [2.40, 5.10] \\
\end{longtable}\normalsize
Note. Ranges in parentheses are the theoretical ranges. Q1 to Q3~=~interquartile range; 95\% CI of M~=~t-based interval of the mean. UX~=~user experience (UEQ-S); acceptability~=~mean of the ten adapted SUTAQ items. The MSSQ-Short sections range from 0 to 27 each. Abbreviations as in Supplementary Table S1.

\Needspace{198pt}
\textbf{Supplementary Table S3.} Sex and reported VR use by age group (counts).

\setlength{\tabcolsep}{2.5pt}\renewcommand{\arraystretch}{1.12}
\small
\begin{longtable}{L{0.400000\dimexpr\textwidth-20pt\relax}L{0.200000\dimexpr\textwidth-20pt\relax}L{0.200000\dimexpr\textwidth-20pt\relax}L{0.200000\dimexpr\textwidth-20pt\relax}}
\toprule
\textbf{\boldmath Characteristic} & \textbf{\boldmath Younger (n = 47)} & \textbf{\boldmath Older (n = 44)} & \textbf{\boldmath All (N = 91)} \\
\midrule\endfirsthead
\multicolumn{4}{l}{\textbf{Supplementary Table S3} (continued)}\\[3pt]\toprule
\textbf{\boldmath Characteristic} & \textbf{\boldmath Younger (n = 47)} & \textbf{\boldmath Older (n = 44)} & \textbf{\boldmath All (N = 91)} \\
\midrule\endhead
\midrule\multicolumn{4}{r}{\footnotesize Continued on next page}\\\endfoot
\bottomrule\endlastfoot
Women & 25 & 24 & 49 \\
Men & 22 & 20 & 42 \\
VR use frequency rating 1 & 46 & 43 & 89 \\
VR use frequency rating 2 & 0 & 1 & 1 \\
VR use frequency rating 3 & 1 & 0 & 1 \\
VR ability rating 1 & 23 & 41 & 64 \\
VR ability rating 2 & 21 & 3 & 24 \\
VR ability rating 3 & 2 & 0 & 2 \\
VR ability rating 4 & 1 & 0 & 1 \\
\end{longtable}\normalsize
Note. The VR items of the technology composite were rated from 1 (lowest) to 6; ratings not listed were given by no participant.

\Needspace{149pt}
\textbf{Supplementary Table S4.} Participants at the minimum and the maximum of each scale.

\setlength{\tabcolsep}{2.5pt}\renewcommand{\arraystretch}{1.12}
\small
\begin{longtable}{L{0.241363\dimexpr\textwidth-20pt\relax}L{0.275912\dimexpr\textwidth-20pt\relax}L{0.241363\dimexpr\textwidth-20pt\relax}L{0.241363\dimexpr\textwidth-20pt\relax}}
\toprule
\textbf{\boldmath Outcome} & \textbf{\boldmath Scale minimum and maximum} & \textbf{\boldmath At minimum, n (\%)} & \textbf{\boldmath At maximum, n (\%)} \\
\midrule\endfirsthead
\multicolumn{4}{l}{\textbf{Supplementary Table S4} (continued)}\\[3pt]\toprule
\textbf{\boldmath Outcome} & \textbf{\boldmath Scale minimum and maximum} & \textbf{\boldmath At minimum, n (\%)} & \textbf{\boldmath At maximum, n (\%)} \\
\midrule\endhead
\midrule\multicolumn{4}{r}{\footnotesize Continued on next page}\\\endfoot
\bottomrule\endlastfoot
SUS & 0 and 100 & 0 (0.0\%) & 12 (13.2\%) \\
Pragmatic UX & \ensuremath{-}3 and +3 & 0 (0.0\%) & 19 (20.9\%) \\
Hedonic UX & \ensuremath{-}3 and +3 & 0 (0.0\%) & 14 (15.4\%) \\
Acceptability & 1 and 6 & 0 (0.0\%) & 0 (0.0\%) \\
CSQ-VR total & 6 and 42 & 38 (41.8\%) & 0 (0.0\%) \\
\end{longtable}\normalsize
Note. N~=~91. Abbreviations as in Supplementary Table S2.

\Needspace{272pt}
\textbf{Supplementary Table S5.} Distribution of the CSQ-VR totals by age group (counts).

\setlength{\tabcolsep}{2.5pt}\renewcommand{\arraystretch}{1.12}
\small
\begin{longtable}{L{0.250000\dimexpr\textwidth-20pt\relax}L{0.250000\dimexpr\textwidth-20pt\relax}L{0.250000\dimexpr\textwidth-20pt\relax}L{0.250000\dimexpr\textwidth-20pt\relax}}
\toprule
\textbf{\boldmath CSQ-VR total} & \textbf{\boldmath Younger (n = 47)} & \textbf{\boldmath Older (n = 44)} & \textbf{\boldmath All (N = 91)} \\
\midrule\endfirsthead
\multicolumn{4}{l}{\textbf{Supplementary Table S5} (continued)}\\[3pt]\toprule
\textbf{\boldmath CSQ-VR total} & \textbf{\boldmath Younger (n = 47)} & \textbf{\boldmath Older (n = 44)} & \textbf{\boldmath All (N = 91)} \\
\midrule\endhead
\midrule\multicolumn{4}{r}{\footnotesize Continued on next page}\\\endfoot
\bottomrule\endlastfoot
6 & 17 & 21 & 38 \\
7 & 9 & 5 & 14 \\
8 & 3 & 5 & 8 \\
9 & 3 & 2 & 5 \\
10 & 1 & 3 & 4 \\
11 & 2 & 2 & 4 \\
12 & 3 & 1 & 4 \\
13 & 1 & 0 & 1 \\
14 & 2 & 0 & 2 \\
15 & 4 & 0 & 4 \\
16 & 1 & 0 & 1 \\
17 & 1 & 2 & 3 \\
18 & 0 & 1 & 1 \\
19 & 0 & 1 & 1 \\
20 & 0 & 1 & 1 \\
\end{longtable}\normalsize
Note. Only observed totals are listed (theoretical range 6--42); 6 denotes no symptom endorsement.

\clearpage\section*{S3 Age-group comparisons with sensitivity tests}
Supplementary Table S6 extends Table 2 of the article with the test statistics, the bootstrap intervals, Hedges' g and the rank-based sensitivity tests. Supplementary Table S7 extends the sample comparisons of Table 1 of the article. After Holm adjustment, the rank-based tests led to the same decisions as the primary tests for every outcome.

\Needspace{176pt}
\textbf{Supplementary Table S6.} Older versus younger participants: ratings, CSQ-VR total and CSQ-VR subscales.

\setlength{\tabcolsep}{2.5pt}\renewcommand{\arraystretch}{1.12}
\fontsize{7.6}{9}\selectfont
\begin{longtable}{L{0.119162\dimexpr\textwidth-65pt\relax}L{0.059543\dimexpr\textwidth-65pt\relax}L{0.051059\dimexpr\textwidth-65pt\relax}L{0.119162\dimexpr\textwidth-65pt\relax}L{0.085148\dimexpr\textwidth-65pt\relax}L{0.093633\dimexpr\textwidth-65pt\relax}L{0.055339\dimexpr\textwidth-65pt\relax}L{0.055339\dimexpr\textwidth-65pt\relax}L{0.046779\dimexpr\textwidth-65pt\relax}L{0.089353\dimexpr\textwidth-65pt\relax}L{0.051059\dimexpr\textwidth-65pt\relax}L{0.059543\dimexpr\textwidth-65pt\relax}L{0.114882\dimexpr\textwidth-65pt\relax}}
\toprule
\textbf{\boldmath Outcome} & \textbf{\boldmath Younger M} & \textbf{\boldmath Older M} & \textbf{\boldmath Difference [95\% CI]} & \textbf{\boldmath Bootstrap 95\% CI} & \textbf{\boldmath Welch t (df)} & \textbf{\boldmath Welch p} & \textbf{\boldmath U} & \textbf{\boldmath Rank p} & \textbf{\boldmath Primary test} & \textbf{\boldmath Holm p} & \textbf{\boldmath Holm p, rank tests} & \textbf{\boldmath Hedges' g [95\% CI]} \\
\midrule\endfirsthead
\multicolumn{13}{l}{\textbf{Supplementary Table S6} (continued)}\\[3pt]\toprule
\textbf{\boldmath Outcome} & \textbf{\boldmath Younger M} & \textbf{\boldmath Older M} & \textbf{\boldmath Difference [95\% CI]} & \textbf{\boldmath Bootstrap 95\% CI} & \textbf{\boldmath Welch t (df)} & \textbf{\boldmath Welch p} & \textbf{\boldmath U} & \textbf{\boldmath Rank p} & \textbf{\boldmath Primary test} & \textbf{\boldmath Holm p} & \textbf{\boldmath Holm p, rank tests} & \textbf{\boldmath Hedges' g [95\% CI]} \\
\midrule\endhead
\midrule\multicolumn{13}{r}{\footnotesize Continued on next page}\\\endfoot
\bottomrule\endlastfoot
SUS & 91.28 & 87.44 & \ensuremath{-}3.83 [\ensuremath{-}7.13, \ensuremath{-}0.54] & [\ensuremath{-}7.02, \ensuremath{-}0.62] & \ensuremath{-}2.32 (76.65) & .023 & 782.5 & .045 & Welch t & .116 & .225 & \ensuremath{-}0.49 [\ensuremath{-}0.91, \ensuremath{-}0.08] \\
Pragmatic UX & 2.09 & 1.64 & \ensuremath{-}0.45 [\ensuremath{-}0.86, \ensuremath{-}0.05] & [\ensuremath{-}0.84, \ensuremath{-}0.06] & \ensuremath{-}2.23 (78.80) & .028 & 802.0 & .064 & Welch t & .116 & .255 & \ensuremath{-}0.47 [\ensuremath{-}0.93, \ensuremath{-}0.07] \\
Hedonic UX & 1.64 & 1.52 & \ensuremath{-}0.12 [\ensuremath{-}0.58, 0.34] & [\ensuremath{-}0.58, 0.30] & \ensuremath{-}0.52 (88.34) & .602 & 971.5 & .621 & Welch t & 1.000 & 1.000 & \ensuremath{-}0.11 [\ensuremath{-}0.53, 0.28] \\
Acceptability & 4.85 & 4.80 & \ensuremath{-}0.05 [\ensuremath{-}0.24, 0.14] & [\ensuremath{-}0.23, 0.14] & \ensuremath{-}0.49 (88.99) & .628 & 1005.5 & .824 & Welch t & 1.000 & 1.000 & \ensuremath{-}0.10 [\ensuremath{-}0.50, 0.32] \\
CSQ-VR total & 8.89 & 8.50 & \ensuremath{-}0.39 [\ensuremath{-}1.93, 1.15] & [\ensuremath{-}1.89, 1.09] & \ensuremath{-}0.51 (86.06) & .613 & 920.5 & .350 & Mann--Whitney & 1.000 & 1.000 & \ensuremath{-}0.11 [\ensuremath{-}0.56, 0.28] \\
CSQ-VR nausea & 2.89 & 2.57 & \ensuremath{-}0.33 [\ensuremath{-}0.80, 0.15] & [\ensuremath{-}0.79, 0.13] & \ensuremath{-}1.37 (88.74) & .175 & 868.5 & .130 & Mann--Whitney & .390 & --- & \ensuremath{-}0.28 [\ensuremath{-}0.72, 0.11] \\
CSQ-VR vestibular & 3.00 & 3.05 & 0.05 [\ensuremath{-}0.60, 0.69] & [\ensuremath{-}0.56, 0.67] & 0.14 (84.81) & .889 & 1022.0 & .920 & Mann--Whitney & .920 & --- & 0.03 [\ensuremath{-}0.41, 0.41] \\
CSQ-VR oculomotor & 3.00 & 2.89 & \ensuremath{-}0.11 [\ensuremath{-}0.85, 0.62] & [\ensuremath{-}0.77, 0.66] & \ensuremath{-}0.31 (73.23) & .759 & 914.5 & .286 & Mann--Whitney & .572 & --- & \ensuremath{-}0.06 [\ensuremath{-}0.62, 0.30] \\
\end{longtable}\normalsize
Note. Differences are older minus younger with Welch 95\% CIs; the bootstrap interval of the difference and the interval of Hedges' g resample participants within each age group (5,000 draws). U~=~Mann--Whitney U with continuity correction. Primary test: Welch's t for the four ratings; Mann--Whitney for the CSQ-VR total and subscales, whose scores were discrete and concentrated at their minimum. Holm p: the five principal outcomes (SUS, pragmatic UX, hedonic UX, acceptability and the CSQ-VR total) form one family, and the three subscales a secondary family. Holm p, rank tests: the five-outcome family re-adjusted with the Mann--Whitney p values of all five outcomes (sensitivity). Abbreviations as in Supplementary Table S2.

\Needspace{135pt}
\textbf{Supplementary Table S7.} Older versus younger participants: sample characteristics.

\setlength{\tabcolsep}{2.5pt}\renewcommand{\arraystretch}{1.12}
\fontsize{7.6}{9}\selectfont
\begin{longtable}{L{0.144145\dimexpr\textwidth-60pt\relax}L{0.091680\dimexpr\textwidth-60pt\relax}L{0.091680\dimexpr\textwidth-60pt\relax}L{0.131048\dimexpr\textwidth-60pt\relax}L{0.065485\dimexpr\textwidth-60pt\relax}L{0.104777\dimexpr\textwidth-60pt\relax}L{0.056780\dimexpr\textwidth-60pt\relax}L{0.052388\dimexpr\textwidth-60pt\relax}L{0.056780\dimexpr\textwidth-60pt\relax}L{0.096071\dimexpr\textwidth-60pt\relax}L{0.052388\dimexpr\textwidth-60pt\relax}L{0.056780\dimexpr\textwidth-60pt\relax}}
\toprule
\textbf{\boldmath Characteristic} & \textbf{\boldmath Younger M (SD)} & \textbf{\boldmath Older M (SD)} & \textbf{\boldmath Difference [95\% CI]} & \textbf{\boldmath Hedges' g} & \textbf{\boldmath Welch t (df)} & \textbf{\boldmath Welch p} & \textbf{\boldmath U} & \textbf{\boldmath Rank p} & \textbf{\boldmath Test in Table 1} & \textbf{\boldmath p} & \textbf{\boldmath Holm p} \\
\midrule\endfirsthead
\multicolumn{12}{l}{\textbf{Supplementary Table S7} (continued)}\\[3pt]\toprule
\textbf{\boldmath Characteristic} & \textbf{\boldmath Younger M (SD)} & \textbf{\boldmath Older M (SD)} & \textbf{\boldmath Difference [95\% CI]} & \textbf{\boldmath Hedges' g} & \textbf{\boldmath Welch t (df)} & \textbf{\boldmath Welch p} & \textbf{\boldmath U} & \textbf{\boldmath Rank p} & \textbf{\boldmath Test in Table 1} & \textbf{\boldmath p} & \textbf{\boldmath Holm p} \\
\midrule\endhead
\midrule\multicolumn{12}{r}{\footnotesize Continued on next page}\\\endfoot
\bottomrule\endlastfoot
Education (years) & 16.64 (3.00) & 16.14 (2.71) & \ensuremath{-}0.50 [\ensuremath{-}1.69, 0.69] & \ensuremath{-}0.17 & \ensuremath{-}0.84 (88.88) & .404 & 866.0 & .174 & Welch t & .404 & 1.000 \\
Gaming (12--72) & 20.00 (8.22) & 12.23 (0.86) & \ensuremath{-}7.77 [\ensuremath{-}10.20, \ensuremath{-}5.35] & \ensuremath{-}1.30 & \ensuremath{-}6.45 (47.07) & < .001 & 302.0 & < .001 & Mann--Whitney & < .001 & < .001 \\
Technology (6--36) & 23.32 (2.04) & 17.98 (3.06) & \ensuremath{-}5.34 [\ensuremath{-}6.44, \ensuremath{-}4.25] & \ensuremath{-}2.05 & \ensuremath{-}9.72 (74.29) & < .001 & 85.0 & < .001 & Welch t & < .001 & < .001 \\
MSSQ-Short total (0--54) & 11.12 (8.72) & 10.90 (8.57) & \ensuremath{-}0.22 [\ensuremath{-}3.82, 3.38] & \ensuremath{-}0.03 & \ensuremath{-}0.12 (88.78) & .903 & 1024.5 & .943 & Welch t & .903 & 1.000 \\
Men, n (\%) & 22 (46.8) & 20 (45.5) & OR 0.95 [0.38, 2.34] & --- & --- & --- & --- & --- & Fisher's exact & 1.000 & 1.000 \\
\end{longtable}\normalsize
Note. Differences are older minus younger with Welch 95\% CIs; for sex, the older-versus-younger odds ratio (OR) of being male with its 95\% CI (Fisher's exact test). Gaming was tested with the Mann--Whitney U test because its scores were concentrated at the minimum, so its p value tests a rank contrast rather than the displayed mean difference. Holm adjustment covers the five sample comparisons as one family. Age defines the groups and is therefore not tested.

\clearpage\section*{S4 Rating models: coefficients, sensitivity specifications and diagnostics}
Supplementary Table S8 reports the four rating models of Table 3 of the article with all their coefficients. Supplementary Tables S9--S11 report the three sensitivity specifications: within-group age added, rank-transformed variables, and continuous age across the whole sample. Supplementary Tables S12 and S13 report the diagnostics. No focal coefficient was statistically significant after correction in any specification. The technology coefficients of the usability and pragmatic ratings were positive in every specification, so the direction of the technology association does not depend on the specification.

\Needspace{355pt}
\textbf{Supplementary Table S8.} Rating models with age group: all coefficients (the models of Table 3 of the article).

\setlength{\tabcolsep}{2.5pt}\renewcommand{\arraystretch}{1.12}
\fontsize{8.5}{10}\selectfont
\begin{longtable}{L{0.136131\dimexpr\textwidth-40pt\relax}L{0.266312\dimexpr\textwidth-40pt\relax}L{0.094686\dimexpr\textwidth-40pt\relax}L{0.082785\dimexpr\textwidth-40pt\relax}L{0.183422\dimexpr\textwidth-40pt\relax}L{0.076939\dimexpr\textwidth-40pt\relax}L{0.076939\dimexpr\textwidth-40pt\relax}L{0.082785\dimexpr\textwidth-40pt\relax}}
\toprule
\textbf{\boldmath Outcome} & \textbf{\boldmath Term} & \textbf{\boldmath b} & \textbf{\boldmath SE} & \textbf{\boldmath 95\% CI} & \textbf{\boldmath t} & \textbf{\boldmath p} & \textbf{\boldmath Holm p} \\
\midrule\endfirsthead
\multicolumn{8}{l}{\textbf{Supplementary Table S8} (continued)}\\[3pt]\toprule
\textbf{\boldmath Outcome} & \textbf{\boldmath Term} & \textbf{\boldmath b} & \textbf{\boldmath SE} & \textbf{\boldmath 95\% CI} & \textbf{\boldmath t} & \textbf{\boldmath p} & \textbf{\boldmath Holm p} \\
\midrule\endhead
\midrule\multicolumn{8}{r}{\footnotesize Continued on next page}\\\endfoot
\bottomrule\endlastfoot
SUS & Intercept & 89.423 & 0.829 & [87.775, 91.071] & 107.89 & < .001 & --- \\
 & Older vs younger & \ensuremath{-}0.907 & 2.712 & [\ensuremath{-}6.299, 4.485] & \ensuremath{-}0.33 & .739 & 1.000 \\
 & Gaming (per SD) & \ensuremath{-}0.592 & 0.826 & [\ensuremath{-}2.235, 1.051] & \ensuremath{-}0.72 & .476 & 1.000 \\
 & Technology (per SD) & 2.323 & 1.132 & [0.072, 4.574] & 2.05 & .043 & .519 \\
 & Education (per year) & 0.474 & 0.292 & [\ensuremath{-}0.106, 1.054] & 1.62 & .108 & --- \\
 & Male & 0.021 & 1.611 & [\ensuremath{-}3.183, 3.225] & 0.01 & .990 & --- \\
Pragmatic UX & Intercept & 1.871 & 0.103 & [1.667, 2.075] & 18.25 & < .001 & --- \\
 & Older vs younger & \ensuremath{-}0.002 & 0.321 & [\ensuremath{-}0.641, 0.637] & \ensuremath{-}0.01 & .995 & 1.000 \\
 & Gaming (per SD) & 0.013 & 0.098 & [\ensuremath{-}0.182, 0.209] & 0.14 & .893 & 1.000 \\
 & Technology (per SD) & 0.316 & 0.159 & [0.000, 0.632] & 1.99 & .050 & .549 \\
 & Education (per year) & \ensuremath{-}0.034 & 0.033 & [\ensuremath{-}0.099, 0.031] & \ensuremath{-}1.04 & .303 & --- \\
 & Male & 0.003 & 0.196 & [\ensuremath{-}0.388, 0.393] & 0.01 & .990 & --- \\
Hedonic UX & Intercept & 1.585 & 0.120 & [1.346, 1.824] & 13.20 & < .001 & --- \\
 & Older vs younger & \ensuremath{-}0.421 & 0.338 & [\ensuremath{-}1.094, 0.251] & \ensuremath{-}1.25 & .216 & 1.000 \\
 & Gaming (per SD) & \ensuremath{-}0.222 & 0.142 & [\ensuremath{-}0.504, 0.061] & \ensuremath{-}1.56 & .122 & 1.000 \\
 & Technology (per SD) & \ensuremath{-}0.033 & 0.186 & [\ensuremath{-}0.403, 0.337] & \ensuremath{-}0.18 & .860 & 1.000 \\
 & Education (per year) & \ensuremath{-}0.019 & 0.043 & [\ensuremath{-}0.104, 0.065] & \ensuremath{-}0.46 & .650 & --- \\
 & Male & \ensuremath{-}0.009 & 0.245 & [\ensuremath{-}0.496, 0.478] & \ensuremath{-}0.04 & .970 & --- \\
Acceptability & Intercept & 4.829 & 0.051 & [4.726, 4.931] & 94.06 & < .001 & --- \\
 & Older vs younger & \ensuremath{-}0.019 & 0.166 & [\ensuremath{-}0.348, 0.310] & \ensuremath{-}0.12 & .907 & 1.000 \\
 & Gaming (per SD) & 0.018 & 0.075 & [\ensuremath{-}0.132, 0.168] & 0.24 & .808 & 1.000 \\
 & Technology (per SD) & 0.000 & 0.064 & [\ensuremath{-}0.127, 0.127] & 0.00 & 1.000 & 1.000 \\
 & Education (per year) & 0.012 & 0.029 & [\ensuremath{-}0.046, 0.071] & 0.42 & .673 & --- \\
 & Male & 0.055 & 0.102 & [\ensuremath{-}0.148, 0.258] & 0.54 & .593 & --- \\
\end{longtable}\normalsize
Note. Ordinary least squares; N~=~91, residual df~=~85; b in outcome units; SE~=~HC3 standard error with a t-based 95\% CI. Holm p: the 12 focal age-group, gaming and technology coefficients of the four models form one family; education and sex are covariates and are not adjusted. Gaming and technology per sample SD; education per year (centred); male~=~centred indicator.

\Needspace{400pt}
\textbf{Supplementary Table S9.} Rating models with within-group age added.

\setlength{\tabcolsep}{2.5pt}\renewcommand{\arraystretch}{1.12}
\fontsize{8.5}{10}\selectfont
\begin{longtable}{L{0.136131\dimexpr\textwidth-40pt\relax}L{0.266312\dimexpr\textwidth-40pt\relax}L{0.094686\dimexpr\textwidth-40pt\relax}L{0.082785\dimexpr\textwidth-40pt\relax}L{0.183422\dimexpr\textwidth-40pt\relax}L{0.076939\dimexpr\textwidth-40pt\relax}L{0.076939\dimexpr\textwidth-40pt\relax}L{0.082785\dimexpr\textwidth-40pt\relax}}
\toprule
\textbf{\boldmath Outcome} & \textbf{\boldmath Term} & \textbf{\boldmath b} & \textbf{\boldmath SE} & \textbf{\boldmath 95\% CI} & \textbf{\boldmath t} & \textbf{\boldmath p} & \textbf{\boldmath Holm p} \\
\midrule\endfirsthead
\multicolumn{8}{l}{\textbf{Supplementary Table S9} (continued)}\\[3pt]\toprule
\textbf{\boldmath Outcome} & \textbf{\boldmath Term} & \textbf{\boldmath b} & \textbf{\boldmath SE} & \textbf{\boldmath 95\% CI} & \textbf{\boldmath t} & \textbf{\boldmath p} & \textbf{\boldmath Holm p} \\
\midrule\endhead
\midrule\multicolumn{8}{r}{\footnotesize Continued on next page}\\\endfoot
\bottomrule\endlastfoot
SUS & Intercept & 89.423 & 0.839 & [87.754, 91.092] & 106.54 & < .001 & --- \\
 & Older vs younger & \ensuremath{-}1.042 & 2.777 & [\ensuremath{-}6.565, 4.481] & \ensuremath{-}0.38 & .709 & 1.000 \\
 & Within-group age (per 10 years) & \ensuremath{-}1.114 & 2.235 & [\ensuremath{-}5.559, 3.330] & \ensuremath{-}0.50 & .619 & 1.000 \\
 & Gaming (per SD) & \ensuremath{-}0.596 & 0.825 & [\ensuremath{-}2.237, 1.045] & \ensuremath{-}0.72 & .472 & --- \\
 & Technology (per SD) & 2.245 & 1.174 & [\ensuremath{-}0.090, 4.580] & 1.91 & .059 & --- \\
 & Education (per year) & 0.442 & 0.313 & [\ensuremath{-}0.180, 1.065] & 1.41 & .161 & --- \\
 & Male & \ensuremath{-}0.139 & 1.592 & [\ensuremath{-}3.304, 3.026] & \ensuremath{-}0.09 & .931 & --- \\
Pragmatic UX & Intercept & 1.871 & 0.104 & [1.664, 2.077] & 18.02 & < .001 & --- \\
 & Older vs younger & 0.005 & 0.329 & [\ensuremath{-}0.650, 0.659] & 0.01 & .988 & 1.000 \\
 & Within-group age (per 10 years) & 0.058 & 0.255 & [\ensuremath{-}0.450, 0.566] & 0.23 & .821 & 1.000 \\
 & Gaming (per SD) & 0.014 & 0.099 & [\ensuremath{-}0.183, 0.210] & 0.14 & .891 & --- \\
 & Technology (per SD) & 0.320 & 0.161 & [0.000, 0.640] & 1.99 & .050 & --- \\
 & Education (per year) & \ensuremath{-}0.032 & 0.035 & [\ensuremath{-}0.102, 0.037] & \ensuremath{-}0.93 & .356 & --- \\
 & Male & 0.011 & 0.200 & [\ensuremath{-}0.387, 0.409] & 0.05 & .957 & --- \\
Hedonic UX & Intercept & 1.585 & 0.122 & [1.342, 1.828] & 12.96 & < .001 & --- \\
 & Older vs younger & \ensuremath{-}0.430 & 0.351 & [\ensuremath{-}1.128, 0.268] & \ensuremath{-}1.23 & .224 & 1.000 \\
 & Within-group age (per 10 years) & \ensuremath{-}0.073 & 0.365 & [\ensuremath{-}0.798, 0.653] & \ensuremath{-}0.20 & .842 & 1.000 \\
 & Gaming (per SD) & \ensuremath{-}0.222 & 0.144 & [\ensuremath{-}0.508, 0.064] & \ensuremath{-}1.55 & .126 & --- \\
 & Technology (per SD) & \ensuremath{-}0.038 & 0.197 & [\ensuremath{-}0.429, 0.353] & \ensuremath{-}0.19 & .847 & --- \\
 & Education (per year) & \ensuremath{-}0.021 & 0.043 & [\ensuremath{-}0.107, 0.065] & \ensuremath{-}0.50 & .622 & --- \\
 & Male & \ensuremath{-}0.020 & 0.256 & [\ensuremath{-}0.530, 0.490] & \ensuremath{-}0.08 & .939 & --- \\
Acceptability & Intercept & 4.829 & 0.052 & [4.725, 4.932] & 93.01 & < .001 & --- \\
 & Older vs younger & \ensuremath{-}0.013 & 0.165 & [\ensuremath{-}0.342, 0.315] & \ensuremath{-}0.08 & .936 & 1.000 \\
 & Within-group age (per 10 years) & 0.050 & 0.126 & [\ensuremath{-}0.200, 0.301] & 0.40 & .690 & 1.000 \\
 & Gaming (per SD) & 0.019 & 0.076 & [\ensuremath{-}0.132, 0.169] & 0.25 & .807 & --- \\
 & Technology (per SD) & 0.004 & 0.065 & [\ensuremath{-}0.125, 0.133] & 0.05 & .957 & --- \\
 & Education (per year) & 0.014 & 0.029 & [\ensuremath{-}0.044, 0.071] & 0.48 & .632 & --- \\
 & Male & 0.062 & 0.108 & [\ensuremath{-}0.153, 0.277] & 0.58 & .567 & --- \\
\end{longtable}\normalsize
Note. As Supplementary Table S8, with within-group age (age minus the mean age of the participant's group, per 10 years) added; residual df~=~84. Holm p: the eight age-group and within-group-age coefficients of the four models form the family of this specification.

\Needspace{355pt}
\textbf{Supplementary Table S10.} Rating models on rank-transformed variables (continuous age).

\setlength{\tabcolsep}{2.5pt}\renewcommand{\arraystretch}{1.12}
\fontsize{8.5}{10}\selectfont
\begin{longtable}{L{0.136131\dimexpr\textwidth-40pt\relax}L{0.266312\dimexpr\textwidth-40pt\relax}L{0.094686\dimexpr\textwidth-40pt\relax}L{0.082785\dimexpr\textwidth-40pt\relax}L{0.183422\dimexpr\textwidth-40pt\relax}L{0.076939\dimexpr\textwidth-40pt\relax}L{0.076939\dimexpr\textwidth-40pt\relax}L{0.082785\dimexpr\textwidth-40pt\relax}}
\toprule
\textbf{\boldmath Outcome} & \textbf{\boldmath Term} & \textbf{\boldmath b} & \textbf{\boldmath SE} & \textbf{\boldmath 95\% CI} & \textbf{\boldmath t} & \textbf{\boldmath p} & \textbf{\boldmath Holm p} \\
\midrule\endfirsthead
\multicolumn{8}{l}{\textbf{Supplementary Table S10} (continued)}\\[3pt]\toprule
\textbf{\boldmath Outcome} & \textbf{\boldmath Term} & \textbf{\boldmath b} & \textbf{\boldmath SE} & \textbf{\boldmath 95\% CI} & \textbf{\boldmath t} & \textbf{\boldmath p} & \textbf{\boldmath Holm p} \\
\midrule\endhead
\midrule\multicolumn{8}{r}{\footnotesize Continued on next page}\\\endfoot
\bottomrule\endlastfoot
SUS & Intercept & 0.000 & 0.104 & [\ensuremath{-}0.206, 0.206] & 0.00 & 1.000 & --- \\
 & Age (rank) & \ensuremath{-}0.028 & 0.170 & [\ensuremath{-}0.367, 0.310] & \ensuremath{-}0.17 & .868 & 1.000 \\
 & Gaming (rank) & \ensuremath{-}0.171 & 0.146 & [\ensuremath{-}0.461, 0.119] & \ensuremath{-}1.17 & .244 & 1.000 \\
 & Technology (rank) & 0.366 & 0.166 & [0.036, 0.696] & 2.21 & .030 & .361 \\
 & Education (rank) & 0.183 & 0.121 & [\ensuremath{-}0.057, 0.422] & 1.51 & .134 & --- \\
 & Male & \ensuremath{-}0.094 & 0.201 & [\ensuremath{-}0.494, 0.305] & \ensuremath{-}0.47 & .640 & --- \\
Pragmatic UX & Intercept & 0.000 & 0.107 & [\ensuremath{-}0.213, 0.213] & 0.00 & 1.000 & --- \\
 & Age (rank) & 0.138 & 0.175 & [\ensuremath{-}0.210, 0.485] & 0.79 & .433 & 1.000 \\
 & Gaming (rank) & 0.117 & 0.155 & [\ensuremath{-}0.192, 0.425] & 0.75 & .455 & 1.000 \\
 & Technology (rank) & 0.293 & 0.171 & [\ensuremath{-}0.047, 0.632] & 1.71 & .090 & .991 \\
 & Education (rank) & \ensuremath{-}0.055 & 0.118 & [\ensuremath{-}0.291, 0.180] & \ensuremath{-}0.47 & .642 & --- \\
 & Male & 0.064 & 0.214 & [\ensuremath{-}0.361, 0.489] & 0.30 & .765 & --- \\
Hedonic UX & Intercept & 0.000 & 0.110 & [\ensuremath{-}0.220, 0.220] & 0.00 & 1.000 & --- \\
 & Age (rank) & \ensuremath{-}0.059 & 0.179 & [\ensuremath{-}0.414, 0.297] & \ensuremath{-}0.33 & .743 & 1.000 \\
 & Gaming (rank) & \ensuremath{-}0.150 & 0.142 & [\ensuremath{-}0.433, 0.134] & \ensuremath{-}1.05 & .297 & 1.000 \\
 & Technology (rank) & 0.098 & 0.186 & [\ensuremath{-}0.271, 0.467] & 0.53 & .600 & 1.000 \\
 & Education (rank) & \ensuremath{-}0.087 & 0.109 & [\ensuremath{-}0.303, 0.129] & \ensuremath{-}0.80 & .424 & --- \\
 & Male & \ensuremath{-}0.023 & 0.228 & [\ensuremath{-}0.476, 0.431] & \ensuremath{-}0.10 & .920 & --- \\
Acceptability & Intercept & 0.000 & 0.111 & [\ensuremath{-}0.220, 0.220] & 0.00 & 1.000 & --- \\
 & Age (rank) & 0.085 & 0.164 & [\ensuremath{-}0.241, 0.410] & 0.52 & .606 & 1.000 \\
 & Gaming (rank) & 0.019 & 0.146 & [\ensuremath{-}0.271, 0.309] & 0.13 & .897 & 1.000 \\
 & Technology (rank) & 0.066 & 0.161 & [\ensuremath{-}0.253, 0.386] & 0.41 & .681 & 1.000 \\
 & Education (rank) & 0.066 & 0.119 & [\ensuremath{-}0.171, 0.303] & 0.55 & .581 & --- \\
 & Male & 0.114 & 0.228 & [\ensuremath{-}0.340, 0.567] & 0.50 & .620 & --- \\
\end{longtable}\normalsize
Note. The rating, age, gaming, technology and education were replaced by their standardised ranks across the whole sample (average ranks for ties). b is therefore in SD units of the ranked outcome per SD of the ranked predictor; sex remained a centred indicator. Age entered as a continuous variable; with no participants aged 39--63 years, its rank mainly contrasts the two groups. Residual df~=~85. Holm p: the 12 age, gaming and technology coefficients form one family.

\Needspace{327pt}
\textbf{Supplementary Table S11.} Rating models with continuous age across the whole sample.

\setlength{\tabcolsep}{2.5pt}\renewcommand{\arraystretch}{1.12}
\fontsize{7.6}{9}\selectfont
\begin{longtable}{L{0.136117\dimexpr\textwidth-45pt\relax}L{0.189353\dimexpr\textwidth-45pt\relax}L{0.076931\dimexpr\textwidth-45pt\relax}L{0.070981\dimexpr\textwidth-45pt\relax}L{0.153862\dimexpr\textwidth-45pt\relax}L{0.153862\dimexpr\textwidth-45pt\relax}L{0.070981\dimexpr\textwidth-45pt\relax}L{0.070981\dimexpr\textwidth-45pt\relax}L{0.076931\dimexpr\textwidth-45pt\relax}}
\toprule
\textbf{\boldmath Outcome} & \textbf{\boldmath Term} & \textbf{\boldmath b} & \textbf{\boldmath SE} & \textbf{\boldmath 95\% CI} & \textbf{\boldmath Bootstrap 95\% CI} & \textbf{\boldmath t} & \textbf{\boldmath p} & \textbf{\boldmath Holm p} \\
\midrule\endfirsthead
\multicolumn{9}{l}{\textbf{Supplementary Table S11} (continued)}\\[3pt]\toprule
\textbf{\boldmath Outcome} & \textbf{\boldmath Term} & \textbf{\boldmath b} & \textbf{\boldmath SE} & \textbf{\boldmath 95\% CI} & \textbf{\boldmath Bootstrap 95\% CI} & \textbf{\boldmath t} & \textbf{\boldmath p} & \textbf{\boldmath Holm p} \\
\midrule\endhead
\midrule\multicolumn{9}{r}{\footnotesize Continued on next page}\\\endfoot
\bottomrule\endlastfoot
SUS & Intercept & 89.423 & 0.829 & [87.775, 91.071] & [87.776, 90.972] & 107.86 & < .001 & --- \\
 & Age (per 10 years) & \ensuremath{-}0.272 & 0.597 & [\ensuremath{-}1.458, 0.914] & [\ensuremath{-}1.432, 0.830] & \ensuremath{-}0.46 & .650 & 1.000 \\
 & Gaming (per SD) & \ensuremath{-}0.636 & 0.822 & [\ensuremath{-}2.271, 1.000] & [\ensuremath{-}2.508, 0.693] & \ensuremath{-}0.77 & .442 & 1.000 \\
 & Technology (per SD) & 2.203 & 1.168 & [\ensuremath{-}0.120, 4.525] & [\ensuremath{-}0.092, 4.373] & 1.89 & .063 & .690 \\
 & Education (per year) & 0.463 & 0.294 & [\ensuremath{-}0.122, 1.049] & [\ensuremath{-}0.095, 1.061] & 1.57 & .119 & --- \\
 & Male & \ensuremath{-}0.044 & 1.597 & [\ensuremath{-}3.218, 3.130] & [\ensuremath{-}3.041, 3.130] & \ensuremath{-}0.03 & .978 & --- \\
Pragmatic UX & Intercept & 1.871 & 0.103 & [1.667, 2.075] & [1.667, 2.057] & 18.24 & < .001 & --- \\
 & Age (per 10 years) & 0.004 & 0.070 & [\ensuremath{-}0.135, 0.143] & [\ensuremath{-}0.127, 0.141] & 0.06 & .953 & 1.000 \\
 & Gaming (per SD) & 0.016 & 0.101 & [\ensuremath{-}0.184, 0.217] & [\ensuremath{-}0.181, 0.220] & 0.16 & .874 & 1.000 \\
 & Technology (per SD) & 0.323 & 0.155 & [0.014, 0.632] & [0.030, 0.627] & 2.08 & .041 & .489 \\
 & Education (per year) & \ensuremath{-}0.034 & 0.033 & [\ensuremath{-}0.100, 0.032] & [\ensuremath{-}0.086, 0.046] & \ensuremath{-}1.02 & .312 & --- \\
 & Male & 0.005 & 0.195 & [\ensuremath{-}0.383, 0.393] & [\ensuremath{-}0.370, 0.363] & 0.02 & .980 & --- \\
Hedonic UX & Intercept & 1.585 & 0.120 & [1.346, 1.824] & [1.355, 1.812] & 13.19 & < .001 & --- \\
 & Age (per 10 years) & \ensuremath{-}0.091 & 0.076 & [\ensuremath{-}0.243, 0.061] & [\ensuremath{-}0.236, 0.051] & \ensuremath{-}1.19 & .236 & 1.000 \\
 & Gaming (per SD) & \ensuremath{-}0.221 & 0.142 & [\ensuremath{-}0.503, 0.061] & [\ensuremath{-}0.499, 0.040] & \ensuremath{-}1.56 & .123 & 1.000 \\
 & Technology (per SD) & \ensuremath{-}0.037 & 0.194 & [\ensuremath{-}0.422, 0.348] & [\ensuremath{-}0.404, 0.316] & \ensuremath{-}0.19 & .848 & 1.000 \\
 & Education (per year) & \ensuremath{-}0.022 & 0.041 & [\ensuremath{-}0.104, 0.061] & [\ensuremath{-}0.109, 0.046] & \ensuremath{-}0.53 & .599 & --- \\
 & Male & \ensuremath{-}0.022 & 0.247 & [\ensuremath{-}0.512, 0.469] & [\ensuremath{-}0.489, 0.459] & \ensuremath{-}0.09 & .930 & --- \\
Acceptability & Intercept & 4.829 & 0.051 & [4.727, 4.930] & [4.731, 4.924] & 94.21 & < .001 & --- \\
 & Age (per 10 years) & 0.000 & 0.032 & [\ensuremath{-}0.064, 0.065] & [\ensuremath{-}0.062, 0.060] & 0.00 & .998 & 1.000 \\
 & Gaming (per SD) & 0.021 & 0.075 & [\ensuremath{-}0.129, 0.171] & [\ensuremath{-}0.103, 0.156] & 0.28 & .782 & 1.000 \\
 & Technology (per SD) & 0.006 & 0.062 & [\ensuremath{-}0.117, 0.129] & [\ensuremath{-}0.105, 0.126] & 0.10 & .923 & 1.000 \\
 & Education (per year) & 0.013 & 0.029 & [\ensuremath{-}0.046, 0.071] & [\ensuremath{-}0.031, 0.063] & 0.43 & .667 & --- \\
 & Male & 0.057 & 0.102 & [\ensuremath{-}0.147, 0.260] & [\ensuremath{-}0.127, 0.253] & 0.55 & .582 & --- \\
\end{longtable}\normalsize
Note. As Supplementary Table S8, with age as a continuous variable across the whole sample (centred, per 10 years) in place of age group; residual df~=~85. Bootstrap 95\% CI: case resampling of participants (5,000 draws; no draw failed). Holm p: the 12 age, gaming and technology coefficients form one family.

\Needspace{358pt}
\textbf{Supplementary Table S12.} Fit and influence diagnostics of the rating models.

\setlength{\tabcolsep}{2.5pt}\renewcommand{\arraystretch}{1.12}
\small
\begin{longtable}{L{0.333403\dimexpr\textwidth-25pt\relax}L{0.166649\dimexpr\textwidth-25pt\relax}L{0.166649\dimexpr\textwidth-25pt\relax}L{0.166649\dimexpr\textwidth-25pt\relax}L{0.166649\dimexpr\textwidth-25pt\relax}}
\toprule
\textbf{\boldmath Statistic} & \textbf{\boldmath SUS} & \textbf{\boldmath Pragmatic UX} & \textbf{\boldmath Hedonic UX} & \textbf{\boldmath Acceptability} \\
\midrule\endfirsthead
\multicolumn{5}{l}{\textbf{Supplementary Table S12} (continued)}\\[3pt]\toprule
\textbf{\boldmath Statistic} & \textbf{\boldmath SUS} & \textbf{\boldmath Pragmatic UX} & \textbf{\boldmath Hedonic UX} & \textbf{\boldmath Acceptability} \\
\midrule\endhead
\midrule\multicolumn{5}{r}{\footnotesize Continued on next page}\\\endfoot
\bottomrule\endlastfoot
\textit{Age group (reported models)} &  &  &  &  \\
R\textsuperscript{2} & .126 & .115 & .035 & .014 \\
Adjusted R\textsuperscript{2} & .074 & .063 & \ensuremath{-}.022 & \ensuremath{-}.044 \\
Residual SD & 7.69 & 0.95 & 1.11 & 0.46 \\
Largest leverage & .261 & .261 & .261 & .261 \\
Largest Cook's distance & .091 & .060 & .100 & .395 \\
Participants with Cook's distance > 4/n & 4 & 5 & 5 & 4 \\
Range of fitted values & 83.049 to 94.988 & 1.049 to 2.560 & 0.804 to 2.077 & 4.702 to 4.975 \\
Fitted values outside the scale & 0 & 0 & 0 & 0 \\
Residual skewness & \ensuremath{-}0.63 & \ensuremath{-}0.60 & \ensuremath{-}0.46 & \ensuremath{-}0.02 \\
Shapiro--Wilk p (residuals) & .036 & .002 & .004 & .261 \\
\textit{Continuous age (sensitivity specification)} &  &  &  &  \\
R\textsuperscript{2} & .127 & .115 & .036 & .014 \\
Adjusted R\textsuperscript{2} & .076 & .063 & \ensuremath{-}.021 & \ensuremath{-}.044 \\
Residual SD & 7.68 & 0.95 & 1.11 & 0.46 \\
Largest leverage & .261 & .261 & .261 & .261 \\
Largest Cook's distance & .112 & .057 & .092 & .396 \\
Participants with Cook's distance > 4/n & 4 & 5 & 6 & 4 \\
Range of fitted values & 83.003 to 94.982 & 1.044 to 2.564 & 0.820 to 2.048 & 4.702 to 4.982 \\
Fitted values outside the scale & 0 & 0 & 0 & 0 \\
Residual skewness & \ensuremath{-}0.61 & \ensuremath{-}0.60 & \ensuremath{-}0.45 & \ensuremath{-}0.01 \\
Shapiro--Wilk p (residuals) & .040 & .002 & .007 & .280 \\
\end{longtable}\normalsize
Note. N~=~91. Cook's distance~>~4/n counts the participants whose Cook's distance exceeded 4/91. Fitted values outside the scale counts the fitted values outside the theoretical range of the rating. Residual skewness is the moment skewness of the residuals (negative~=~left-skewed). No participant was excluded on the basis of these diagnostics.

\Needspace{174pt}
\textbf{Supplementary Table S13.} Variance inflation factors of the predictors of the rating models and the symptom model.

\setlength{\tabcolsep}{2.5pt}\renewcommand{\arraystretch}{1.12}
\small
\begin{longtable}{L{0.265826\dimexpr\textwidth-25pt\relax}L{0.183544\dimexpr\textwidth-25pt\relax}L{0.183544\dimexpr\textwidth-25pt\relax}L{0.183544\dimexpr\textwidth-25pt\relax}L{0.183544\dimexpr\textwidth-25pt\relax}}
\toprule
\textbf{\boldmath Predictor} & \textbf{\boldmath Rating models, age group} & \textbf{\boldmath Rating models, continuous age} & \textbf{\boldmath Symptom model, age group} & \textbf{\boldmath Symptom model, continuous age} \\
\midrule\endfirsthead
\multicolumn{5}{l}{\textbf{Supplementary Table S13} (continued)}\\[3pt]\toprule
\textbf{\boldmath Predictor} & \textbf{\boldmath Rating models, age group} & \textbf{\boldmath Rating models, continuous age} & \textbf{\boldmath Symptom model, age group} & \textbf{\boldmath Symptom model, continuous age} \\
\midrule\endhead
\midrule\multicolumn{5}{r}{\footnotesize Continued on next page}\\\endfoot
\bottomrule\endlastfoot
MSSQ-Short total (per SD) & --- & --- & 1.06 & 1.06 \\
Older vs younger & 2.43 & --- & 2.43 & --- \\
Age (per 10 years) & --- & 2.44 & --- & 2.45 \\
Gaming (per SD) & 1.48 & 1.46 & 1.49 & 1.48 \\
Technology (per SD) & 2.23 & 2.25 & 2.23 & 2.25 \\
Education (per year) & 1.01 & 1.02 & 1.05 & 1.06 \\
Male & 1.06 & 1.07 & 1.07 & 1.08 \\
\end{longtable}\normalsize
Note. The same predictors enter the four rating models, so their variance inflation factors are common to them; the symptom model of the CSQ-VR total adds the MSSQ-Short total. ---~=~not in the model.

In the age-group models, the predictors explained between 1\% and 13\% of the variance of the ratings (adjusted R\textsuperscript{2} from \ensuremath{-}.044 to .074). No fitted value lay outside its scale, the largest Cook's distance was .395, and the variance inflation factors were at most 2.43. The residuals of the usability and user-experience models were left-skewed, as expected for ratings concentrated near the top of their scales (Shapiro--Wilk p~\ensuremath{\leq}~.036). The rank-transformed specification, which does not depend on the shape of the residual distribution, leads to the same conclusions, and the continuous-age models gave closely similar diagnostics.

\clearpage\section*{S5 Symptom model: coefficients and diagnostics}
Supplementary Table S14 reports the model of the CSQ-VR total with all its coefficients, and Supplementary Table S15 its diagnostics; its variance inflation factors are in Supplementary Table S13. One fitted value lay below the scale minimum of 6 (5.67), and the residuals were right-skewed (skewness 1.13; Shapiro--Wilk p~<~.001), as expected for symptom scores concentrated at the minimum. The rank-transformed and grouped specifications of Section S6 therefore accompany the linear model.

\Needspace{165pt}
\textbf{Supplementary Table S14.} Model of the CSQ-VR total with age group: all coefficients.

\setlength{\tabcolsep}{2.5pt}\renewcommand{\arraystretch}{1.12}
\fontsize{8.5}{10}\selectfont
\begin{longtable}{L{0.136131\dimexpr\textwidth-40pt\relax}L{0.266312\dimexpr\textwidth-40pt\relax}L{0.094686\dimexpr\textwidth-40pt\relax}L{0.082785\dimexpr\textwidth-40pt\relax}L{0.183422\dimexpr\textwidth-40pt\relax}L{0.076939\dimexpr\textwidth-40pt\relax}L{0.076939\dimexpr\textwidth-40pt\relax}L{0.082785\dimexpr\textwidth-40pt\relax}}
\toprule
\textbf{\boldmath Outcome} & \textbf{\boldmath Term} & \textbf{\boldmath b} & \textbf{\boldmath SE} & \textbf{\boldmath 95\% CI} & \textbf{\boldmath t} & \textbf{\boldmath p} & \textbf{\boldmath Holm p} \\
\midrule\endfirsthead
\multicolumn{8}{l}{\textbf{Supplementary Table S14} (continued)}\\[3pt]\toprule
\textbf{\boldmath Outcome} & \textbf{\boldmath Term} & \textbf{\boldmath b} & \textbf{\boldmath SE} & \textbf{\boldmath 95\% CI} & \textbf{\boldmath t} & \textbf{\boldmath p} & \textbf{\boldmath Holm p} \\
\midrule\endhead
\midrule\multicolumn{8}{r}{\footnotesize Continued on next page}\\\endfoot
\bottomrule\endlastfoot
CSQ-VR total & Intercept & 8.703 & 0.386 & [7.935, 9.472] & 22.52 & < .001 & --- \\
 & MSSQ-Short total (per SD) & 0.535 & 0.430 & [\ensuremath{-}0.319, 1.390] & 1.25 & .216 & .648 \\
 & Older vs younger & \ensuremath{-}2.043 & 0.988 & [\ensuremath{-}4.007, \ensuremath{-}0.078] & \ensuremath{-}2.07 & .042 & .167 \\
 & Gaming (per SD) & 0.173 & 0.460 & [\ensuremath{-}0.742, 1.088] & 0.38 & .708 & 1.000 \\
 & Technology (per SD) & \ensuremath{-}1.251 & 0.499 & [\ensuremath{-}2.242, \ensuremath{-}0.259] & \ensuremath{-}2.51 & .014 & .084 \\
 & Education (per year) & \ensuremath{-}0.062 & 0.138 & [\ensuremath{-}0.337, 0.214] & \ensuremath{-}0.45 & .657 & 1.000 \\
 & Male & \ensuremath{-}1.815 & 0.756 & [\ensuremath{-}3.318, \ensuremath{-}0.311] & \ensuremath{-}2.40 & .019 & .093 \\
\end{longtable}\normalsize
Note. Ordinary least squares; N~=~91, residual df~=~84; b in CSQ-VR points; SE~=~HC3 standard error with a t-based 95\% CI. Holm p: the six coefficients form one family. MSSQ-Short total, gaming and technology per sample SD; education per year (centred); male~=~centred indicator.

\Needspace{211pt}
\textbf{Supplementary Table S15.} Fit and influence diagnostics of the model of the CSQ-VR total.

\setlength{\tabcolsep}{2.5pt}\renewcommand{\arraystretch}{1.12}
\small
\begin{longtable}{L{0.454570\dimexpr\textwidth-15pt\relax}L{0.272715\dimexpr\textwidth-15pt\relax}L{0.272715\dimexpr\textwidth-15pt\relax}}
\toprule
\textbf{\boldmath Statistic} & \textbf{\boldmath Age group (reported model)} & \textbf{\boldmath Continuous age (sensitivity)} \\
\midrule\endfirsthead
\multicolumn{3}{l}{\textbf{Supplementary Table S15} (continued)}\\[3pt]\toprule
\textbf{\boldmath Statistic} & \textbf{\boldmath Age group (reported model)} & \textbf{\boldmath Continuous age (sensitivity)} \\
\midrule\endhead
\midrule\multicolumn{3}{r}{\footnotesize Continued on next page}\\\endfoot
\bottomrule\endlastfoot
R\textsuperscript{2} & .120 & .113 \\
Adjusted R\textsuperscript{2} & .057 & .050 \\
Residual SD & 3.56 & 3.57 \\
Largest leverage & .263 & .263 \\
Largest Cook's distance & .090 & .089 \\
Participants with Cook's distance > 4/n & 7 & 7 \\
Range of fitted values & 5.668 to 12.049 & 5.997 to 12.118 \\
Fitted values outside the scale & 1 & 1 \\
Residual skewness & 1.13 & 1.15 \\
Shapiro--Wilk p (residuals) & < .001 & < .001 \\
\end{longtable}\normalsize
Note. As Supplementary Table S12; the theoretical range of the CSQ-VR total is 6--42, and positive skewness denotes right-skewed residuals. The variance inflation factors of this model are in Supplementary Table S13.

\clearpage\section*{S6 Symptom model: sensitivity specifications}
Supplementary Tables S16--S21 report the symptom model with within-group age added, with continuous age across the whole sample, with the recent-adulthood section of the MSSQ-Short in place of the total, on rank-transformed variables, and across susceptibility tertiles. No coefficient was statistically significant after correction in any specification. The rank-transformed specification was less supportive than the linear model of the pointwise associations of age, technology experience and sex with symptoms. In every specification, the association of susceptibility with symptoms remained small, as the article reports.

\Needspace{176pt}
\textbf{Supplementary Table S16.} Model of the CSQ-VR total with within-group age added.

\setlength{\tabcolsep}{2.5pt}\renewcommand{\arraystretch}{1.12}
\fontsize{8.5}{10}\selectfont
\begin{longtable}{L{0.136131\dimexpr\textwidth-40pt\relax}L{0.266312\dimexpr\textwidth-40pt\relax}L{0.094686\dimexpr\textwidth-40pt\relax}L{0.082785\dimexpr\textwidth-40pt\relax}L{0.183422\dimexpr\textwidth-40pt\relax}L{0.076939\dimexpr\textwidth-40pt\relax}L{0.076939\dimexpr\textwidth-40pt\relax}L{0.082785\dimexpr\textwidth-40pt\relax}}
\toprule
\textbf{\boldmath Outcome} & \textbf{\boldmath Term} & \textbf{\boldmath b} & \textbf{\boldmath SE} & \textbf{\boldmath 95\% CI} & \textbf{\boldmath t} & \textbf{\boldmath p} & \textbf{\boldmath Holm p} \\
\midrule\endfirsthead
\multicolumn{8}{l}{\textbf{Supplementary Table S16} (continued)}\\[3pt]\toprule
\textbf{\boldmath Outcome} & \textbf{\boldmath Term} & \textbf{\boldmath b} & \textbf{\boldmath SE} & \textbf{\boldmath 95\% CI} & \textbf{\boldmath t} & \textbf{\boldmath p} & \textbf{\boldmath Holm p} \\
\midrule\endhead
\midrule\multicolumn{8}{r}{\footnotesize Continued on next page}\\\endfoot
\bottomrule\endlastfoot
CSQ-VR total & Intercept & 8.703 & 0.391 & [7.926, 9.480] & 22.28 & < .001 & --- \\
 & MSSQ-Short total (per SD) & 0.524 & 0.432 & [\ensuremath{-}0.334, 1.383] & 1.21 & .228 & .912 \\
 & Older vs younger & \ensuremath{-}1.998 & 1.022 & [\ensuremath{-}4.030, 0.035] & \ensuremath{-}1.95 & .054 & .270 \\
 & Within-group age (per 10 years) & 0.358 & 0.895 & [\ensuremath{-}1.422, 2.139] & 0.40 & .690 & 1.000 \\
 & Gaming (per SD) & 0.176 & 0.463 & [\ensuremath{-}0.745, 1.096] & 0.38 & .705 & 1.000 \\
 & Technology (per SD) & \ensuremath{-}1.226 & 0.514 & [\ensuremath{-}2.249, \ensuremath{-}0.202] & \ensuremath{-}2.38 & .019 & .136 \\
 & Education (per year) & \ensuremath{-}0.051 & 0.140 & [\ensuremath{-}0.330, 0.228] & \ensuremath{-}0.36 & .719 & 1.000 \\
 & Male & \ensuremath{-}1.765 & 0.748 & [\ensuremath{-}3.253, \ensuremath{-}0.278] & \ensuremath{-}2.36 & .021 & .136 \\
\end{longtable}\normalsize
Note. As Supplementary Table S14, with within-group age added; residual df~=~83. Holm p: the seven coefficients form one family.

\Needspace{156pt}
\textbf{Supplementary Table S17.} Model of the CSQ-VR total with continuous age across the whole sample.

\setlength{\tabcolsep}{2.5pt}\renewcommand{\arraystretch}{1.12}
\fontsize{7.6}{9}\selectfont
\begin{longtable}{L{0.136117\dimexpr\textwidth-45pt\relax}L{0.189353\dimexpr\textwidth-45pt\relax}L{0.076931\dimexpr\textwidth-45pt\relax}L{0.070981\dimexpr\textwidth-45pt\relax}L{0.153862\dimexpr\textwidth-45pt\relax}L{0.153862\dimexpr\textwidth-45pt\relax}L{0.070981\dimexpr\textwidth-45pt\relax}L{0.070981\dimexpr\textwidth-45pt\relax}L{0.076931\dimexpr\textwidth-45pt\relax}}
\toprule
\textbf{\boldmath Outcome} & \textbf{\boldmath Term} & \textbf{\boldmath b} & \textbf{\boldmath SE} & \textbf{\boldmath 95\% CI} & \textbf{\boldmath Bootstrap 95\% CI} & \textbf{\boldmath t} & \textbf{\boldmath p} & \textbf{\boldmath Holm p} \\
\midrule\endfirsthead
\multicolumn{9}{l}{\textbf{Supplementary Table S17} (continued)}\\[3pt]\toprule
\textbf{\boldmath Outcome} & \textbf{\boldmath Term} & \textbf{\boldmath b} & \textbf{\boldmath SE} & \textbf{\boldmath 95\% CI} & \textbf{\boldmath Bootstrap 95\% CI} & \textbf{\boldmath t} & \textbf{\boldmath p} & \textbf{\boldmath Holm p} \\
\midrule\endhead
\midrule\multicolumn{9}{r}{\footnotesize Continued on next page}\\\endfoot
\bottomrule\endlastfoot
CSQ-VR total & Intercept & 8.703 & 0.388 & [7.932, 9.475] & [8.006, 9.471] & 22.43 & < .001 & --- \\
 & MSSQ-Short total (per SD) & 0.543 & 0.427 & [\ensuremath{-}0.307, 1.393] & [\ensuremath{-}0.240, 1.362] & 1.27 & .207 & .622 \\
 & Age (per 10 years) & \ensuremath{-}0.386 & 0.217 & [\ensuremath{-}0.817, 0.046] & [\ensuremath{-}0.801, 0.037] & \ensuremath{-}1.78 & .079 & .316 \\
 & Gaming (per SD) & 0.208 & 0.459 & [\ensuremath{-}0.703, 1.120] & [\ensuremath{-}0.535, 1.256] & 0.45 & .651 & 1.000 \\
 & Technology (per SD) & \ensuremath{-}1.189 & 0.501 & [\ensuremath{-}2.185, \ensuremath{-}0.193] & [\ensuremath{-}2.172, \ensuremath{-}0.237] & \ensuremath{-}2.37 & .020 & .099 \\
 & Education (per year) & \ensuremath{-}0.071 & 0.137 & [\ensuremath{-}0.342, 0.201] & [\ensuremath{-}0.361, 0.180] & \ensuremath{-}0.52 & .607 & 1.000 \\
 & Male & \ensuremath{-}1.845 & 0.754 & [\ensuremath{-}3.345, \ensuremath{-}0.346] & [\ensuremath{-}3.288, \ensuremath{-}0.452] & \ensuremath{-}2.45 & .016 & .099 \\
\end{longtable}\normalsize
Note. As Supplementary Table S14, with continuous age (centred, per 10 years) in place of age group; residual df~=~84. Bootstrap 95\% CI: case resampling (5,000 draws; no draw failed). Holm p: the six coefficients form one family.

\Needspace{165pt}
\textbf{Supplementary Table S18.} Model of the CSQ-VR total with the recent-adulthood MSSQ-Short section (continuous age).

\setlength{\tabcolsep}{2.5pt}\renewcommand{\arraystretch}{1.12}
\fontsize{8.5}{10}\selectfont
\begin{longtable}{L{0.136131\dimexpr\textwidth-40pt\relax}L{0.266312\dimexpr\textwidth-40pt\relax}L{0.094686\dimexpr\textwidth-40pt\relax}L{0.082785\dimexpr\textwidth-40pt\relax}L{0.183422\dimexpr\textwidth-40pt\relax}L{0.076939\dimexpr\textwidth-40pt\relax}L{0.076939\dimexpr\textwidth-40pt\relax}L{0.082785\dimexpr\textwidth-40pt\relax}}
\toprule
\textbf{\boldmath Outcome} & \textbf{\boldmath Term} & \textbf{\boldmath b} & \textbf{\boldmath SE} & \textbf{\boldmath 95\% CI} & \textbf{\boldmath t} & \textbf{\boldmath p} & \textbf{\boldmath Holm p} \\
\midrule\endfirsthead
\multicolumn{8}{l}{\textbf{Supplementary Table S18} (continued)}\\[3pt]\toprule
\textbf{\boldmath Outcome} & \textbf{\boldmath Term} & \textbf{\boldmath b} & \textbf{\boldmath SE} & \textbf{\boldmath 95\% CI} & \textbf{\boldmath t} & \textbf{\boldmath p} & \textbf{\boldmath Holm p} \\
\midrule\endhead
\midrule\multicolumn{8}{r}{\footnotesize Continued on next page}\\\endfoot
\bottomrule\endlastfoot
CSQ-VR total & Intercept & 8.703 & 0.390 & [7.929, 9.478] & 22.34 & < .001 & --- \\
 & MSSQ-Short recent adulthood (per SD) & 0.458 & 0.441 & [\ensuremath{-}0.419, 1.335] & 1.04 & .302 & .905 \\
 & Age (per 10 years) & \ensuremath{-}0.387 & 0.217 & [\ensuremath{-}0.819, 0.045] & \ensuremath{-}1.78 & .079 & .315 \\
 & Gaming (per SD) & 0.228 & 0.458 & [\ensuremath{-}0.683, 1.139] & 0.50 & .620 & 1.000 \\
 & Technology (per SD) & \ensuremath{-}1.285 & 0.502 & [\ensuremath{-}2.284, \ensuremath{-}0.287] & \ensuremath{-}2.56 & .012 & .074 \\
 & Education (per year) & \ensuremath{-}0.063 & 0.135 & [\ensuremath{-}0.332, 0.207] & \ensuremath{-}0.46 & .645 & 1.000 \\
 & Male & \ensuremath{-}1.881 & 0.759 & [\ensuremath{-}3.391, \ensuremath{-}0.371] & \ensuremath{-}2.48 & .015 & .076 \\
\end{longtable}\normalsize
Note. As Supplementary Table S17, with the recent-adulthood (last ten years) section of the MSSQ-Short (per sample SD) in place of the total. Holm p: the six coefficients form one family.

\Needspace{165pt}
\textbf{Supplementary Table S19.} Model of the CSQ-VR total on rank-transformed variables (continuous age).

\setlength{\tabcolsep}{2.5pt}\renewcommand{\arraystretch}{1.12}
\fontsize{8.5}{10}\selectfont
\begin{longtable}{L{0.136131\dimexpr\textwidth-40pt\relax}L{0.266312\dimexpr\textwidth-40pt\relax}L{0.094686\dimexpr\textwidth-40pt\relax}L{0.082785\dimexpr\textwidth-40pt\relax}L{0.183422\dimexpr\textwidth-40pt\relax}L{0.076939\dimexpr\textwidth-40pt\relax}L{0.076939\dimexpr\textwidth-40pt\relax}L{0.082785\dimexpr\textwidth-40pt\relax}}
\toprule
\textbf{\boldmath Outcome} & \textbf{\boldmath Term} & \textbf{\boldmath b} & \textbf{\boldmath SE} & \textbf{\boldmath 95\% CI} & \textbf{\boldmath t} & \textbf{\boldmath p} & \textbf{\boldmath Holm p} \\
\midrule\endfirsthead
\multicolumn{8}{l}{\textbf{Supplementary Table S19} (continued)}\\[3pt]\toprule
\textbf{\boldmath Outcome} & \textbf{\boldmath Term} & \textbf{\boldmath b} & \textbf{\boldmath SE} & \textbf{\boldmath 95\% CI} & \textbf{\boldmath t} & \textbf{\boldmath p} & \textbf{\boldmath Holm p} \\
\midrule\endhead
\midrule\multicolumn{8}{r}{\footnotesize Continued on next page}\\\endfoot
\bottomrule\endlastfoot
CSQ-VR total & Intercept & 0.000 & 0.109 & [\ensuremath{-}0.217, 0.217] & 0.00 & 1.000 & --- \\
 & MSSQ-Short total (rank) & 0.099 & 0.112 & [\ensuremath{-}0.124, 0.322] & 0.89 & .379 & 1.000 \\
 & Age (rank) & \ensuremath{-}0.226 & 0.159 & [\ensuremath{-}0.542, 0.090] & \ensuremath{-}1.42 & .158 & .791 \\
 & Gaming (rank) & 0.095 & 0.149 & [\ensuremath{-}0.201, 0.391] & 0.64 & .525 & 1.000 \\
 & Technology (rank) & \ensuremath{-}0.302 & 0.173 & [\ensuremath{-}0.645, 0.041] & \ensuremath{-}1.75 & .084 & .501 \\
 & Education (rank) & \ensuremath{-}0.008 & 0.113 & [\ensuremath{-}0.232, 0.217] & \ensuremath{-}0.07 & .946 & 1.000 \\
 & Male & \ensuremath{-}0.294 & 0.219 & [\ensuremath{-}0.730, 0.142] & \ensuremath{-}1.34 & .183 & .791 \\
\end{longtable}\normalsize
Note. The CSQ-VR total, MSSQ-Short total, age, gaming, technology and education were replaced by their standardised ranks across the whole sample (average ranks for ties); sex remained a centred indicator. Residual df~=~84. Holm p: the six coefficients form one family.

\Needspace{115pt}
\textbf{Supplementary Table S20.} CSQ-VR total by MSSQ-Short tertile.

\setlength{\tabcolsep}{2.5pt}\renewcommand{\arraystretch}{1.12}
\fontsize{7.6}{9}\selectfont
\begin{longtable}{L{0.099386\dimexpr\textwidth-45pt\relax}L{0.055884\dimexpr\textwidth-45pt\relax}L{0.198773\dimexpr\textwidth-45pt\relax}L{0.074512\dimexpr\textwidth-45pt\relax}L{0.074512\dimexpr\textwidth-45pt\relax}L{0.074512\dimexpr\textwidth-45pt\relax}L{0.142888\dimexpr\textwidth-45pt\relax}L{0.118014\dimexpr\textwidth-45pt\relax}L{0.161517\dimexpr\textwidth-45pt\relax}}
\toprule
\textbf{\boldmath Tertile} & \textbf{\boldmath n} & \textbf{\boldmath MSSQ-Short range} & \textbf{\boldmath M} & \textbf{\boldmath SD} & \textbf{\boldmath Mdn} & \textbf{\boldmath Q1 to Q3} & \textbf{\boldmath Min to max} & \textbf{\boldmath 95\% CI of M} \\
\midrule\endfirsthead
\multicolumn{9}{l}{\textbf{Supplementary Table S20} (continued)}\\[3pt]\toprule
\textbf{\boldmath Tertile} & \textbf{\boldmath n} & \textbf{\boldmath MSSQ-Short range} & \textbf{\boldmath M} & \textbf{\boldmath SD} & \textbf{\boldmath Mdn} & \textbf{\boldmath Q1 to Q3} & \textbf{\boldmath Min to max} & \textbf{\boldmath 95\% CI of M} \\
\midrule\endhead
\midrule\multicolumn{9}{r}{\footnotesize Continued on next page}\\\endfoot
\bottomrule\endlastfoot
Lower & 31 & 0.000 to 5.625 & 8.35 & 3.10 & 7.00 & 6.00 to 9.00 & 6 to 17 & [7.22, 9.49] \\
Middle & 30 & 6.000 to 14.143 & 8.83 & 4.14 & 6.50 & 6.00 to 9.75 & 6 to 18 & [7.29, 10.38] \\
Higher & 30 & 14.304 to 29.475 & 8.93 & 3.79 & 7.00 & 6.00 to 11.00 & 6 to 20 & [7.52, 10.35] \\
\end{longtable}\normalsize
Note. The tertiles are relative to this sample; the cut points of the MSSQ-Short total were 5.625 and 14.143 (the 33rd and 67th percentiles). No participant had a total above 30.

\Needspace{176pt}
\textbf{Supplementary Table S21.} Model of the CSQ-VR total across the MSSQ-Short tertiles (continuous age).

\setlength{\tabcolsep}{2.5pt}\renewcommand{\arraystretch}{1.12}
\fontsize{8.5}{10}\selectfont
\begin{longtable}{L{0.136131\dimexpr\textwidth-40pt\relax}L{0.266312\dimexpr\textwidth-40pt\relax}L{0.094686\dimexpr\textwidth-40pt\relax}L{0.082785\dimexpr\textwidth-40pt\relax}L{0.183422\dimexpr\textwidth-40pt\relax}L{0.076939\dimexpr\textwidth-40pt\relax}L{0.076939\dimexpr\textwidth-40pt\relax}L{0.082785\dimexpr\textwidth-40pt\relax}}
\toprule
\textbf{\boldmath Outcome} & \textbf{\boldmath Term} & \textbf{\boldmath b} & \textbf{\boldmath SE} & \textbf{\boldmath 95\% CI} & \textbf{\boldmath t} & \textbf{\boldmath p} & \textbf{\boldmath Holm p} \\
\midrule\endfirsthead
\multicolumn{8}{l}{\textbf{Supplementary Table S21} (continued)}\\[3pt]\toprule
\textbf{\boldmath Outcome} & \textbf{\boldmath Term} & \textbf{\boldmath b} & \textbf{\boldmath SE} & \textbf{\boldmath 95\% CI} & \textbf{\boldmath t} & \textbf{\boldmath p} & \textbf{\boldmath Holm p} \\
\midrule\endhead
\midrule\multicolumn{8}{r}{\footnotesize Continued on next page}\\\endfoot
\bottomrule\endlastfoot
CSQ-VR total & Intercept & 8.542 & 0.597 & [7.355, 9.730] & 14.31 & < .001 & --- \\
 & Middle vs lower tertile & 0.178 & 0.949 & [\ensuremath{-}1.710, 2.065] & 0.19 & .852 & 1.000 \\
 & Higher vs lower tertile & 0.311 & 0.930 & [\ensuremath{-}1.539, 2.161] & 0.33 & .739 & 1.000 \\
 & Age (per 10 years) & \ensuremath{-}0.370 & 0.220 & [\ensuremath{-}0.808, 0.068] & \ensuremath{-}1.68 & .097 & --- \\
 & Gaming (per SD) & 0.251 & 0.468 & [\ensuremath{-}0.679, 1.181] & 0.54 & .593 & --- \\
 & Technology (per SD) & \ensuremath{-}1.181 & 0.508 & [\ensuremath{-}2.191, \ensuremath{-}0.171] & \ensuremath{-}2.33 & .022 & --- \\
 & Education (per year) & \ensuremath{-}0.037 & 0.133 & [\ensuremath{-}0.303, 0.228] & \ensuremath{-}0.28 & .779 & --- \\
 & Male & \ensuremath{-}1.915 & 0.767 & [\ensuremath{-}3.441, \ensuremath{-}0.389] & \ensuremath{-}2.50 & .015 & --- \\
\end{longtable}\normalsize
Note. The lower tertile is the reference category, so the intercept is the predicted CSQ-VR total of the lower tertile at the means of the other predictors; residual df~=~83. Holm p: the two tertile contrasts form one family; the covariates are not adjusted.

\clearpage\section*{S7 Cybersickness subscale models}
Supplementary Table S22 reports the models of the three CSQ-VR subscales, the secondary symptom outcomes. They were fitted with continuous age, and their 18 coefficients were Holm-adjusted as one family. No coefficient was statistically significant after correction (smallest Holm p~=~.248), so the subscales add no association that the total did not show.

\Needspace{297pt}
\textbf{Supplementary Table S22.} Models of the CSQ-VR subscales (continuous age).

\setlength{\tabcolsep}{2.5pt}\renewcommand{\arraystretch}{1.12}
\fontsize{7.6}{9}\selectfont
\begin{longtable}{L{0.152984\dimexpr\textwidth-45pt\relax}L{0.200000\dimexpr\textwidth-45pt\relax}L{0.070576\dimexpr\textwidth-45pt\relax}L{0.064660\dimexpr\textwidth-45pt\relax}L{0.147068\dimexpr\textwidth-45pt\relax}L{0.147068\dimexpr\textwidth-45pt\relax}L{0.070576\dimexpr\textwidth-45pt\relax}L{0.070576\dimexpr\textwidth-45pt\relax}L{0.076492\dimexpr\textwidth-45pt\relax}}
\toprule
\textbf{\boldmath Outcome} & \textbf{\boldmath Term} & \textbf{\boldmath b} & \textbf{\boldmath SE} & \textbf{\boldmath 95\% CI} & \textbf{\boldmath Bootstrap 95\% CI} & \textbf{\boldmath t} & \textbf{\boldmath p} & \textbf{\boldmath Holm p} \\
\midrule\endfirsthead
\multicolumn{9}{l}{\textbf{Supplementary Table S22} (continued)}\\[3pt]\toprule
\textbf{\boldmath Outcome} & \textbf{\boldmath Term} & \textbf{\boldmath b} & \textbf{\boldmath SE} & \textbf{\boldmath 95\% CI} & \textbf{\boldmath Bootstrap 95\% CI} & \textbf{\boldmath t} & \textbf{\boldmath p} & \textbf{\boldmath Holm p} \\
\midrule\endhead
\midrule\multicolumn{9}{r}{\footnotesize Continued on next page}\\\endfoot
\bottomrule\endlastfoot
CSQ-VR nausea & Intercept & 2.736 & 0.121 & [2.496, 2.977] & [2.519, 2.984] & 22.63 & < .001 & --- \\
 & MSSQ-Short total (per SD) & 0.095 & 0.120 & [\ensuremath{-}0.144, 0.334] & [\ensuremath{-}0.129, 0.334] & 0.79 & .431 & 1.000 \\
 & Age (per 10 years) & \ensuremath{-}0.192 & 0.076 & [\ensuremath{-}0.344, \ensuremath{-}0.040] & [\ensuremath{-}0.338, \ensuremath{-}0.042] & \ensuremath{-}2.52 & .014 & .248 \\
 & Gaming (per SD) & \ensuremath{-}0.094 & 0.136 & [\ensuremath{-}0.365, 0.176] & [\ensuremath{-}0.345, 0.207] & \ensuremath{-}0.69 & .490 & 1.000 \\
 & Technology (per SD) & \ensuremath{-}0.331 & 0.152 & [\ensuremath{-}0.632, \ensuremath{-}0.029] & [\ensuremath{-}0.631, \ensuremath{-}0.050] & \ensuremath{-}2.18 & .032 & .452 \\
 & Education (per year) & \ensuremath{-}0.029 & 0.042 & [\ensuremath{-}0.113, 0.056] & [\ensuremath{-}0.115, 0.052] & \ensuremath{-}0.68 & .501 & 1.000 \\
 & Male & \ensuremath{-}0.533 & 0.235 & [\ensuremath{-}1.001, \ensuremath{-}0.065] & [\ensuremath{-}0.996, \ensuremath{-}0.080] & \ensuremath{-}2.26 & .026 & .419 \\
CSQ-VR vestibular & Intercept & 3.022 & 0.167 & [2.691, 3.353] & [2.736, 3.360] & 18.15 & < .001 & --- \\
 & MSSQ-Short total (per SD) & 0.169 & 0.162 & [\ensuremath{-}0.154, 0.492] & [\ensuremath{-}0.129, 0.489] & 1.04 & .300 & 1.000 \\
 & Age (per 10 years) & \ensuremath{-}0.077 & 0.105 & [\ensuremath{-}0.285, 0.131] & [\ensuremath{-}0.267, 0.128] & \ensuremath{-}0.74 & .463 & 1.000 \\
 & Gaming (per SD) & 0.121 & 0.205 & [\ensuremath{-}0.287, 0.529] & [\ensuremath{-}0.203, 0.592] & 0.59 & .557 & 1.000 \\
 & Technology (per SD) & \ensuremath{-}0.392 & 0.233 & [\ensuremath{-}0.854, 0.071] & [\ensuremath{-}0.814, 0.066] & \ensuremath{-}1.68 & .096 & 1.000 \\
 & Education (per year) & \ensuremath{-}0.041 & 0.063 & [\ensuremath{-}0.167, 0.084] & [\ensuremath{-}0.181, 0.064] & \ensuremath{-}0.65 & .514 & 1.000 \\
 & Male & \ensuremath{-}0.500 & 0.325 & [\ensuremath{-}1.146, 0.146] & [\ensuremath{-}1.112, 0.104] & \ensuremath{-}1.54 & .128 & 1.000 \\
CSQ-VR oculomotor & Intercept & 2.945 & 0.184 & [2.579, 3.311] & [2.635, 3.335] & 16.02 & < .001 & --- \\
 & MSSQ-Short total (per SD) & 0.279 & 0.253 & [\ensuremath{-}0.224, 0.781] & [\ensuremath{-}0.131, 0.816] & 1.10 & .274 & 1.000 \\
 & Age (per 10 years) & \ensuremath{-}0.117 & 0.095 & [\ensuremath{-}0.305, 0.072] & [\ensuremath{-}0.285, 0.084] & \ensuremath{-}1.23 & .222 & 1.000 \\
 & Gaming (per SD) & 0.182 & 0.177 & [\ensuremath{-}0.170, 0.533] & [\ensuremath{-}0.137, 0.576] & 1.03 & .306 & 1.000 \\
 & Technology (per SD) & \ensuremath{-}0.467 & 0.206 & [\ensuremath{-}0.877, \ensuremath{-}0.056] & [\ensuremath{-}0.887, \ensuremath{-}0.066] & \ensuremath{-}2.26 & .026 & .419 \\
 & Education (per year) & 0.000 & 0.051 & [\ensuremath{-}0.103, 0.102] & [\ensuremath{-}0.100, 0.106] & \ensuremath{-}0.01 & .993 & 1.000 \\
 & Male & \ensuremath{-}0.813 & 0.331 & [\ensuremath{-}1.471, \ensuremath{-}0.154] & [\ensuremath{-}1.485, \ensuremath{-}0.234] & \ensuremath{-}2.45 & .016 & .275 \\
\end{longtable}\normalsize
Note. Ordinary least squares; N~=~91, residual df~=~84; b in subscale points (range 2--14); SE~=~HC3 standard error with a t-based 95\% CI; bootstrap 95\% CI from case resampling (5,000 draws; no draw failed). Holm p: the 18 coefficients of the three models form one family.

\clearpage\section*{S8 Correlations among the predictors}
Supplementary Table S23 gives the correlations among the predictors. Age group overlapped strongly with technology experience (r~=~\ensuremath{-}.72) and moderately with gaming experience (r~=~\ensuremath{-}.55). This overlap accounts for the near-zero adjusted age-group differences in usability and pragmatic user experience once technology experience was held constant (Table 3 of the article). The variance inflation factors nevertheless remained low (at most 2.45; Supplementary Table S13). Continuous age and age group were almost interchangeable across the whole sample (r~=~.98), because no participant was aged 39--63 years.

\Needspace{165pt}
\textbf{Supplementary Table S23.} Pearson correlations among the predictors.

\setlength{\tabcolsep}{2.5pt}\renewcommand{\arraystretch}{1.12}
\fontsize{8.5}{10}\selectfont
\begin{longtable}{L{0.300177\dimexpr\textwidth-40pt\relax}L{0.099975\dimexpr\textwidth-40pt\relax}L{0.099975\dimexpr\textwidth-40pt\relax}L{0.099975\dimexpr\textwidth-40pt\relax}L{0.099975\dimexpr\textwidth-40pt\relax}L{0.099975\dimexpr\textwidth-40pt\relax}L{0.099975\dimexpr\textwidth-40pt\relax}L{0.099975\dimexpr\textwidth-40pt\relax}}
\toprule
\textbf{\boldmath Predictor} & \textbf{\boldmath 1} & \textbf{\boldmath 2} & \textbf{\boldmath 3} & \textbf{\boldmath 4} & \textbf{\boldmath 5} & \textbf{\boldmath 6} & \textbf{\boldmath 7} \\
\midrule\endfirsthead
\multicolumn{8}{l}{\textbf{Supplementary Table S23} (continued)}\\[3pt]\toprule
\textbf{\boldmath Predictor} & \textbf{\boldmath 1} & \textbf{\boldmath 2} & \textbf{\boldmath 3} & \textbf{\boldmath 4} & \textbf{\boldmath 5} & \textbf{\boldmath 6} & \textbf{\boldmath 7} \\
\midrule\endhead
\midrule\multicolumn{8}{r}{\footnotesize Continued on next page}\\\endfoot
\bottomrule\endlastfoot
1. Age group (older) & --- &  &  &  &  &  &  \\
2. Age (continuous) & .98 & --- &  &  &  &  &  \\
3. Gaming & \ensuremath{-}.55 & \ensuremath{-}.55 & --- &  &  &  &  \\
4. Technology & \ensuremath{-}.72 & \ensuremath{-}.72 & .47 & --- &  &  &  \\
5. Education & \ensuremath{-}.09 & \ensuremath{-}.12 & .04 & .04 & --- &  &  \\
6. Male & \ensuremath{-}.01 & \ensuremath{-}.04 & .07 & \ensuremath{-}.14 & .00 & --- &  \\
7. MSSQ-Short total & \ensuremath{-}.01 & \ensuremath{-}.01 & .07 & .03 & .20 & \ensuremath{-}.10 & --- \\
\end{longtable}\normalsize
Note. N~=~91. Age group: older~=~1, younger~=~0; male: male~=~1. Correlations are unchanged by the centring and standardisation of the predictors.

\clearpage\section*{S9 Ratings and TMT-VR performance: whole sample}
Supplementary Table S24 extends Table 4 of the article with Fisher-z intervals and the raw and Benjamini--Hochberg-adjusted p values of the 30 whole-sample correlations. Two of the 30 raw p values fell below .05, against the 1.5 expected by chance, and none survived adjustment (smallest adjusted p~=~.659). The largest coefficient was small (|r|~=~.23).

\Needspace{422pt}
\textbf{Supplementary Table S24.} Correlations between the ratings and the TMT-VR scores: whole sample (N~=~91).

\setlength{\tabcolsep}{2.5pt}\renewcommand{\arraystretch}{1.12}
\fontsize{8.5}{10}\selectfont
\begin{longtable}{L{0.222209\dimexpr\textwidth-35pt\relax}L{0.169972\dimexpr\textwidth-35pt\relax}L{0.084986\dimexpr\textwidth-35pt\relax}L{0.176430\dimexpr\textwidth-35pt\relax}L{0.176430\dimexpr\textwidth-35pt\relax}L{0.084986\dimexpr\textwidth-35pt\relax}L{0.084986\dimexpr\textwidth-35pt\relax}}
\toprule
\textbf{\boldmath TMT-VR score} & \textbf{\boldmath Rating} & \textbf{\boldmath r} & \textbf{\boldmath Fisher-z 95\% CI} & \textbf{\boldmath Bootstrap 95\% CI} & \textbf{\boldmath p} & \textbf{\boldmath BH p} \\
\midrule\endfirsthead
\multicolumn{7}{l}{\textbf{Supplementary Table S24} (continued)}\\[3pt]\toprule
\textbf{\boldmath TMT-VR score} & \textbf{\boldmath Rating} & \textbf{\boldmath r} & \textbf{\boldmath Fisher-z 95\% CI} & \textbf{\boldmath Bootstrap 95\% CI} & \textbf{\boldmath p} & \textbf{\boldmath BH p} \\
\midrule\endhead
\midrule\multicolumn{7}{r}{\footnotesize Continued on next page}\\\endfoot
\bottomrule\endlastfoot
Part A time & SUS & \ensuremath{-}.01 & [\ensuremath{-}.21, .20] & [\ensuremath{-}.23, .17] & .932 & .978 \\
 & Pragmatic UX & \ensuremath{-}.06 & [\ensuremath{-}.27, .14] & [\ensuremath{-}.29, .14] & .543 & .978 \\
 & Hedonic UX & .10 & [\ensuremath{-}.11, .30] & [\ensuremath{-}.13, .29] & .338 & .978 \\
 & Acceptability & \ensuremath{-}.10 & [\ensuremath{-}.30, .11] & [\ensuremath{-}.28, .12] & .370 & .978 \\
 & CSQ-VR total & .00 & [\ensuremath{-}.20, .21] & [\ensuremath{-}.24, .25] & .982 & .982 \\
Part B time & SUS & \ensuremath{-}.23 & [\ensuremath{-}.42, \ensuremath{-}.02] & [\ensuremath{-}.39, \ensuremath{-}.03] & .028 & .659 \\
 & Pragmatic UX & \ensuremath{-}.09 & [\ensuremath{-}.29, .12] & [\ensuremath{-}.34, .10] & .382 & .978 \\
 & Hedonic UX & \ensuremath{-}.02 & [\ensuremath{-}.22, .19] & [\ensuremath{-}.22, .19] & .868 & .978 \\
 & Acceptability & .09 & [\ensuremath{-}.12, .29] & [\ensuremath{-}.09, .26] & .390 & .978 \\
 & CSQ-VR total & \ensuremath{-}.04 & [\ensuremath{-}.25, .16] & [\ensuremath{-}.21, .15] & .677 & .978 \\
Total time & SUS & \ensuremath{-}.18 & [\ensuremath{-}.37, .03] & [\ensuremath{-}.36, .02] & .091 & .902 \\
 & Pragmatic UX & \ensuremath{-}.10 & [\ensuremath{-}.30, .11] & [\ensuremath{-}.32, .10] & .360 & .978 \\
 & Hedonic UX & .03 & [\ensuremath{-}.18, .23] & [\ensuremath{-}.17, .24] & .788 & .978 \\
 & Acceptability & .03 & [\ensuremath{-}.18, .23] & [\ensuremath{-}.17, .22] & .779 & .978 \\
 & CSQ-VR total & \ensuremath{-}.03 & [\ensuremath{-}.24, .17] & [\ensuremath{-}.22, .17] & .759 & .978 \\
Part A adjusted time & SUS & .01 & [\ensuremath{-}.20, .21] & [\ensuremath{-}.23, .18] & .934 & .978 \\
 & Pragmatic UX & \ensuremath{-}.03 & [\ensuremath{-}.23, .18] & [\ensuremath{-}.28, .16] & .778 & .978 \\
 & Hedonic UX & .13 & [\ensuremath{-}.08, .33] & [\ensuremath{-}.10, .31] & .226 & .978 \\
 & Acceptability & \ensuremath{-}.11 & [\ensuremath{-}.31, .10] & [\ensuremath{-}.28, .11] & .296 & .978 \\
 & CSQ-VR total & \ensuremath{-}.03 & [\ensuremath{-}.24, .17] & [\ensuremath{-}.24, .21] & .746 & .978 \\
Part B adjusted time & SUS & \ensuremath{-}.21 & [\ensuremath{-}.40, \ensuremath{-}.01] & [\ensuremath{-}.37, \ensuremath{-}.03] & .044 & .659 \\
 & Pragmatic UX & \ensuremath{-}.05 & [\ensuremath{-}.26, .15] & [\ensuremath{-}.34, .14] & .606 & .978 \\
 & Hedonic UX & \ensuremath{-}.01 & [\ensuremath{-}.21, .20] & [\ensuremath{-}.21, .17] & .946 & .978 \\
 & Acceptability & .07 & [\ensuremath{-}.14, .27] & [\ensuremath{-}.10, .23] & .492 & .978 \\
 & CSQ-VR total & \ensuremath{-}.06 & [\ensuremath{-}.26, .15] & [\ensuremath{-}.20, .15] & .581 & .978 \\
Total adjusted time & SUS & \ensuremath{-}.16 & [\ensuremath{-}.36, .04] & [\ensuremath{-}.35, .04] & .120 & .902 \\
 & Pragmatic UX & \ensuremath{-}.06 & [\ensuremath{-}.27, .14] & [\ensuremath{-}.33, .13] & .550 & .978 \\
 & Hedonic UX & .04 & [\ensuremath{-}.16, .25] & [\ensuremath{-}.17, .23] & .681 & .978 \\
 & Acceptability & .01 & [\ensuremath{-}.19, .22] & [\ensuremath{-}.18, .20] & .911 & .978 \\
 & CSQ-VR total & \ensuremath{-}.05 & [\ensuremath{-}.26, .15] & [\ensuremath{-}.23, .16] & .618 & .978 \\
\end{longtable}\normalsize
Note. Pearson r with Fisher-z and paired-participant bootstrap 95\% CIs (5,000 draws). BH p~=~Benjamini--Hochberg-adjusted p across the 30 correlations. Adjusted time~=~completion time \ensuremath{\times} (1 + wrong-target selections/24) for each part and \ensuremath{\times} (1 + wrong-target selections/48) for the total. Abbreviations as in Supplementary Table S2.

\clearpage\section*{S10 Ratings and TMT-VR performance: within-group and Spearman families}
Supplementary Tables S25 and S26 report the same family within the younger and the older group, and Supplementary Table S27 the Spearman sensitivity families of the three populations. No correlation survived adjustment in any family (smallest adjusted p: younger .446, older .951; Spearman: whole sample .511, younger .927, older .977). The largest within-group coefficients linked slower Part A times with slightly higher hedonic ratings in the younger group (r~=~.30 and .34 for the completion and adjusted times); neither survived adjustment.

\Needspace{422pt}
\textbf{Supplementary Table S25.} Correlations between the ratings and the TMT-VR scores: younger group (n~=~47).

\setlength{\tabcolsep}{2.5pt}\renewcommand{\arraystretch}{1.12}
\fontsize{8.5}{10}\selectfont
\begin{longtable}{L{0.222209\dimexpr\textwidth-35pt\relax}L{0.169972\dimexpr\textwidth-35pt\relax}L{0.084986\dimexpr\textwidth-35pt\relax}L{0.176430\dimexpr\textwidth-35pt\relax}L{0.176430\dimexpr\textwidth-35pt\relax}L{0.084986\dimexpr\textwidth-35pt\relax}L{0.084986\dimexpr\textwidth-35pt\relax}}
\toprule
\textbf{\boldmath TMT-VR score} & \textbf{\boldmath Rating} & \textbf{\boldmath r} & \textbf{\boldmath Fisher-z 95\% CI} & \textbf{\boldmath Bootstrap 95\% CI} & \textbf{\boldmath p} & \textbf{\boldmath BH p} \\
\midrule\endfirsthead
\multicolumn{7}{l}{\textbf{Supplementary Table S25} (continued)}\\[3pt]\toprule
\textbf{\boldmath TMT-VR score} & \textbf{\boldmath Rating} & \textbf{\boldmath r} & \textbf{\boldmath Fisher-z 95\% CI} & \textbf{\boldmath Bootstrap 95\% CI} & \textbf{\boldmath p} & \textbf{\boldmath BH p} \\
\midrule\endhead
\midrule\multicolumn{7}{r}{\footnotesize Continued on next page}\\\endfoot
\bottomrule\endlastfoot
Part A time & SUS & .28 & [\ensuremath{-}.01, .52] & [\ensuremath{-}.04, .53] & .059 & .446 \\
 & Pragmatic UX & .20 & [\ensuremath{-}.10, .46] & [\ensuremath{-}.06, .41] & .184 & .586 \\
 & Hedonic UX & .30 & [.01, .54] & [.03, .51] & .044 & .446 \\
 & Acceptability & \ensuremath{-}.04 & [\ensuremath{-}.32, .25] & [\ensuremath{-}.30, .23] & .808 & .858 \\
 & CSQ-VR total & \ensuremath{-}.19 & [\ensuremath{-}.45, .10] & [\ensuremath{-}.38, .03] & .195 & .586 \\
Part B time & SUS & .03 & [\ensuremath{-}.26, .32] & [\ensuremath{-}.40, .35] & .830 & .858 \\
 & Pragmatic UX & \ensuremath{-}.11 & [\ensuremath{-}.39, .18] & [\ensuremath{-}.51, .23] & .455 & .758 \\
 & Hedonic UX & .14 & [\ensuremath{-}.15, .41] & [\ensuremath{-}.15, .36] & .343 & .719 \\
 & Acceptability & .12 & [\ensuremath{-}.18, .39] & [\ensuremath{-}.10, .33] & .430 & .758 \\
 & CSQ-VR total & \ensuremath{-}.05 & [\ensuremath{-}.33, .24] & [\ensuremath{-}.24, .23] & .733 & .853 \\
Total time & SUS & .15 & [\ensuremath{-}.14, .42] & [\ensuremath{-}.23, .40] & .301 & .719 \\
 & Pragmatic UX & .00 & [\ensuremath{-}.29, .29] & [\ensuremath{-}.33, .26] & .988 & .988 \\
 & Hedonic UX & .25 & [\ensuremath{-}.04, .50] & [\ensuremath{-}.03, .47] & .090 & .449 \\
 & Acceptability & .08 & [\ensuremath{-}.21, .36] & [\ensuremath{-}.16, .31] & .606 & .853 \\
 & CSQ-VR total & \ensuremath{-}.13 & [\ensuremath{-}.40, .16] & [\ensuremath{-}.32, .12] & .384 & .719 \\
Part A adjusted time & SUS & .29 & [.00, .53] & [\ensuremath{-}.04, .53] & .049 & .446 \\
 & Pragmatic UX & .20 & [\ensuremath{-}.09, .46] & [\ensuremath{-}.06, .43] & .169 & .586 \\
 & Hedonic UX & .34 & [.05, .57] & [.09, .55] & .021 & .446 \\
 & Acceptability & \ensuremath{-}.06 & [\ensuremath{-}.34, .23] & [\ensuremath{-}.31, .20] & .698 & .853 \\
 & CSQ-VR total & \ensuremath{-}.20 & [\ensuremath{-}.46, .10] & [\ensuremath{-}.39, .03] & .187 & .586 \\
Part B adjusted time & SUS & .07 & [\ensuremath{-}.23, .35] & [\ensuremath{-}.38, .36] & .660 & .853 \\
 & Pragmatic UX & \ensuremath{-}.07 & [\ensuremath{-}.35, .23] & [\ensuremath{-}.51, .26] & .657 & .853 \\
 & Hedonic UX & .14 & [\ensuremath{-}.16, .41] & [\ensuremath{-}.18, .35] & .362 & .719 \\
 & Acceptability & .09 & [\ensuremath{-}.20, .37] & [\ensuremath{-}.12, .32] & .533 & .841 \\
 & CSQ-VR total & \ensuremath{-}.07 & [\ensuremath{-}.35, .22] & [\ensuremath{-}.26, .21] & .637 & .853 \\
Total adjusted time & SUS & .18 & [\ensuremath{-}.12, .44] & [\ensuremath{-}.21, .42] & .236 & .645 \\
 & Pragmatic UX & .03 & [\ensuremath{-}.26, .32] & [\ensuremath{-}.30, .28] & .824 & .858 \\
 & Hedonic UX & .26 & [\ensuremath{-}.03, .51] & [\ensuremath{-}.03, .47] & .079 & .449 \\
 & Acceptability & .05 & [\ensuremath{-}.24, .33] & [\ensuremath{-}.18, .28] & .739 & .853 \\
 & CSQ-VR total & \ensuremath{-}.14 & [\ensuremath{-}.41, .15] & [\ensuremath{-}.32, .11] & .349 & .719 \\
\end{longtable}\normalsize
Note. As Supplementary Table S24; Benjamini--Hochberg adjustment across the 30 correlations of the younger group.

\Needspace{422pt}
\textbf{Supplementary Table S26.} Correlations between the ratings and the TMT-VR scores: older group (n~=~44).

\setlength{\tabcolsep}{2.5pt}\renewcommand{\arraystretch}{1.12}
\fontsize{8.5}{10}\selectfont
\begin{longtable}{L{0.222209\dimexpr\textwidth-35pt\relax}L{0.169972\dimexpr\textwidth-35pt\relax}L{0.084986\dimexpr\textwidth-35pt\relax}L{0.176430\dimexpr\textwidth-35pt\relax}L{0.176430\dimexpr\textwidth-35pt\relax}L{0.084986\dimexpr\textwidth-35pt\relax}L{0.084986\dimexpr\textwidth-35pt\relax}}
\toprule
\textbf{\boldmath TMT-VR score} & \textbf{\boldmath Rating} & \textbf{\boldmath r} & \textbf{\boldmath Fisher-z 95\% CI} & \textbf{\boldmath Bootstrap 95\% CI} & \textbf{\boldmath p} & \textbf{\boldmath BH p} \\
\midrule\endfirsthead
\multicolumn{7}{l}{\textbf{Supplementary Table S26} (continued)}\\[3pt]\toprule
\textbf{\boldmath TMT-VR score} & \textbf{\boldmath Rating} & \textbf{\boldmath r} & \textbf{\boldmath Fisher-z 95\% CI} & \textbf{\boldmath Bootstrap 95\% CI} & \textbf{\boldmath p} & \textbf{\boldmath BH p} \\
\midrule\endhead
\midrule\multicolumn{7}{r}{\footnotesize Continued on next page}\\\endfoot
\bottomrule\endlastfoot
Part A time & SUS & .11 & [\ensuremath{-}.20, .39] & [\ensuremath{-}.25, .36] & .488 & .951 \\
 & Pragmatic UX & .03 & [\ensuremath{-}.27, .33] & [\ensuremath{-}.30, .30] & .839 & .966 \\
 & Hedonic UX & .09 & [\ensuremath{-}.21, .38] & [\ensuremath{-}.28, .37] & .552 & .951 \\
 & Acceptability & \ensuremath{-}.12 & [\ensuremath{-}.40, .19] & [\ensuremath{-}.40, .24] & .457 & .951 \\
 & CSQ-VR total & .16 & [\ensuremath{-}.14, .44] & [\ensuremath{-}.18, .52] & .297 & .951 \\
Part B time & SUS & \ensuremath{-}.20 & [\ensuremath{-}.47, .11] & [\ensuremath{-}.42, .04] & .201 & .951 \\
 & Pragmatic UX & .07 & [\ensuremath{-}.23, .36] & [\ensuremath{-}.24, .32] & .650 & .951 \\
 & Hedonic UX & \ensuremath{-}.05 & [\ensuremath{-}.35, .25] & [\ensuremath{-}.32, .21] & .729 & .951 \\
 & Acceptability & .16 & [\ensuremath{-}.15, .43] & [\ensuremath{-}.14, .39] & .313 & .951 \\
 & CSQ-VR total & \ensuremath{-}.01 & [\ensuremath{-}.31, .29] & [\ensuremath{-}.24, .30] & .953 & .966 \\
Total time & SUS & \ensuremath{-}.12 & [\ensuremath{-}.40, .18] & [\ensuremath{-}.39, .16] & .440 & .951 \\
 & Pragmatic UX & .07 & [\ensuremath{-}.23, .36] & [\ensuremath{-}.26, .32] & .643 & .951 \\
 & Hedonic UX & \ensuremath{-}.01 & [\ensuremath{-}.30, .29] & [\ensuremath{-}.30, .25] & .966 & .966 \\
 & Acceptability & .08 & [\ensuremath{-}.22, .37] & [\ensuremath{-}.24, .38] & .597 & .951 \\
 & CSQ-VR total & .06 & [\ensuremath{-}.24, .35] & [\ensuremath{-}.17, .35] & .702 & .951 \\
Part A adjusted time & SUS & .12 & [\ensuremath{-}.19, .40] & [\ensuremath{-}.27, .35] & .450 & .951 \\
 & Pragmatic UX & .07 & [\ensuremath{-}.23, .36] & [\ensuremath{-}.28, .31] & .634 & .951 \\
 & Hedonic UX & .13 & [\ensuremath{-}.18, .41] & [\ensuremath{-}.28, .38] & .416 & .951 \\
 & Acceptability & \ensuremath{-}.13 & [\ensuremath{-}.41, .17] & [\ensuremath{-}.40, .24] & .391 & .951 \\
 & CSQ-VR total & .07 & [\ensuremath{-}.23, .36] & [\ensuremath{-}.20, .45] & .641 & .951 \\
Part B adjusted time & SUS & \ensuremath{-}.18 & [\ensuremath{-}.45, .12] & [\ensuremath{-}.40, .04] & .244 & .951 \\
 & Pragmatic UX & .09 & [\ensuremath{-}.21, .38] & [\ensuremath{-}.28, .32] & .547 & .951 \\
 & Hedonic UX & \ensuremath{-}.02 & [\ensuremath{-}.32, .27] & [\ensuremath{-}.33, .20] & .876 & .966 \\
 & Acceptability & .13 & [\ensuremath{-}.17, .41] & [\ensuremath{-}.14, .35] & .397 & .951 \\
 & CSQ-VR total & \ensuremath{-}.03 & [\ensuremath{-}.33, .27] & [\ensuremath{-}.22, .29] & .840 & .966 \\
Total adjusted time & SUS & \ensuremath{-}.11 & [\ensuremath{-}.39, .20] & [\ensuremath{-}.38, .16] & .495 & .951 \\
 & Pragmatic UX & .10 & [\ensuremath{-}.20, .39] & [\ensuremath{-}.27, .35] & .519 & .951 \\
 & Hedonic UX & .03 & [\ensuremath{-}.27, .32] & [\ensuremath{-}.29, .26] & .867 & .966 \\
 & Acceptability & .06 & [\ensuremath{-}.25, .35] & [\ensuremath{-}.28, .35] & .720 & .951 \\
 & CSQ-VR total & .01 & [\ensuremath{-}.28, .31] & [\ensuremath{-}.18, .32] & .926 & .966 \\
\end{longtable}\normalsize
Note. As Supplementary Table S24; Benjamini--Hochberg adjustment across the 30 correlations of the older group.

\Needspace{387pt}
\textbf{Supplementary Table S27.} Spearman correlations between the ratings and the TMT-VR scores (sensitivity families).

\setlength{\tabcolsep}{2.5pt}\renewcommand{\arraystretch}{1.12}
\fontsize{7.6}{9}\selectfont
\begin{longtable}{L{0.134011\dimexpr\textwidth-70pt\relax}L{0.098243\dimexpr\textwidth-70pt\relax}L{0.049082\dimexpr\textwidth-70pt\relax}L{0.098243\dimexpr\textwidth-70pt\relax}L{0.049082\dimexpr\textwidth-70pt\relax}L{0.049082\dimexpr\textwidth-70pt\relax}L{0.066966\dimexpr\textwidth-70pt\relax}L{0.098243\dimexpr\textwidth-70pt\relax}L{0.049082\dimexpr\textwidth-70pt\relax}L{0.049082\dimexpr\textwidth-70pt\relax}L{0.062475\dimexpr\textwidth-70pt\relax}L{0.098243\dimexpr\textwidth-70pt\relax}L{0.049082\dimexpr\textwidth-70pt\relax}L{0.049082\dimexpr\textwidth-70pt\relax}}
\toprule
\textbf{\boldmath TMT-VR score} & \textbf{\boldmath Rating} & \textbf{\boldmath All \ensuremath{\uprho}} & \textbf{\boldmath Bootstrap 95\% CI} & \textbf{\boldmath p} & \textbf{\boldmath BH p} & \textbf{\boldmath Younger \ensuremath{\uprho}} & \textbf{\boldmath Bootstrap 95\% CI} & \textbf{\boldmath p} & \textbf{\boldmath BH p} & \textbf{\boldmath Older \ensuremath{\uprho}} & \textbf{\boldmath Bootstrap 95\% CI} & \textbf{\boldmath p} & \textbf{\boldmath BH p} \\
\midrule\endfirsthead
\multicolumn{14}{l}{\textbf{Supplementary Table S27} (continued)}\\[3pt]\toprule
\textbf{\boldmath TMT-VR score} & \textbf{\boldmath Rating} & \textbf{\boldmath All \ensuremath{\uprho}} & \textbf{\boldmath Bootstrap 95\% CI} & \textbf{\boldmath p} & \textbf{\boldmath BH p} & \textbf{\boldmath Younger \ensuremath{\uprho}} & \textbf{\boldmath Bootstrap 95\% CI} & \textbf{\boldmath p} & \textbf{\boldmath BH p} & \textbf{\boldmath Older \ensuremath{\uprho}} & \textbf{\boldmath Bootstrap 95\% CI} & \textbf{\boldmath p} & \textbf{\boldmath BH p} \\
\midrule\endhead
\midrule\multicolumn{14}{r}{\footnotesize Continued on next page}\\\endfoot
\bottomrule\endlastfoot
Part A time & SUS & \ensuremath{-}.05 & [\ensuremath{-}.26, .15] & .609 & .950 & .17 & [\ensuremath{-}.15, .46] & .257 & .927 & \ensuremath{-}.01 & [\ensuremath{-}.31, .30] & .966 & .985 \\
 & Pragmatic UX & \ensuremath{-}.08 & [\ensuremath{-}.28, .13] & .445 & .950 & .13 & [\ensuremath{-}.16, .40] & .367 & .927 & \ensuremath{-}.03 & [\ensuremath{-}.31, .25] & .829 & .977 \\
 & Hedonic UX & .08 & [\ensuremath{-}.13, .29] & .460 & .950 & .19 & [\ensuremath{-}.12, .47] & .213 & .927 & .02 & [\ensuremath{-}.30, .35] & .904 & .977 \\
 & Acceptability & \ensuremath{-}.04 & [\ensuremath{-}.25, .17] & .691 & .950 & \ensuremath{-}.04 & [\ensuremath{-}.32, .26] & .798 & .927 & \ensuremath{-}.05 & [\ensuremath{-}.36, .26] & .750 & .977 \\
 & CSQ-VR total & \ensuremath{-}.08 & [\ensuremath{-}.30, .14] & .459 & .950 & \ensuremath{-}.15 & [\ensuremath{-}.41, .13] & .315 & .927 & .17 & [\ensuremath{-}.17, .48] & .270 & .977 \\
Part B time & SUS & \ensuremath{-}.19 & [\ensuremath{-}.38, .02] & .068 & .511 & .02 & [\ensuremath{-}.26, .29] & .916 & .927 & \ensuremath{-}.16 & [\ensuremath{-}.40, .13] & .315 & .977 \\
 & Pragmatic UX & \ensuremath{-}.20 & [\ensuremath{-}.40, .01] & .052 & .511 & \ensuremath{-}.13 & [\ensuremath{-}.40, .15] & .389 & .927 & \ensuremath{-}.07 & [\ensuremath{-}.36, .23] & .670 & .977 \\
 & Hedonic UX & \ensuremath{-}.02 & [\ensuremath{-}.22, .20] & .883 & .950 & .04 & [\ensuremath{-}.24, .32] & .787 & .927 & \ensuremath{-}.02 & [\ensuremath{-}.30, .27] & .912 & .977 \\
 & Acceptability & .04 & [\ensuremath{-}.17, .25] & .695 & .950 & .09 & [\ensuremath{-}.19, .37] & .528 & .927 & .06 & [\ensuremath{-}.27, .38] & .690 & .977 \\
 & CSQ-VR total & .05 & [\ensuremath{-}.15, .25] & .612 & .950 & .10 & [\ensuremath{-}.20, .39] & .497 & .927 & .22 & [\ensuremath{-}.12, .52] & .160 & .977 \\
Total time & SUS & \ensuremath{-}.16 & [\ensuremath{-}.35, .05] & .127 & .564 & .07 & [\ensuremath{-}.23, .35] & .626 & .927 & \ensuremath{-}.15 & [\ensuremath{-}.41, .15] & .331 & .977 \\
 & Pragmatic UX & \ensuremath{-}.17 & [\ensuremath{-}.36, .05] & .118 & .564 & \ensuremath{-}.03 & [\ensuremath{-}.30, .25] & .839 & .927 & \ensuremath{-}.07 & [\ensuremath{-}.35, .23] & .672 & .977 \\
 & Hedonic UX & .03 & [\ensuremath{-}.18, .25] & .772 & .950 & .15 & [\ensuremath{-}.15, .43] & .315 & .927 & \ensuremath{-}.04 & [\ensuremath{-}.33, .26] & .803 & .977 \\
 & Acceptability & .02 & [\ensuremath{-}.18, .23] & .876 & .950 & .03 & [\ensuremath{-}.26, .32] & .820 & .927 & .06 & [\ensuremath{-}.29, .40] & .698 & .977 \\
 & CSQ-VR total & .00 & [\ensuremath{-}.21, .20] & .996 & .996 & .03 & [\ensuremath{-}.25, .31] & .850 & .927 & .23 & [\ensuremath{-}.07, .49] & .140 & .977 \\
Part A adjusted time & SUS & \ensuremath{-}.05 & [\ensuremath{-}.26, .17] & .656 & .950 & .17 & [\ensuremath{-}.16, .47] & .255 & .927 & .00 & [\ensuremath{-}.32, .30] & .985 & .985 \\
 & Pragmatic UX & \ensuremath{-}.07 & [\ensuremath{-}.27, .13] & .509 & .950 & .17 & [\ensuremath{-}.12, .44] & .262 & .927 & \ensuremath{-}.02 & [\ensuremath{-}.30, .26] & .874 & .977 \\
 & Hedonic UX & .11 & [\ensuremath{-}.11, .31] & .317 & .950 & .25 & [\ensuremath{-}.05, .51] & .092 & .927 & .02 & [\ensuremath{-}.31, .36] & .874 & .977 \\
 & Acceptability & \ensuremath{-}.04 & [\ensuremath{-}.24, .16] & .722 & .950 & \ensuremath{-}.05 & [\ensuremath{-}.33, .23] & .729 & .927 & \ensuremath{-}.03 & [\ensuremath{-}.32, .28] & .864 & .977 \\
 & CSQ-VR total & \ensuremath{-}.08 & [\ensuremath{-}.30, .13] & .430 & .950 & \ensuremath{-}.12 & [\ensuremath{-}.38, .16] & .424 & .927 & .12 & [\ensuremath{-}.22, .44] & .451 & .977 \\
Part B adjusted time & SUS & \ensuremath{-}.20 & [\ensuremath{-}.39, .01] & .056 & .511 & .02 & [\ensuremath{-}.27, .29] & .911 & .927 & \ensuremath{-}.17 & [\ensuremath{-}.43, .13] & .282 & .977 \\
 & Pragmatic UX & \ensuremath{-}.21 & [\ensuremath{-}.41, \ensuremath{-}.01] & .044 & .511 & \ensuremath{-}.12 & [\ensuremath{-}.41, .16] & .406 & .927 & \ensuremath{-}.08 & [\ensuremath{-}.38, .22] & .605 & .977 \\
 & Hedonic UX & \ensuremath{-}.03 & [\ensuremath{-}.24, .18] & .781 & .950 & .03 & [\ensuremath{-}.26, .30] & .867 & .927 & \ensuremath{-}.03 & [\ensuremath{-}.31, .26] & .861 & .977 \\
 & Acceptability & .02 & [\ensuremath{-}.19, .22] & .883 & .950 & .07 & [\ensuremath{-}.21, .35] & .619 & .927 & .04 & [\ensuremath{-}.30, .36] & .807 & .977 \\
 & CSQ-VR total & .05 & [\ensuremath{-}.16, .26] & .645 & .950 & .10 & [\ensuremath{-}.19, .39] & .497 & .927 & .19 & [\ensuremath{-}.13, .49] & .210 & .977 \\
Total adjusted time & SUS & \ensuremath{-}.15 & [\ensuremath{-}.35, .06] & .144 & .564 & .09 & [\ensuremath{-}.21, .37] & .533 & .927 & \ensuremath{-}.16 & [\ensuremath{-}.43, .13] & .293 & .977 \\
 & Pragmatic UX & \ensuremath{-}.15 & [\ensuremath{-}.35, .06] & .150 & .564 & .01 & [\ensuremath{-}.26, .29] & .927 & .927 & \ensuremath{-}.07 & [\ensuremath{-}.36, .23] & .659 & .977 \\
 & Hedonic UX & .04 & [\ensuremath{-}.17, .26] & .689 & .950 & .18 & [\ensuremath{-}.12, .45] & .237 & .927 & \ensuremath{-}.04 & [\ensuremath{-}.33, .26] & .812 & .977 \\
 & Acceptability & .01 & [\ensuremath{-}.20, .22] & .949 & .981 & .02 & [\ensuremath{-}.28, .30] & .917 & .927 & .06 & [\ensuremath{-}.27, .39] & .687 & .977 \\
 & CSQ-VR total & \ensuremath{-}.02 & [\ensuremath{-}.22, .19] & .886 & .950 & .02 & [\ensuremath{-}.26, .29] & .868 & .927 & .18 & [\ensuremath{-}.12, .47] & .239 & .977 \\
\end{longtable}\normalsize
Note. Spearman \ensuremath{\uprho} with paired-participant bootstrap 95\% CIs (5,000 draws); Benjamini--Hochberg adjustment (BH p) across the 30 correlations of each population. The Fisher-z intervals of these coefficients are in the archived outputs.

\clearpage\section*{S11 Ratings and TMT-VR performance: partial correlations}
Supplementary Table S28 reports the 30 pooled partial correlations, adjusted for age group, within-group age, education and sex. None survived adjustment (smallest adjusted p~=~.783), and the coefficients remained small (|r|~\ensuremath{\leq}~.19).

\Needspace{422pt}
\textbf{Supplementary Table S28.} Partial correlations between the ratings and the TMT-VR scores (N~=~91).

\setlength{\tabcolsep}{2.5pt}\renewcommand{\arraystretch}{1.12}
\fontsize{8.5}{10}\selectfont
\begin{longtable}{L{0.257182\dimexpr\textwidth-30pt\relax}L{0.199975\dimexpr\textwidth-30pt\relax}L{0.114289\dimexpr\textwidth-30pt\relax}L{0.214340\dimexpr\textwidth-30pt\relax}L{0.107107\dimexpr\textwidth-30pt\relax}L{0.107107\dimexpr\textwidth-30pt\relax}}
\toprule
\textbf{\boldmath TMT-VR score} & \textbf{\boldmath Rating} & \textbf{\boldmath Partial r} & \textbf{\boldmath Bootstrap 95\% CI} & \textbf{\boldmath p} & \textbf{\boldmath BH p} \\
\midrule\endfirsthead
\multicolumn{6}{l}{\textbf{Supplementary Table S28} (continued)}\\[3pt]\toprule
\textbf{\boldmath TMT-VR score} & \textbf{\boldmath Rating} & \textbf{\boldmath Partial r} & \textbf{\boldmath Bootstrap 95\% CI} & \textbf{\boldmath p} & \textbf{\boldmath BH p} \\
\midrule\endhead
\midrule\multicolumn{6}{r}{\footnotesize Continued on next page}\\\endfoot
\bottomrule\endlastfoot
Part A time & SUS & .17 & [\ensuremath{-}.10, .37] & .123 & .783 \\
 & Pragmatic UX & .08 & [\ensuremath{-}.16, .28] & .458 & .783 \\
 & Hedonic UX & .17 & [\ensuremath{-}.06, .38] & .111 & .783 \\
 & Acceptability & \ensuremath{-}.09 & [\ensuremath{-}.27, .13] & .430 & .783 \\
 & CSQ-VR total & \ensuremath{-}.02 & [\ensuremath{-}.25, .24] & .862 & .999 \\
Part B time & SUS & \ensuremath{-}.11 & [\ensuremath{-}.31, .14] & .324 & .783 \\
 & Pragmatic UX & .00 & [\ensuremath{-}.25, .21] & 1.000 & 1.000 \\
 & Hedonic UX & .01 & [\ensuremath{-}.19, .22] & .954 & 1.000 \\
 & Acceptability & .15 & [\ensuremath{-}.04, .31] & .173 & .783 \\
 & CSQ-VR total & \ensuremath{-}.07 & [\ensuremath{-}.27, .14] & .533 & .783 \\
Total time & SUS & \ensuremath{-}.02 & [\ensuremath{-}.26, .23] & .866 & .999 \\
 & Pragmatic UX & .03 & [\ensuremath{-}.20, .24] & .749 & .977 \\
 & Hedonic UX & .08 & [\ensuremath{-}.13, .29] & .463 & .783 \\
 & Acceptability & .09 & [\ensuremath{-}.13, .28] & .421 & .783 \\
 & CSQ-VR total & \ensuremath{-}.07 & [\ensuremath{-}.25, .14] & .548 & .783 \\
Part A adjusted time & SUS & .17 & [\ensuremath{-}.09, .37] & .107 & .783 \\
 & Pragmatic UX & .10 & [\ensuremath{-}.16, .28] & .342 & .783 \\
 & Hedonic UX & .19 & [\ensuremath{-}.04, .36] & .074 & .783 \\
 & Acceptability & \ensuremath{-}.10 & [\ensuremath{-}.27, .11] & .347 & .783 \\
 & CSQ-VR total & \ensuremath{-}.07 & [\ensuremath{-}.25, .19] & .548 & .783 \\
Part B adjusted time & SUS & \ensuremath{-}.08 & [\ensuremath{-}.29, .14] & .443 & .783 \\
 & Pragmatic UX & .03 & [\ensuremath{-}.25, .22] & .803 & .999 \\
 & Hedonic UX & .01 & [\ensuremath{-}.19, .20] & .906 & 1.000 \\
 & Acceptability & .13 & [\ensuremath{-}.05, .29] & .229 & .783 \\
 & CSQ-VR total & \ensuremath{-}.10 & [\ensuremath{-}.28, .12] & .366 & .783 \\
Total adjusted time & SUS & .00 & [\ensuremath{-}.24, .23] & .998 & 1.000 \\
 & Pragmatic UX & .06 & [\ensuremath{-}.20, .25] & .585 & .798 \\
 & Hedonic UX & .09 & [\ensuremath{-}.12, .29] & .397 & .783 \\
 & Acceptability & .07 & [\ensuremath{-}.14, .27] & .543 & .783 \\
 & CSQ-VR total & \ensuremath{-}.10 & [\ensuremath{-}.27, .12] & .352 & .783 \\
\end{longtable}\normalsize
Note. The TMT-VR score and the rating were each residualised on age group, within-group age, education and sex, and the residuals were correlated (df~=~85). The bootstrap refitted the adjustment within each of 5,000 participant resamples. BH p~=~Benjamini--Hochberg-adjusted p across the 30 correlations.

\clearpage\section*{Appendix. R code for the model diagnostics}
The code below computes the fit and influence diagnostics, variance inflation factors and predictor correlations of Supplementary Tables S12, S13, S15 and S23 from the processed data set of the repository (\path{01_Data/processed_analysis_data_v04.csv}). It was run in R 4.4.1.

{\footnotesize
\begin{verbatim}
# Diagnostics of the rating and symptom models (Supplementary Tables S12, S13, S15 and S23)
# Usage: Rscript diagnostics_agegroup.R <processed_analysis_data_v04.csv> <output folder>
# Requires R (4.4.1 used); no resampling, so the results are exact.
args <- commandArgs(trailingOnly = TRUE)
DATA <- args[1]; OUT <- args[2]
dir.create(OUT, showWarnings = FALSE, recursive = TRUE)

# 1. The processed data set of the repository, identified by its SHA-256 checksum
sha <- toupper(sub(" .*$", "", system2("sha256sum", shQuote(DATA), stdout = TRUE)))
stopifnot(sha == "380226AEECEABC0BED0C8CB34ED7752D6B8BAEA3397AFBDF48A90486CCD96671")
d <- read.csv(DATA, check.names = FALSE)
stopifnot(nrow(d) == 91)

# 2. Models. Predictors (all centred): OlderC = age group (older = 1, younger = 0);
#    Age10 = age per 10 years; GSQz, Techz, MSSQz = gaming, technology and motion sickness
#    susceptibility per sample SD; EduC = years of education; MaleC = male indicator.
#    The rating models use five predictors; the symptom model adds MSSQz.
ys    <- c("SUS_Total_0_100", "UEQS_Pragmatic", "UEQS_Hedonic", "Acceptability_Mean_1_6")
scale <- list(SUS_Total_0_100 = c(0, 100), UEQS_Pragmatic = c(-3, 3),
              UEQS_Hedonic = c(-3, 3), Acceptability_Mean_1_6 = c(1, 6),
              CSQ_Total_Post = c(6, 42))                  # theoretical range of each score
spec  <- list(age_group      = c("OlderC", "GSQz", "Techz", "EduC", "MaleC"),  # reported
              continuous_age = c("Age10",  "GSQz", "Techz", "EduC", "MaleC"))  # sensitivity
form  <- function(y, x) reformulate(x, response = y)
fit   <- function(y, s) {
  x <- if (y == "CSQ_Total_Post") c("MSSQz", spec[[s]]) else spec[[s]]
  lm(form(y, x), d)
}

# 3. Fit and influence diagnostics of one model
skew <- function(e) { e <- e - mean(e); mean(e^3) / mean(e^2)^1.5 }  # > 0 = right-skewed
diagnose <- function(y, s) {
  m <- fit(y, s); r <- scale[[y]]; f <- fitted(m); cd <- cooks.distance(m)
  data.frame(specification = s, outcome = y, n = nobs(m),
             R2 = summary(m)$r.squared, adjustedR2 = summary(m)$adj.r.squared,
             residual_SD = sigma(m), max_leverage = max(hatvalues(m)),
             max_Cook = max(cd), n_Cook_gt_4_n = sum(cd > 4 / nobs(m)),
             fitted_min = min(f), fitted_max = max(f),
             outside_scale = sum(f < r[1] | f > r[2]),
             residual_skewness = skew(residuals(m)),
             shapiro_p = shapiro.test(residuals(m))$p.value)
}
outcomes <- c(ys, "CSQ_Total_Post")
diag <- do.call(rbind, lapply(names(spec), function(s)
  do.call(rbind, lapply(outcomes, diagnose, s = s))))
write.csv(diag, file.path(OUT, "model_diagnostics_both_specifications.csv"), row.names = FALSE)

# 4. Variance inflation factors: 1 / (1 - R2) of each predictor regressed on the others
vif <- function(pred)
  sapply(pred, function(x) 1 / (1 - summary(lm(form(x, setdiff(pred, x)), d))$r.squared))
v <- do.call(rbind, lapply(names(spec), function(s) rbind(
  data.frame(specification = s, model = "ratings", predictor = spec[[s]],
             VIF = vif(spec[[s]])),
  data.frame(specification = s, model = "CSQ_Total_Post", predictor = c("MSSQz", spec[[s]]),
             VIF = vif(c("MSSQz", spec[[s]]))))))
write.csv(v, file.path(OUT, "VIF_both_specifications.csv"), row.names = FALSE)

# 5. Pearson correlations among the predictors
pv <- c("OlderC", "Age10", "GSQz", "Techz", "EduC", "MaleC", "MSSQz")
write.csv(data.frame(variable = pv, cor(d[pv])), file.path(OUT, "predictor_correlations.csv"),
          row.names = FALSE)
print(diag, digits = 4)
\end{verbatim}}